\documentclass[onefignum,onetabnum]{siamart171218}
\usepackage{lipsum}
\usepackage{amsfonts}
\usepackage{graphicx}
\usepackage{epstopdf}
\usepackage[figuresright]{rotating}
\usepackage{subcaption}
\ifpdf
\DeclareGraphicsExtensions{.eps,.pdf,.png,.jpg}
\else
\DeclareGraphicsExtensions{.eps}
\fi

\usepackage{lipsum}
\usepackage{amsfonts}
\usepackage{graphicx}
\usepackage{epstopdf}
\usepackage[figuresright]{rotating}
\usepackage{subcaption}
\ifpdf
\DeclareGraphicsExtensions{.eps,.pdf,.png,.jpg}
\else
\DeclareGraphicsExtensions{.eps}
\fi

\newsiamremark{remark}{Remark}
\newsiamremark{hypothesis}{Hypothesis}
\crefname{hypothesis}{Hypothesis}{Hypotheses}
\newsiamthm{claim}{Claim}

\title{Massively Parallel Simulations of Binary Black Hole Intermediate-Mass-Ratio Inspirals\thanks{Submitted to the editors 06-27-2018.
		\funding{This work was funded by National Science Foundation grants ACI-1464244, CCF-1643056 and  PHY-1607356. This research used resources of the Oak Ridge Leadership Computing Facility, which is a DOE Office of Science User Facility supported under Contract DE-AC05-00OR22725 and the Extreme Science and Engineering Discovery Environment (XSEDE) allocation TG-PHY180002.}}}

\author{Milinda Fernando\thanks{School of Computing, University of Utah.
		(\email{milinda@cs.utah.edu})}
	\and David Neilsen \thanks{Department of Physics and Astronomy, Brigham Young University.
		(\email{david.neilsen@byu.edu})}
	\and Hyun Lim \thanks{Department of Physics and Astronomy, Brigham Young University.
		(\email{hyun.lim@byu.edu})}
	\and Eric Hirschmann  \thanks{Department of Physics and Astronomy, Brigham Young University.
		(\email{ehirsch@physics.byu.edu})}
	\and Hari Sundar \thanks{School of Computing, University of Utah.
		(\email{hari@cs.utah.edu})}
}

\usepackage[noend]{algpseudocode}
\usepackage{array}
\usepackage{amsfonts}
\usepackage{graphicx}
\usepackage{epstopdf}
\usepackage{amssymb}
\usepackage{tikz,pgfplots}
\usetikzlibrary{snakes,arrows,shapes,trees}
\usepackage{xcolor, graphicx}
\usepackage{amsopn}
\usepackage{listings}
\usepackage{adjustbox}
\usepackage{longtable}
\usepackage{multirow}

\usepackage[draft]{minted}
\usepackage{amssymb}
\usepackage{pgfplots}
\usepackage{pgfplotstable}
\usetikzlibrary{arrows,shapes,plotmarks}
\pgfplotsset{compat=1.8}
\usepackage{soul}
\usepackage{rotating}
\usepackage{tabularx}

\input{default.pygstyle}
\input{default-pyg-prefix.pygstyle}

\definecolor{bgblue}{RGB}{245,243,253}

\newcommand{\dendro}{\textsc{Dendro}}
\newcommand{\dendrogr}{\textsc{Dendro-GR}}
\newcommand{\HAD}{\textsc{HAD}}
\newcommand{\et}{\textsc{ET}}
\newcommand{\bsolver}{\texttt{bssnSolver}}
\newcommand{\BSSN}{BSSNOK}
\newcommand{\NLSM}{NLSM}

\newcommand{\oTo}{\textsc{o2o}}
\newcommand{\oTn}{\textsc{o2n}}

\pgfplotsset{compat=1.13}

\newcommand{\tsort}{\textsc{TreeSort}}
\newcommand{\tsearch}{\textsc{TreeSearch}}

\newcommand{\tghost}{\textsc{ComputeGhostOctants}}
\newcommand{\teToe}{\textsc{BuildOctantToOctant}}
\newcommand{\teTon}{\textsc{BuildOctantToNodal}}

\newcommand{\dgn}{\textit{octant local nodes}}
\newcommand{\cgn}{\textit{shared octant nodes}}
\newcommand{\unzip}{\textit{unzip}}
\newcommand{\zip}{\textit{zip}}
\newcommand{\remesh}{\textit{re-mesh}}
\newcommand{\igt}{\textit{inter-grid transfer}}

\newcommand{\maxDepth}{\textsc{maxdepth}}

\algrenewcommand\algorithmicrequire{\textbf{Input:}}
\algrenewcommand\algorithmicensure{\textbf{Output:}}
\algrenewcommand\algorithmicforall{\textbf{parallel for}}

\newcommand{\Stampede}{\href{https://portal.tacc.utexas.edu/user-guides/stampede2}{Stampede2}}
\newcommand{\Titan}{\href{https://www.olcf.ornl.gov/titan/}{Titan}}
\newcommand{\ET}{\href{https://einsteintoolkit.org/}{\textsc{Einstein Toolkit}}}

\newcommand{\norm}[1]{\left\lVert#1\right\rVert}

\newdimen\HilbertLastX
\newdimen\HilbertLastY
\newcounter{HilbertOrder}

\def\DrawToNext#1#2{%
	\advance \HilbertLastX by #1
	\advance \HilbertLastY by #2
	\pgfpathlineto{\pgfqpoint{\HilbertLastX}{\HilbertLastY}}
}

\def\Hilbert[#1,#2,#3,#4,#5,#6,#7,#8] {
	\ifnum\value{HilbertOrder} > 0%
	\addtocounter{HilbertOrder}{-1}
	\Hilbert[#5,#6,#7,#8,#1,#2,#3,#4]
	\DrawToNext {#1} {#2}
	\Hilbert[#1,#2,#3,#4,#5,#6,#7,#8]
	\DrawToNext {#5} {#6}
	\Hilbert[#1,#2,#3,#4,#5,#6,#7,#8]
	\DrawToNext {#3} {#4}
	\Hilbert[#7,#8,#5,#6,#3,#4,#1,#2]
	\addtocounter{HilbertOrder}{1}
	\fi
}

\def\hilbert((#1,#2),#3){%
	\advance \HilbertLastX by #1
	\advance \HilbertLastY by #2
	\pgfpathmoveto{\pgfqpoint{\HilbertLastX}{\HilbertLastY}}
	\setcounter{HilbertOrder}{#3}
	\Hilbert[1mm,0mm,-1mm,0mm,0mm,1mm,0mm,-1mm]
	\pgfusepath{stroke}%
}

\ifpdf
\DeclareGraphicsExtensions{.eps,.pdf,.png,.jpg}
\else
\DeclareGraphicsExtensions{.eps}
\fi

\definecolor{cpu3}{HTML}{F44336}
\definecolor{cpu4}{HTML}{2196F3}
\definecolor{cpu1}{HTML}{4CAF50}
\definecolor{cpu2}{HTML}{FFC107}
\definecolor{gpu3}{HTML}{EF9A9A}
\definecolor{gpu4}{HTML}{90CAF9}
\definecolor{gpu1}{HTML}{A5D6A7}
\definecolor{gpu2}{HTML}{FFE082}

\definecolor{cpu5}{HTML}{9932CC}

\definecolor{sq_b1}{RGB}{37,52,148}
\definecolor{sq_b2}{RGB}{44,127,184}
\definecolor{sq_b3}{RGB}{65,182,196}
\definecolor{sq_b4}{RGB}{127,205,187}
\definecolor{sq_b5}{RGB}{199,233,180}
\definecolor{sq_b5}{RGB}{255,255,204}

\definecolor{sq_r1}{RGB}{189,0,38}
\definecolor{sq_r2}{RGB}{240,59,32}
\definecolor{sq_r3}{RGB}{253,141,60}
\definecolor{sq_r4}{RGB}{254,178,76}
\definecolor{sq_r5}{RGB}{254,217,118}
\definecolor{sq_r6}{RGB}{255,255,178}

\definecolor{sq_g1}{RGB}{0,104,55}
\definecolor{sq_g2}{RGB}{49,163,84}
\definecolor{sq_g3}{RGB}{120,198,121}
\definecolor{sq_g4}{RGB}{173,221,142}
\definecolor{sq_g5}{RGB}{217,240,163}
\definecolor{sq_g6}{RGB}{255,255,204}

\definecolor{div_c1}{RGB}{230,171,2}
\definecolor{div_c2}{RGB}{102,166,30}
\definecolor{div_c3}{RGB}{231,41,138}
\definecolor{div_c4}{RGB}{117,112,179}
\definecolor{div_c5}{RGB}{217,95,2}
\definecolor{div_c6}{RGB}{27,158,119}
\definecolor{div_c7}{RGB}{215,48,39}

\definecolor{lineclr}{RGB}{0,0,0}
\definecolor{utorange}{RGB}{0,0,255}
\definecolor{utsecblue}{RGB}{255,255,0}
\definecolor{utsecgreen}{RGB}{255,0,0}
\definecolor{red!15}{RGB}{0,255,255}
\definecolor{fillclr5}{RGB}{0,255,0}
\definecolor{fillclr6}{RGB}{255,0,255}
\definecolor{fillclr7}{RGB}{255,255,255}
\definecolor{fillclr8}{RGB}{0,0,0}

\def\drawcubeI(#1,#2,#3,#4,#5){ 
	\coordinate (O) at (#1,#2,#3);
	\coordinate (A) at (#1,#2+#4,#3);
	\coordinate (B) at (#1,#2+#4,#3+#4);
	\coordinate (C) at (#1,#2,#3+#4);
	\coordinate (D) at (#1+#4,#2,#3);
	\coordinate (E) at (#1+#4,#2+#4,#3);
	\coordinate (F) at (#1+#4,#2+#4,#3+#4);
	\coordinate (G) at (#1+#4,#2,#3+#4);
	\draw[#5] (O) -- (C) -- (G) -- (D) -- cycle;
	\draw[#5] (O) -- (A) -- (E) -- (D) -- cycle;
	\draw[#5] (O) -- (A) -- (B) -- (C) -- cycle;
	\draw[#5] (D) -- (E) -- (F) -- (G) -- cycle;
	\draw[#5] (C) -- (B) -- (F) -- (G) -- cycle;
	\draw[#5] (A) -- (B) -- (F) -- (E) -- cycle;
}

\def\drawcubeII(#1,#2,#3,#4,#5,#6,#7){ 
	\coordinate (O) at (#1,#2,#3);
	\coordinate (A) at (#1,#2+#4,#3);
	\coordinate (B) at (#1,#2+#4,#3+#4);
	\coordinate (C) at (#1,#2,#3+#4);
	\coordinate (D) at (#1+#4,#2,#3);
	\coordinate (E) at (#1+#4,#2+#4,#3);
	\coordinate (F) at (#1+#4,#2+#4,#3+#4);
	\coordinate (G) at (#1+#4,#2,#3+#4);
	\draw[#5,fill=#6,opacity=#7] (O) -- (C) -- (G) -- (D) -- cycle;
	\draw[#5,fill=#6,opacity=#7] (O) -- (A) -- (E) -- (D) -- cycle;
	\draw[#5,fill=#6,opacity=#7] (O) -- (A) -- (B) -- (C) -- cycle;
	\draw[#5,fill=#6,opacity=#7] (D) -- (E) -- (F) -- (G) -- cycle;
	\draw[#5,fill=#6,opacity=#7] (C) -- (B) -- (F) -- (G) -- cycle;
	\draw[#5,fill=#6,opacity=#7] (A) -- (B) -- (F) -- (E) -- cycle;
}

\def\drawNodes(#1,#2,#3,#4,#5,#6,#7){ 
	\foreach \x in {#1,#7,...,#2}{
		\foreach \y in {#3,#7,...,#4}{
			\foreach \z in {#5,#7,...,#6}{
				\draw[fill=red!60] (\x,\y,\z) circle (0.15);
			}
		}
	}				
	
}

\makeatletter
\newcommand\resetstackedplots{
	\makeatletter
	\pgfplots@stacked@isfirstplottrue
	\makeatother
	\addplot [forget plot,draw=none] coordinates{(16,0) (32,0) (64,0) (128,0) (256,0) (512,0) (1024,0) (2048,0) (4096,0) (8192,0) (16384,0) (32768,0) (65536,0) (131072,0) (262144,0)};
}
\makeatother

\makeatletter
\newcommand\resetstackedplotsOne{
	\makeatletter
	\pgfplots@stacked@isfirstplottrue
	\makeatother
	\addplot [forget plot,draw=none] coordinates{(16,0) (32,0) (64,0) (128,0) (256,0) (512,0) (1024,0) (2048,0) (4096,0)};
}
\makeatother

\makeatletter
\newcommand\resetstackedplotsTwo{
	\makeatletter
	\pgfplots@stacked@isfirstplottrue
	\makeatother
	\addplot [forget plot,draw=none] coordinates{(32,0) (64,0) (128,0) (256,0) (512,0) (1024,0) (2048,0) (4096,0) (8192,0) (16384,0) (32768,0) (65536,0) };
}
\makeatother

\makeatletter
\newcommand\resetstackedplotsThree{
	\makeatletter
	\pgfplots@stacked@isfirstplottrue
	\makeatother
	\addplot [forget plot,draw=none] coordinates{(2,0) (4,0) (8,0) (16,0) (32,0) (64,0)};
}
\makeatother

\makeatletter
\newcommand\resetstackedplotsFour{
	\makeatletter
	\pgfplots@stacked@isfirstplottrue
	\makeatother
	\addplot [forget plot,draw=none] coordinates{(4,0) (8,0) (16,0) (32,0) (64,0)};
}
\makeatother

\ifpdf
\hypersetup{
	pdftitle={Massively Parallel Simulations of Binary Black Hole Intermediate-Mass-Ratio Inspirals},
}
\fi

\begin{document}
	
	\maketitle
	\begin{abstract}
		We present a highly-scalable framework that targets problems of interest to the numerical relativity and broader astrophysics communities. 
		This framework combines a parallel octree-refined adaptive mesh with a wavelet adaptive multiresolution and a physics module to solve the Einstein equations of general relativity in the \BSSN~formulation. The goal of this work is to perform advanced, massively parallel numerical simulations of Intermediate Mass Ratio Inspirals (IMRIs) of binary black holes with mass ratios on the order of 100:1. These studies will be used to generate waveforms as used in LIGO data analysis and to calibrate semi-analytical approximate methods. 
		Our framework consists of a distributed memory octree-based adaptive meshing framework in conjunction with a node-local code generator. The code generator makes our code portable across different architectures. The equations corresponding to the target application are written in symbolic notation
		and generators for different architectures can be added independent of the application. Additionally, this symbolic interface also makes our code extensible, and as such has been  
		designed to easily accommodate many existing algorithms in astrophysics for plasma dynamics and radiation hydrodynamics. 
		Our adaptive meshing algorithms and data-structures have been optimized for modern architectures with deep memory hierarchies. This enables our framework to have achieve excellent performance and scalability on modern leadership architectures. 
		We 
		demonstrate excellent weak scalability up to $131K$ cores on ORNL's Titan for binary mergers for mass ratios up to $100$. 
	\end{abstract}
	
	
	
	\section{Introduction} 
	\label{sec:introduction}
	
	\begin{figure}[t]
		\includegraphics[width=1.0\textwidth]{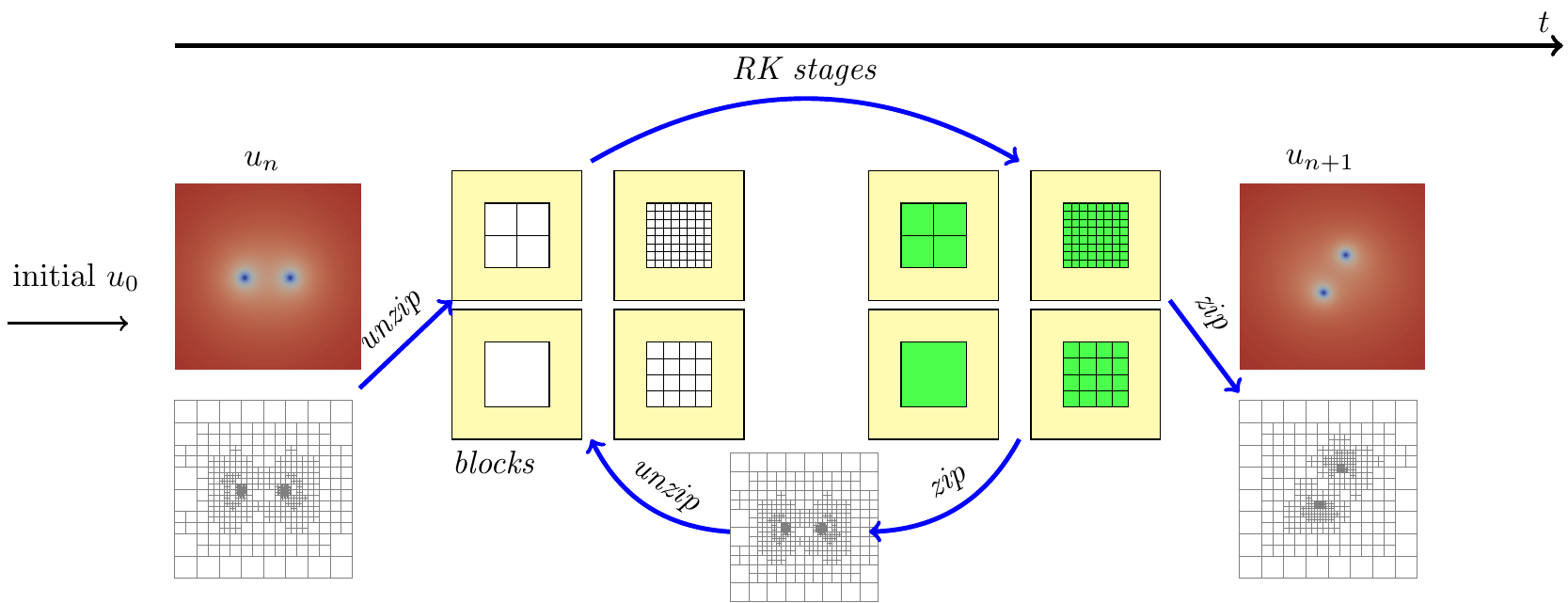}
		\caption{\small This figure illustrates the calculation of
			a single Runge-Kutta
			time step, computing the solution at the advanced time, $u_{n+1}$, 
			from data at the previous time step, $u_{n}$. For computational 
			efficiency, spatial and time derivatives are evaluated on equispaced
			blocks \textit{(unzipped)}; 
			a sparse grid constructed from wavelet coefficients is used
			for communication and to store the final solution \textit{(zipped)}.
			For each RK stage $s$ we perform the \textit{unzip} operation which results in a sequence of blocks which are used to compute the solution on the internal block (\textcolor{green!70}{$\blacksquare$}), using the padding values at the block boundary (\textcolor{yellow!30}{$\blacksquare$}) followed by a \textit{zip} operation in between RK stages while the final update (i.e. next time step)  performed using the \textit{zip} version of the variables. Note that the re-meshing is performed as needed based on the wavelet expansion of the current solution (see \S\ref{sec:alltogether}).  \label{fig:overview}}
		\vspace{-0.2in} 
	\end{figure}

	In 2015, shortly after beginning its first observing run, the Laser
	Interferometer Gravitational-Wave Observatory (LIGO)~\cite{LIGOweb,LIGOtdr} 
	made the first direct
	detection of gravitational waves from the
	merger of two black holes~\cite{Abbott:2016blz}.  
	Since that time, gravitational waves from four other 
	binary black hole mergers~\cite{PhysRevLett.116.061102,PhysRevLett.116.241103,Abbott:2017vtc,Abbott:2017oio}
	have been detected by 
	LIGO and the European Virgo detectors~\cite{VIRGOweb,TheVirgo:2014hva}.
	In August 2017, LIGO and Virgo detected gravitational waves from
	the merger of a neutron star binary~\cite{TheLIGOScientific:2017qsa}.  This 
	latter detection was particularly exciting because electromagnetic radiation
	from the resulting gamma-ray burst was detected by the Fermi Gamma-Ray
	Burst Monitor~\cite{Goldstein:2017mmi} and INTEGRAL~\cite{Savchenko:2017ffs},
	as well as by several other observatories~\cite{Monitor:2017mdv}.
	The combination of gravitational and
	electromagnetic observations of binary mergers will give new insight into the
	physics of black holes (BHs) and neutron 
	stars~\cite{GBM:2017lvd,ANTARES:2017bia,Abbott:2017wuw}.
	As the sensitivity of the LIGO detectors improves, gravitational wave
	detections will increase in frequency 
	and open a new era of gravitational wave astronomy.
	
	Gravitational waves carry the imprint of their origins within the complicated
	pattern of their waveform.  
	The information therein can be untangled through
	a careful comparison of the gravitational wave signal with a library of possible
	waveforms constructed using approximate methods and
	results from numerical simulations. 
	Indeed, waveform information from numerical
	relativity is particularly important for certain binary black hole   
	configurations. Examples include binary black holes with 
	arbitrary spin configurations~\cite{Babak:2016tgq}, binaries with orbital eccentricity,
	and binaries for which the black holes have very different 
	masses~\cite{Graff:2015bba,Keitel:2016krm}. 

	In this paper, we use $q$ to denote the mass ratio of a binary as 
	$q \equiv m_1/m_2$, where $m_1 \ge m_2$.
	At this time, very few large mass-ratio BH binaries ($q \gg 1$) have been
	studied in numerical relativity, compared to studies with nearly equal 
	mass ($q\approx 1$)~\cite{Lehner:2014asa,Sperhake:2014wpa}.
	Codes developed for $q\approx 1$ binaries are accurate and well tuned, so the problem
	is well-understood and numerical results are confidently used in the LIGO 
	data-analysis pipeline.  However, configurations with large $q$ 
	remain largely beyond the capabilities of current techniques in numerical 
	relativity.  Examples include Intermediate Mass-Ratio Inspiral (IMRI) binaries 
	and are characterized roughly by $q \simeq 100$.  It is estimated 
	that about 5\% of the detections in LIGO might come from 
	IMRIs~\cite{Abadie:2010cf,Fregeau:2006yz}.

	For an IMRI, the size of the smaller black hole adds an
	extra length-scale to the problem, compared to the $q\approx 1$ case. 
	The need to resolve this scale, together with the 
	large range other important length scales for the binary system, 
	makes this a very challenging computational problem.
	It requires a highly adaptive and efficient computational algorithm tuned to handle binaries with large mass ratios.
	While BH evolutions with $q=100$ have been 
	performed~\cite{Lousto:2010ut,Lousto:2010tb,Sperhake:2011ik} previously, they were not completed till the merger event or simulated direct head-on collisions. Therefore the above simulations are not satisfactory to be useful toward gravitational wave analysis.
	
	{\em 
	
		The central goal of our effort is to create a general purpose framework to study the evolution of spacetimes with black holes or neutron stars, including binary black holes with large mass ratios up to $q\simeq 100$. 
	}
	Here we present our portable,
	highly-scalable, extensible, and easy-to-use framework for
	general relativity (GR) simulations that will be forward-compatible with
	next-generation heterogeneous clusters. 
	
	We build on our octree-based adaptive mesh refinement (AMR) framework \dendro~ \cite{SundarSampathBiros08, SampathAdavaniSundarEtAl08} 
	to support Wavelet Adaptive Multiresolution 
	(WAMR)~\cite{Paolucci1,Paolucci2,DeBuhr:2015jqk}. 
	The fast wavelet transform can be used to create a
	sparse representation of functions that retains sharp features and an
	\textit{a priori} error bound.
	
	The high-level overview of our approach is illustrated in Figure \ref{fig:overview}.  
	We use an efficient block-decomposition of the distributed octree to produce a collection of overlapping regular grids (at different levels of refinement, see \S\ref{sec:unzip_and_zip} ). 
	
	The Einstein equations of general relativity describe the spacetime
	geometry and, expressed in terms of the \BSSN~formulation~\cite{1995PhRvD..52.5428S,1987PThPS..90....1N}, where each spatial grid point is associated with $24$ unknowns. 
	Given their complexity and a desire for portable code, we auto-generate the core computational kernels automatically from the equations written in symbolic Python (\texttt{SymPy}~ \cite{joyner2012open}, see \S\ref{sec:symbolic}).  The auto-generated code is applied at the block-level and is therefore very efficient and
	enables portability. The equations are integrated in time using the method of lines with a Runge-Kutta (RK) integrator\footnote{Currently $3$rd and $4$th order RK are supported.}.

	\noindent The key \ul{contributions} of this work include:
	
	\vspace{0.1in}    
	\noindent \textbf{Wavelet Adaptive GR}.
	To the best of our knowledge, comparable codes use simple models of adaptivity, i.e.,
	structured adaptivity, block adaptivity, or logically uniform grids \cite{EINSTEIN_TOOLKIT,Neilsen:2014hha,Bruegmann:2006at,Yamamoto:2008js}.	
	We present a novel computational GR framework (\dendrogr)  which uses octree-based Adaptive Multiresolution (AMR) grids, where the adaptivity is guided by wavelet expansion \cite{Vasilyev1995,Vasilyev1996,Paolucci1,Paolucci2,Holmstrom1999}
	of the functions represented in the underlying grid. We refer this as Wavelet Adaptive Multiresolution (WAMR). This is the first \ul{highly adaptive} fully relativistic---i.e., including the full Einstein equations---code with an arbitrary localized 
	adaptive mesh. For example for a mass ratio of $q=1$, we use approximately \texttt{7x} fewer degrees of freedom for the same accuracy compared to the block adaptivity (via Carpet \cite{CARPET}). 
	\\

	\noindent \textbf{Automatic code generation}. Given the complexity of the Einstein equations, we have developed an automatic code generation framework for GR using \texttt{SymPy} that automatically generates architecture-optimized codes. This greatly improves code portability, use by domain scientists and the ability to add additional constraints and checks to validate the code. \\
	
	\noindent \textbf{Performance}. We developed a new parallel search algorithm \tsearch\ (see \S\ref{sec:meshing}),  to improve the efficiency of octree meshing. We also developed efficient \texttt{unzip} and \texttt{zip} operations (block-decomposition of WAMR grid to produce collection of overlapping regular grids, see \S\ref{sec:unzip_and_zip}) to allow the application of the core computational kernels to small process-local regular blocks. This improves performance as well as performance portability in conjunction with our automatic code generation. Local calculations on regular blocks allows us to use established, existing numerical methods for the Einstein equations, and in future work, the relativistic fluid equations and radiation hydrodynamics equations. \\
	
	\noindent \textbf{Simulations}. We demonstrate the ability to scale to large mass ratios, enabling simulations and extraction of gravitational waves for mass ratios as high as $100$. \\
	
	\noindent \textbf{Implementation} \dendrogr ~is implemented in \textsc{C++} using MPI except for the automatic code generation framework which is implemented using \texttt{SymPy}. Our code is freely available at \texttt{https://github.com/paralab/Dendro-GR} under the MIT license. \\

	\par \textit{Organization of the paper:} The rest of the paper is organized as follows. In \S\ref{sec:background}, we give a brief motivation on the importance of numerical simulations of BH binary configurations and a quick overview of the existing state-of-the-art approaches in the field of numerical relativity. In \S\ref{sec:methodology}, we present the methodology used in \dendrogr~ in detail and how it is efficient compared to the existing approaches. In \S\ref{sec:results}, we discuss the experimental setup, strong and weak scalability of our approach, including a detailed comparison study with the state-of-the-art Einstein Toolkit \cite{EINSTEIN_TOOLKIT} package. In \S\ref{sec:conclusion}, we conclude with directions for future work. 

	
	\section{Background} 
	\label{sec:background}
	
	In this section, we discuss the motivating applications and summarize
	the most relevant work of other groups in this area.
	As gravitational waves pass through the Earth, their effect on matter is
	extremely small.  LIGO searches for gravitational waves by using laser
	interferometry 
	to detect changes in the relative position 
	of mirrors, to a precision of
	four orders of magnitude 
	smaller than an atomic nucleus. 
	The gravitational wavesignals in the
	detector are often smaller in magnitude than noise from other sources, but
	the signals can be extracted using 
	matched filtering~\cite{Sathyaprakash:1991mt}, which
	uses a library of hundreds of candidate waveforms that are convolved with
	the data. 
	Including complete numerical waveforms in the waveform library is 
	important to maximize the detection
	rate of IMRIs in LIGO-class detectors~\cite{PhysRevD.88.044010}.
	
	\begin{figure}[tbh]
		\centering
		\includegraphics[width=0.45\textwidth,height=0.18\textheight]{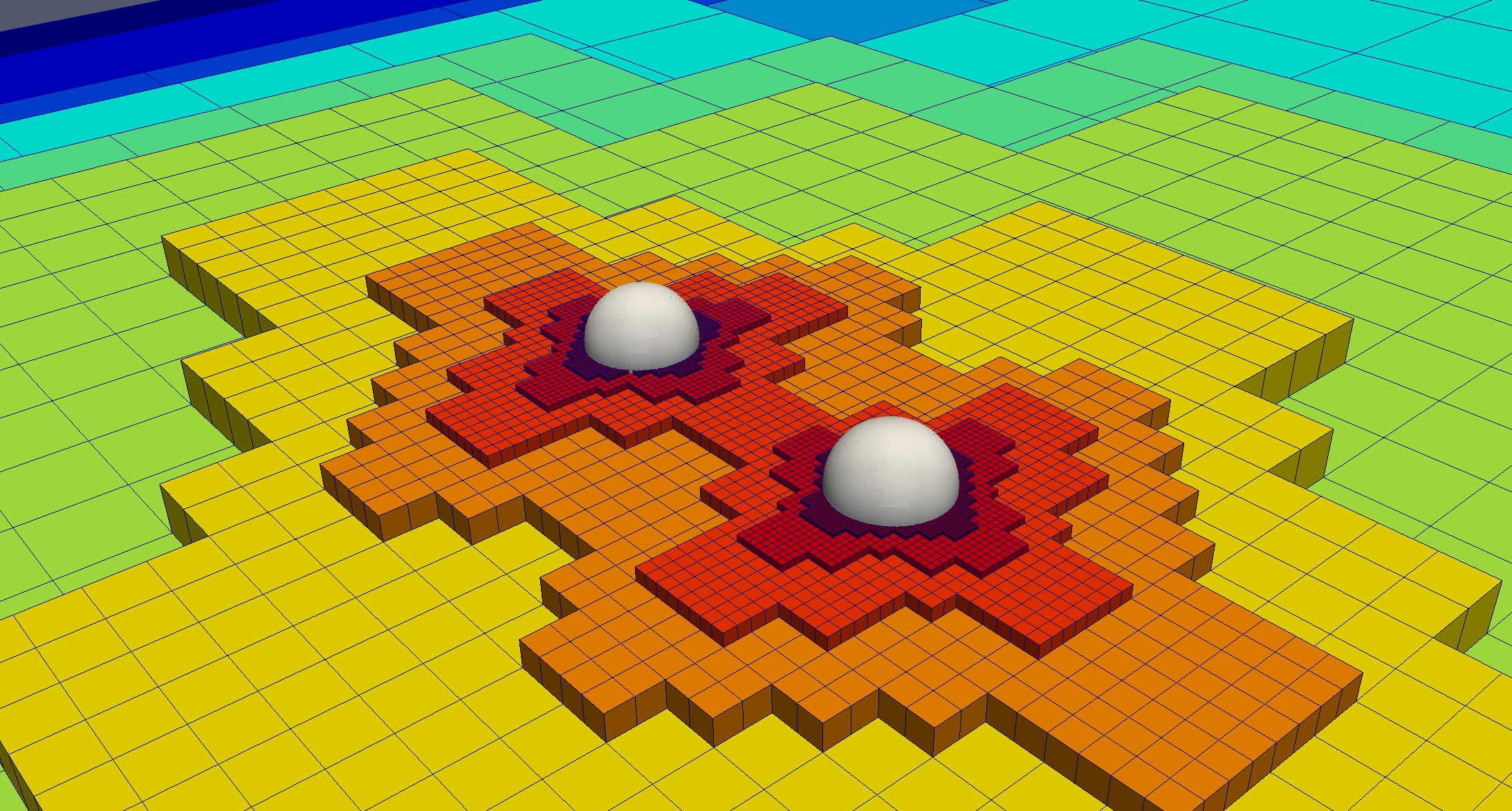}
		\includegraphics[trim={10cm 6cm 10cm 8cm},clip,height=0.18\textheight,width=0.45\textwidth]{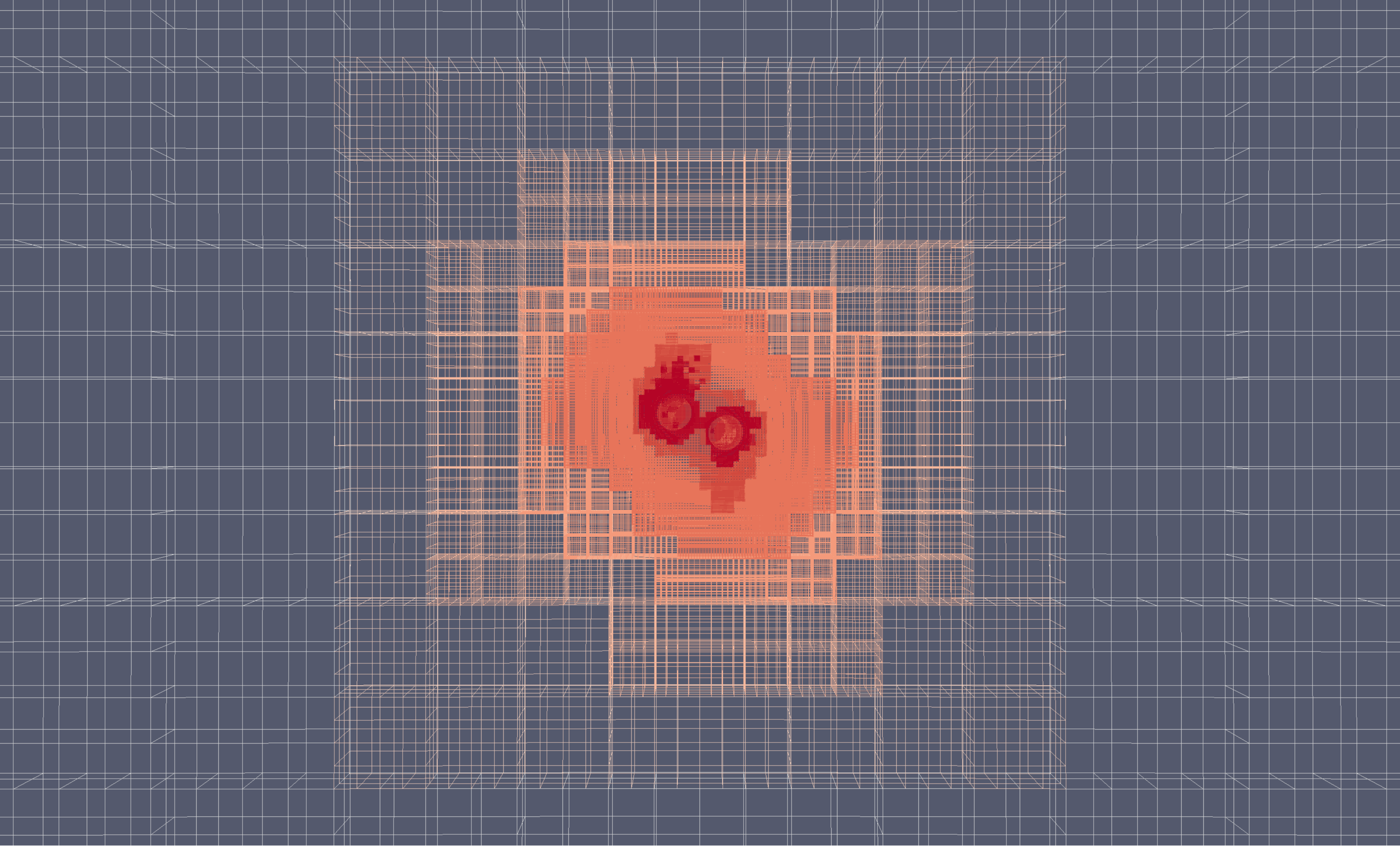}
		\caption{\label{fig:bhole} \small ({\bf left}) A example of the adaptive mesh created by \dendro ~for the binary black-hole system. ({\bf right}) the hierarchical wavelet grids generated for the binary black hole system. }
		\label{f:bh_image1}
		\vspace{-0.2in}
	\end{figure}

	The Einstein equations of general relativity describe how the geometry
	of spacetime curves in response to the presence and motion of matter and
	energy.
	The Einstein equations contain both hyperbolic evolution equations
	and elliptic constraint equations. Commonly the hyperbolic equations
	are solved, and the elliptic equations are used to "monitor" the quality
	of the solution~\cite{Alcubierre:1138167,Shibata:2015:NR:2904075} (see Appendix \ref{sec:bssn}). The solution at time $t$ for evolution equations 
	should satisfy Hamiltonian and momentum constraints (i.e. compute the $l_2$ norm of the constraint violation) in order to verify the solution is physically valid.  
	While the equations can be formulated in many different ways,
	few formulations are well-suited for numerical work.
	One such formulation is the \BSSN ~formulation~\cite{Baumgarte:1998te}.
	The \BSSN\ evolution equations are a set of 
	strongly hyperbolic~\cite{HyperbolicBSSNKO} 
	coupled PDEs that are first-order in time and second-order in space. A brief summary of the \BSSN~ with constraint equations is provided in the Appendix \ref{sec:bssn}.

	Several computer codes have been developed to solve the Einstein equations
	for binary BH and neutron star systems. 
	One of the oldest open source projects in this community
	is the Cactus Computational Toolkit~\cite{CACTUS,Cactus_Goodale03a}, 
	that provides a modular infrastructure for solving
	time-dependent PDEs in parallel using structured 
	grids. Modules for specific tasks, known as \textit{thorns} in Cactus parlance, 
	can be shared and combined to produce a sophisticated evolution code.
	The \ET\ (\et) is a 
	suite of community-developed thorns for relativistic 
	physics~\cite{EINSTEIN_TOOLKIT}.
	
	It includes thorns for constructing binary BH 
	initial data, for evolving the Einstein equations and/or the relativistic fluid 
	equations, and for data analysis.  
	Similar codes include~\cite{Thierfelder:2011yi,Kiuchi:2012qv,Kiuchi:2014hja,Neilsen:2014hha,Etienne:2015cea}.
	Further, the SXS collaboration has developed SpEC~\cite{Szilagyi:2009qz}, 
	a spectral code for solving the Einstein equations that has
	produced the longest and most-accurate binary waveforms to date.
	
	The challenge of running on modern massively parallel computers is pushing
	new developments in numerical relativity. The use of structured grids with 
	block-based AMR, such as used by Cactus/\et\ and similar codes, 
	is not ideal for new massively parallel architectures and 
	can lead to inefficient refinement (refined in the regions where coarser representation is sufficient), especially for 
	$q \gg 1$.
	One new approach for the \et\  is the 
	SENR project~\cite{Ruchlin:2017com,Mewes:2018szi},
	that uses coordinate systems adapted to the binary BHs to eliminate the need
	for AMR. Another approach is to use discontinuous Galerkin (DG) methods, 
	that requires less communication between processes.
	The first three-dimensional ADER-DG simulations of the 
	Einstein equations were 
	performed by Dumbser et al.~\cite{Dumbser:2017okk}. 
	
	The SXS collaboration is developing the 
	SpECTRE code~\cite{Kidder:2016hev,Roberts:2016lzn} that uses
	task-based parallelism and DG. Thus far only results for the relativistic MHD
	equations 
	have been published.
	
	We have chosen to focus on one type of BH 
	binary that is particularly difficult to study both numerically
	and with semi-analytical approximations.
	These are IMRIs, BH 
	binaries with mass ratios with $50 \lesssim q \lesssim 1000$.
	The successful numerical simulation of IMRIs
	and their predicted gravitational
	wave signal is difficult because of the large difference in the two mass-scales in the problem. 
	Gravitational waves must be resolvable far from
	the binary system while the region around both black holes must also be
	accurately simulated.  Standard approaches \cite{EINSTEIN_TOOLKIT,Neilsen:2014hha,Bruegmann:2006at,Yamamoto:2008js} to black hole simulations often
	include mesh adaptivity by which necessary resolution is concentrated
	in dynamic regions. 
	
	The \dendrogr ~code uses octree grids based on
	the wavelet expansion~\cite{Vasilyev1995,Vasilyev1996,Paolucci1,Paolucci2,Holmstrom1999}, that produces refinement regions adapted
	to features in the solution with a minimum number of points. 
	This is 
	important for problems with fine-scale features that are not spatially
	localized (i.e. adaptivity of the grid is not pre-determined based on the spatial locations of BHs), or problems with widely disparate scales, such as IMRIs. 
	Moreover, 
	the numerical methods used in this paper are based on the conventional finite difference
	methods that have been widely used and tested (see, \S\ref{sec:nm}). This allows
	existing numerical approaches to be more easily adapted to the \dendrogr~\ through symbolic code generation framework.
	Given the scale of our problem, even with adaptivity, massively parallel
	computing resources are required. We build our \dendrogr~ framework based on our parallel adaptive meshing framework
	\dendro~\cite{SundarSampathAdavaniEtAl07,SundarSampathBiros08} and extend it to support numerical relativity codes with finite differencing. 
	
	A key reason to develop scalable codes is that
	as the relative differences in masses becomes larger ($\sim 100\times$), the computational
	requirements will grow significantly, potentially requiring exascale resources.
	A simple calculation illustrates how spatial resolution requirements increase
	with 
	$q$.  A convenient measure for the size of a black hole
	is the radius of its event horizon, which is proportional to its mass.
	In a black hole binary, the mass of the smallest black hole effectively sets
	the minimum length scale for the simulation. The total mass of the binary
	$M = m_1 + m_2$ is a global scaling parameter and is typically fixed to 
	a constant value. 
	Then the mass of the smaller black hole can
	be written $m_2 = M/(q+1)$, showing that the minimum resolution scale for the
	binary is inversely proportional to the mass ratio. 
	In three spatial dimensions
	the number of points required to resolve the smallest black hole
	grows as $q^3$.
	This presents both a challenging problem in computational relativity
	as well as a challenge for high-performance computing.
	The successful scaling of our code is a first step in this direction.

	\section{Methodology} 
	\label{sec:methodology}
	
	
	
	Research in relativistic astrophysics requires
	specialized computational models for gravitational, plasma,
	and nuclear physics.
	The massively parallel infrastructure that we propose 
	is compatible with the 
	standard finite difference or finite volume discretizations 
	that are currently used in these communities.
	We solve the \BSSN ~equations using conventional 
	finite-difference discretizations, 
	standard gauges, and puncture initial data (see \S\ref{sec:nm}). We 
	also adapt them to specific computer hardware and cache sizes
	using our new symbolic interface (see \S\ref{sec:symbolic}). 
	While this paper focuses on the vacuum Einstein equations,
	we are currently working to add modules for the relativistic MHD equations, 
	nuclear equations of state, and neutrino leakage.
	

	
	\subsection{Numerical Methods}
	\label{sec:nm}
	
	There is extensive literature on solving the \linebreak \BSSN\ equations in
	general relativity, and some general reviews 
	include~\cite{Alcubierre:1138167,Shibata:2015:NR:2904075,Rezzolla_book}.
	In this section we briefly outline our particular choices for 
	solving \BSSN~ equations.
	We write the \BSSN\ equations in terms of the conformal 
	factor $\chi$~\cite{lousto}. We use the parameterization of the
	``$1+\log$'' slicing condition and the $\Gamma$-driver shift used 
	in~\cite{Neilsen:2014hha}.
	Spatial derivatives are calculated using
	finite difference operators that are ${\rm O}(h^4)$ in the grid spacing, $h$, 
	with upwind derivatives for Lie derivative terms~\cite{Zlochower:2005bj}. 
	We calculate derivatives for the Ricci tensor and enforce the algebraic
	constraints as described in~\cite{Bruegmann:2006at}. 
	Outgoing radiative boundary conditions are applied to each \BSSN\ function.
	The \BSSN\ equations are integrated in time using an explicit RK
	scheme. The solution at each point is integrated
	with a single global times step, that is set by the smallest grid spacing
	and the Courant condition~\cite{CFL}. While we support $3$rd and $4$ order RK, the tests in this paper were done 
	using $3$rd order RK with Courant factor $\lambda=0.1$. 
	Kreiss-Oliger dissipation is added~\cite{kodissipation,Alcubierre:1138167} 
	to the solution to eliminate high-frequency noise that might be generated 
	near the black hole singularities.

	\subsection{Wavelet Adaptive Multiresolution}
	\label{sec:wamr}
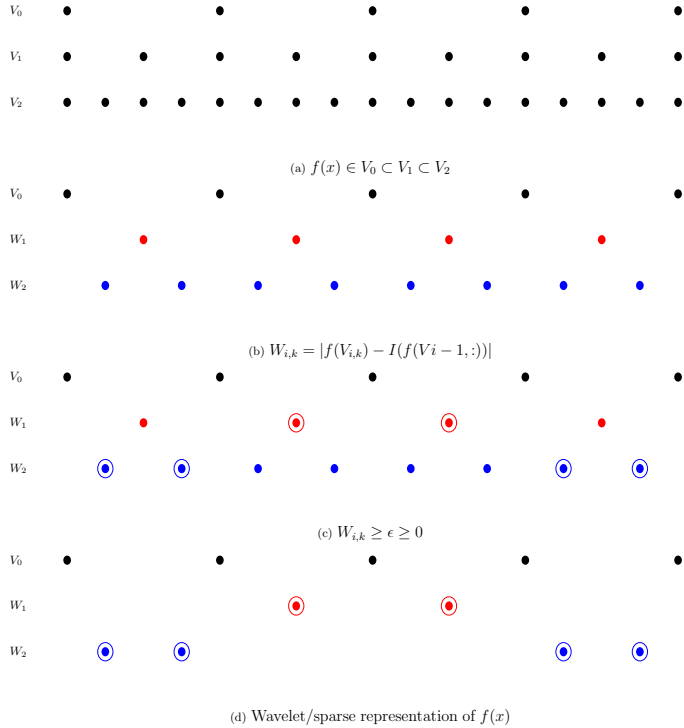
\begin{figure}
	\centering
\resizebox{0.7\textwidth}{!}{
	\begin{tikzpicture}[yscale=-1.2]
	\begin{scope}
	\foreach \y in {0,1,2}
	\node[text width=1cm] at (-1,\y) {\small $V_\y$};
	\foreach \x in {0,4,8,12,16}
	\fill (\x,0) circle (0.1cm);
	
	\foreach \x in {0,2,4,6,8,10,12,14,16}
	\fill (\x,1) circle (0.1cm);		
	
	\foreach \x in {0,1,2,3,4,5,6,7,8,9,10,11,12,13,14,15,16}
	\fill (\x,2) circle (0.1cm);
	
	\node[text width=8cm,align=center] at (8,3.5) {\subcaption{\large $f(x) \in V_0 \subset V_1 \subset V_2$ \label{fig:w:a}} };
	\end{scope}
	
	\begin{scope}[yshift=4cm]
	
	\node[text width=1cm] at (-1,0) {\small $V_0$};
	
	\foreach \y in {1,2}
	\node[text width=1cm] at (-1,\y) {\small $W_\y$};
	\foreach \x in {0,4,8,12,16}
	\fill (\x,0) circle (0.1cm);
	
	\foreach \x in {2,6,10,14}
	\fill[red] (\x,1) circle (0.1cm);		
	
	\foreach \x in {1,3,5,7,9,11,13,15}
	\fill[blue] (\x,2) circle (0.1cm);
	
	\node[text width=8cm,align=center] at (8,3.5) {\subcaption{ \large $W_{i,k}=|f(V_{i,k})-I(f(V{i-1,:}))|$ \label{fig:w:b}}};
	
	\end{scope}
	
	\begin{scope}[yshift=8cm]
	
	\node[text width=1cm] at (-1,0) {\small $V_0$};
	
	\foreach \y in {1,2}
	\node[text width=1cm] at (-1,\y) {\small $W_\y$};
	\foreach \x in {0,4,8,12,16}
	\fill (\x,0) circle (0.1cm);
	
	\foreach \x in {2,6,10,14}
	\fill[red] (\x,1) circle (0.1cm);		
	
	\draw[red] (6,1) circle (0.2cm);
	\draw[red] (10,1) circle (0.2cm);
	
	\foreach \x in {1,3,5,7,9,11,13,15}
	\fill[blue] (\x,2) circle (0.1cm);
	
	\draw[blue] (1,2) circle (0.2cm);
	\draw[blue] (3,2) circle (0.2cm);
	\draw[blue] (13,2) circle (0.2cm);
	\draw[blue] (15,2) circle (0.2cm);
	
	\node[text width=8cm,align=center] at (8,3.5) {\subcaption{\large $W_{i,k}\geq \epsilon \geq 0$ \label{fig:w:c}}};
	
	\end{scope}
	
	\begin{scope}[yshift=12cm]
	
	\node[text width=1cm] at (-1,0) {\small $V_0$};
	
	\foreach \y in {1,2}
	\node[text width=1cm] at (-1,\y) {\small $W_\y$};
	\foreach \x in {0,4,8,12,16}
	\fill (\x,0) circle (0.1cm);
	
	\foreach \x in {6,10}
	\fill[red] (\x,1) circle (0.1cm);		
	
	\draw[red] (6,1) circle (0.2cm);
	\draw[red] (10,1) circle (0.2cm);
	
	\foreach \x in {1,3,13,15}
	\fill[blue] (\x,2) circle (0.1cm);
	
	\draw[blue] (1,2) circle (0.2cm);
	\draw[blue] (3,2) circle (0.2cm);
	\draw[blue] (13,2) circle (0.2cm);
	\draw[blue] (15,2) circle (0.2cm);
	
	\node[text width=8cm,align=center] at (8,3.5) {\subcaption{\large Wavelet/sparse representation of $f(x)$ \label{fig:w:d}}};
	
	\end{scope}
	\end{tikzpicture}
}
		\caption{\small For a given function $f:V\rightarrow \mathcal{R}$ let $V_i \subset V$ be the finite dimensional approximation of $f$ (see Figure \ref{fig:w:a}). As number of nodes increases (i.e. going from $V_i$ to $V_{i+1}$) for each additional node introduced, we compute wavelet coefficients based on the absolute difference between $f(V_{i,k})$ and interpolated value from previous level $f(V_{i-1,:})$ (see Figure \ref{fig:w:b}). In Figure \ref{fig:w:c} shows the chosen nodes that violate specified wavelet tolerance $epsilon$ and these nodal wavelets are stored as the sparse/wavelet representation of function $f$ (see Figure \ref{fig:w:d}).\label{fig:wavelets}}
	\end{figure}

	WAMR uses a basis of interpolating wavelets to create
	a sparse, quasi-structured grid  that naturally adapts to the features of the 
	solution~\cite{Paolucci1,Paolucci2,DeBuhr:2015jqk}.
	This grid adaptivity is realized by expanding functions using the fast wavelet transform~\cite{Holmstrom1999},
	and thresholding (i.e. based on a user-specified tolerance $\epsilon > 0$) the solution to create a sparse representation
	that retains small-scale features~\cite{Donoho1992}. In WAMR, we start with the coarsest representation ($V_0$) of a given function $f$.  Then we compute wavelet coefficients based on the interpolation error as a result of the interpolation of $f$ from the coarser representation to the next finer representation ($V_1$). Hence the wavelet coefficient denotes the interpolation error that occurs when $f$ is constructed from the immediate coarser level. The sparse representation of the function $f$ is computed based on the user-specified tolerance and removing the spatial points whose wavelet coefficients are within the specified threshold (see Figure \ref{fig:wavelets}). 
	
	Wavelet basis functions are localized (i.e. have compact support) both spatially and with respect to scale.
	In comparison, spectral bases are infinitely differentiable, but have global support; basis functions used in
	finite difference or finite element methods have small compact support, but poor
	continuity properties. As an example, in Figure~\ref{f:bh_image1} we show
	a binary black hole spacetime generated with WAMR using \dendrogr. 
	

	Wavelets encode solution features at different scales 
	very efficiently, a characteristic that leads to many applications in
	data and image compression~\cite{Akansu}.  Studies
	of WAMR 
	have shown
	the method to be significantly more efficient in terms of computational cost
	when compared with traditional numerical schemes~\cite{Rastigejev}.
	The wavelet amplitudes also provide a direct measure of the
	local approximation error and serve as a refinement
	criterion. 
	We work simultaneously with both the point and wavelet 
	representations~\cite{Bertoluzza1996, Vasilyev2000, Regele2009,
		Vasilyev1995,Vasilyev1996, Vasilyev1997}.
	This gives wavelet methods some of the same advantages
	as DG~\cite{Teukolsky:2015ega,Bugner:2015gqa}, 
	including exponential convergence.  Combining the sparse 
	grid generated by the truncated wavelet expansion
	with 
	\dendro\ 
	yields a wavelet adaptive multiresolution 
	method that enables a promising improvement for simulating the mergers of compact 
	object binaries. 
	

	\subsection{Computational Framework}
	\label{sec:dendro}
	
	We now give 
	an overview of the approach that \dendrogr ~takes in order to obtain 
	excellent scalability in the context of the WAMR 
	method.  
	%
	Our parallel WAMR framework is based on adaptive spatial octrees \cite{SundarBirosBurstedde12, Fernando:2017} where the adaptivity is determined by the hierarchical computation of wavelet coefficients and a user-specified tolerance. 
	The construction of adaptive octrees is similar to other octree-codes 
	such that every element at level $l$ 
	gets replaced by eight finer (smaller) elements (i.e. level $l+1$) if the computed wavelet coefficient is larger than the user-specified tolerance. The main steps in building the parallel octree-WAMR framework are partitioning, construction and enforcement of constraints on the relative sizes of neighboring octants, and meshing. By meshing, we refer to the process of building required data structures to perform numerical computations on a topological octree (see \S \ref{sec:meshing}). 
	
	\subsubsection{Preliminaries} 
	Octree based spatial subdivisions are fairly common in computational science applications \cite{BursteddeWilcoxGhattas11,SampathAdavaniSundarEtAl08,AhimianLashukVeerapaneniEtAl10,peanoPaper,bonsai}, due to their simplicity and scalability. 
	Here we present some basic concepts and notation related to octrees used in this paper. A distributed octree $\mathcal{T}$ consists of $p$ subtrees $\tau_i, i=1,\ldots,p$, where $p$ is the number of processes and $\mathcal{T}=\cup\tau_i$. For an octant $e$, $F(e)$ denotes the faces, $E(e)$ denotes the edges and $V(e)$ denotes the vertices of $e$. The neighbors of $e$ are given by $N(e) = N_F(e) \cup N_E(e) \cup N_V(e)$, where $N_F(e)$ 
	denotes the octants that share only a face, $N_E(e)$ denotes the octants 
	that share only an edge and $N_V(e)$ denotes the octants that share only a 
	vertex with $e$. If $\tau_i$ spans the sub-domain $\omega_i \subset \Omega$, the boundary octant set, $bdy(\tau_i)$, consists of those octants that share 
	faces, edges and vertices with $\partial \omega_i$.  
	Correspondingly, the set of interior octants
	is given by $int(\tau_i)=\tau_i \setminus bdy(\tau_i)$. Finally, the octree $\tau$ is said to 
	be $2:1$ balanced if and only if for any $e_k\in \tau$ where $level(e_k)=l_k$, $\forall e \in N(e_k)$ then $level(e)=max(l_k\pm 1,0)$. In this work, we enforce a $2:1$ balance constraint on our octrees.
	
	\subsubsection{Octree partitioning} 
	The problem size or the local number of octants varies significantly during WAMR based octree construction (in WAMR we start with the coarsest representation and add grid points until all the wavelet coefficients are within the specified tolerance ), balancing, meshing and during the simulation as well.
	
\begin{figure}
		\centering
		\includegraphics[width=\textwidth]{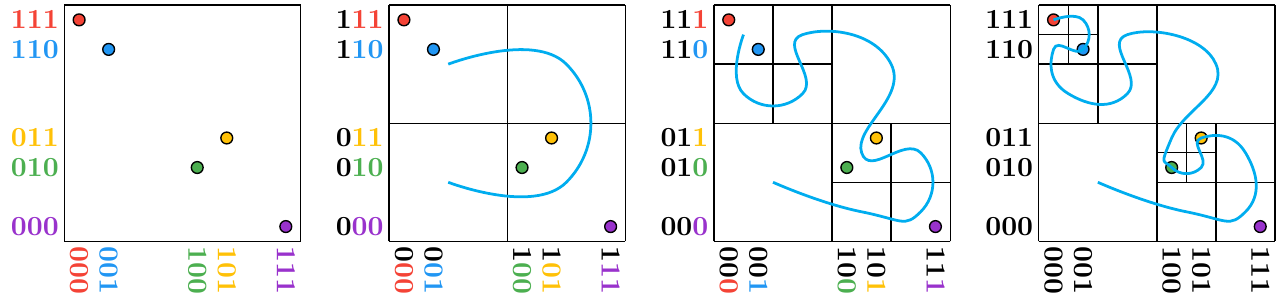}
		\caption{\label{fig:tsort} Bucketing each point and reordering the buckets based on the SFC ordering at each level $l$ with top-down traversal. Each color-coded point is represented by its $x$ and $y$ coordinates. From the MSD-Radix perspective, we start with the most-significant bit for both the $x$ and $y$ coordinates and progressively bucket (order) the points based on these. The bits are colored based on the points and turn black as they get used to (partially) order the points.}
	\end{figure}

	This necessitates efficient partitioning of the octree to make it load balanced. We use a space-filling curve (SFC) \cite{sfcIntro} based partitioning scheme \cite{Fernando:2017}, specifically the 
	Hilbert-curve.  
	An SFC specifies a surjective mapping from the one-dimensional space to higher dimensional space. This can be used to enforce an SFC based ordering operator on higher dimensional space. The Hilbert ordering maps higher dimensional data (octants) to a 1D curve which makes the process of partitioning trivial.  The key challenge is to order the octants or regions according to the SFC, usually performed using an ordering function and sorting algorithm. This approach is easily parallelized using efficient parallel sorting algorithms such
	as SampleSort \cite{Frazer:1970:SSA:321592.321600} and BitonicSort \cite{bitonicSort}  which is the approach used by several state-of-the-art packages \cite{BursteddeWilcoxGhattas11,SundarSampathBiros08,SundarSampathAdavaniEtAl07}. We use a comparison-free SFC sorting algorithm \tsort, based on the radix sort. 
	In \tsort, we start with the \textit{root} octant and hierarchically split each dimension while bucketing points for each octant and reordering the buckets, based on the specified SFC ordering (see Figure \ref{fig:tsort}). This is performed recursively on depth-first traversal until we reach all the leaf nodes (see Algorithm \ref{alg:tsort}). Additional details on our partitioning algorithm can be found in \cite{Fernando:2017}.

	\begin{algorithm}[t]
		\caption{\tsort}\label{alg:tsort}
		\footnotesize
		\begin{algorithmic}[1]
			\Require A list of points or regions $W$, the starting level $l_1$ and the ending level $l_2$
			\Ensure $W$ is reordered according to the SFC.
			\State $counts [] \leftarrow 0$
			\Comment $|counts| = 2^{d}$, 8 for $3D$
			\For{$w \in W$}
			\State increment $counts[child\_num(w)]$
			\EndFor
			\State $counts [] \leftarrow R_h(counts)$ 
			\Comment Permute counts using SFC ordering
			\State offsets [] $\leftarrow scan(counts)$
			\For{$w \in W$}
			\State $i\leftarrow child\_num(w)$
			\State append $w$ to $W_i$ at offsets$[i]$
			\State increment offset$[i]$
			\EndFor 
			\If{$l_1 > l_2$} 
			\For{$i:=1:2^{d}$}
			\State \tsort ($W_i, l_1-1, l_2$)
			\Comment local sort
			\EndFor 
			\EndIf
			\State \Return $W$
		\end{algorithmic}
	\end{algorithm}

	\subsubsection{Octree Construction and Refinement} 

	The octree construction is based on expanding user-specified functions (e.g. initial conditions for a hyperbolic differential equation ) via the 
	wavelet transformation and truncating the expansion (i.e. stopping the 
	refinement at that level) when the coefficients are smaller than a 
	user-specified tolerance $\epsilon>0$. Intuitively, the wavelet coefficient 
	measures the failure of the field to be interpolated from the 
	coarser level. In distributed memory, all processes start from the root and refine until at least $p^2$ octants are produced, where $p$ denotes the number of processors. 
	These are equally partitioned across all processes.
	Subsequent refinements happen in an element-local fashion, and are embarrassingly parallel. A re-partition is performed at the end of construction to load-balance.

	\subsubsection{$2:1$ Balancing} 
	Following the octree construction, we enforce a $2:1$ balance condition. This makes subsequent operations (\S\ref{sec:meshing}, \S\ref{sec:unzip_and_zip} and \S\ref{sec:remesh_and_gridTransfer} ) simpler without affecting the adaptive properties. 
	Our balancing algorithm 
	is an updated version of the algorithm presented in \cite{SundarSampathBiros08}. The octree is divided into small blocks, that are independently balanced by preemptively generating all balancing octants \cite{bern1999parallel}, followed by ripple propagation for balancing across the blocks. The ripple propagation inter-block balancing approach performs poorly at large levels of parallelism, and we instead extend the generation algorithm \cite{bern1999parallel}. The basic idea is to generate all balancing octants for a given octant and then to remove duplicates. While this approach can generate up to \texttt{8x} the number of total octants, it is very simple and highly parallel. We ensure that the overall algorithm works efficiently, by relying on our \tsort\ algorithm to sort and remove duplicates periodically, ensuring that the number of octants generated remains small.

	\subsubsection{Meshing}
	\label{sec:meshing} 
	By meshing or mesh generation we simply refer to the construction of data structures required to perform numerical computations on topological octree data. 
	As mentioned in \S\ref{sec:nm}, we use $4^{th}$ order Finite Differences (FD) 
	with 5-point stencils 
	for $\partial_i, \partial_{ij} ^2$, and 7-point stencils for $\partial_i, i,j\in [1,2,3]$ with upwind/downwind and Kriess-Oliger derivatives.
	We use a $RK$ time integrator with the method of lines to solve the \BSSN~ equations. In this section, we present the data structure choices that we have made and how everything comes together to perform numerical computations on adaptive octree data to evolve the \BSSN~ equations in time.   
	
\begin{figure}
		\centering
		\includegraphics[width=0.9\textwidth]{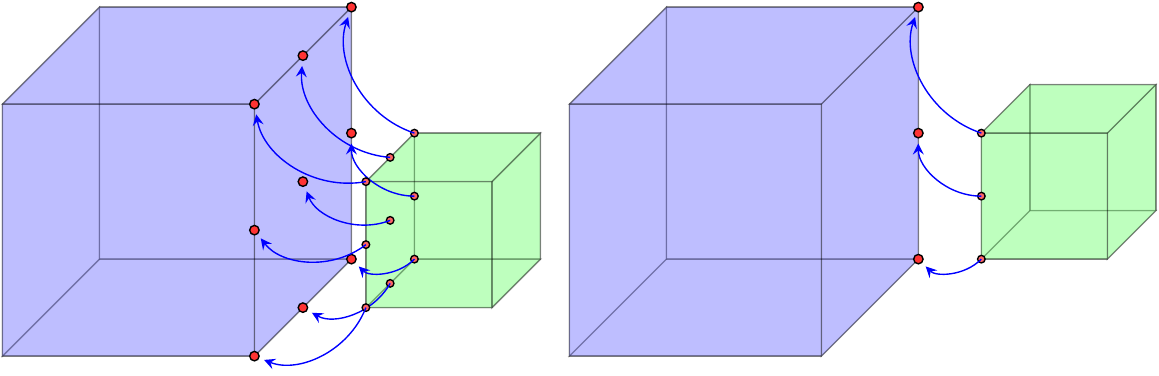}
		\caption{\label{fig:hangingElementsNodes} \small An example of a hanging face  and a hanging edge where in both cases octant (\textcolor{green!50}{$\blacksquare$})  has a hanging face (left figure) and  a hanging edge (right figure) with octant (\textcolor{blue!50}{$\blacksquare$}). Nodes on the hanging face/edge are mapped to the larger octant and the hanging nodal values are obtained via interpolation. 
			Note that for illustrative purposes, the two octants are drawn separately, but are contiguous.}
		\vspace{-0.15in}
	\end{figure}

	\textbf{Embedding nodal information on an octree}:  In order to perform FD computations on an octree, we need to embed spatial/nodal points for each octant. Assuming that we want to perform $d^{th}$ order FD computations, we uniformly place $(d+1)\times (d+1) \times (d+1)$ points for each octant. In our simulations, we have used $d=4$ since we are performing $4^{th}$ order FD computations, but the meshing algorithms presented in the paper are valid for any integer value of $d$. The nodes obtained by uniform node placement are referred to as \dgn~ ($V_D$). Octants that share a face or an edge will have duplicate nodal points in $V_D$, and we need to remove the duplicate nodes from $V_D$ to get \cgn~ ($V_S$) for several reasons. 1).$V_S$ has a lower memory footprint compared to $V_D$. 2). A function representation, on a $V_D$, can be discontinuous due to node duplications, unless function values of duplicate nodes are synchronized.  Due to the above reasons, we use $V_S$ as our prime nodal representation (also referred as \textit{zipped} representation) of the octree. Since the octrees are generated with WAMR, finding the duplicate and hanging nodes (see Figure \ref{fig:hangingElementsNodes}) from $V_D$ becomes non-trivial and requires an octant level neighborhood data structure which is referred to as octant to octant (\oTo) mapping. Also, we need an additional mapping to map the octants to the $V_S$ representation which is referred to as octant to nodal (\oTn) mapping.

	\begin{figure}[bth]
		\centering
		\includegraphics[width=0.9\textwidth]{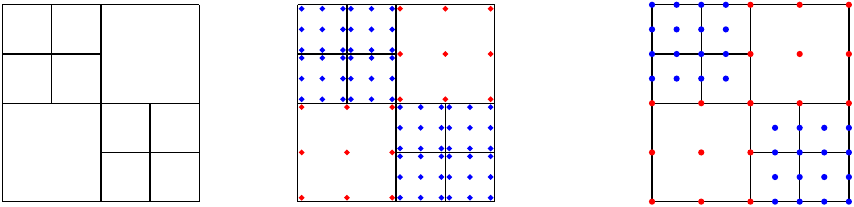}
		\caption{\label{fig:dg_to_cg} \small A $2D$ example of \dgn~ (in the center) and \cgn~(the rightmost figure) nodal representation (with $d=2$, where $d$ denotes the order of FD computations) of the adaptive quadtree shown in the leftmost figure. Note that in \dgn~ representation nodes are local to each octant and contain duplicate nodes. By removing all the duplicate and hanging nodes by the rule of nodal ownership we get the \cgn~ representation. Note that the nodes are color coded based on the octant level.}
		\vspace{-0.15in}
	\end{figure}

	We now describe the methods for building these maps. Note that for the mesh generation, we assume the input is a complete, ordered and 2:1 balanced octree. 
	
	\begin{figure}[tbh]
		\centering
		\includegraphics[width=0.9\textwidth]{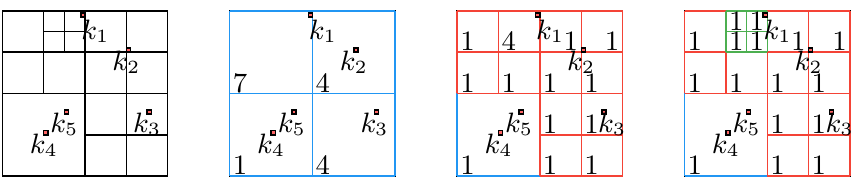}
		\caption{\label{fig:sfcSearch}\small For a given ordered octree $\tau$  and a set of keys (leftmost figure), \tsearch~ performs the traversal in a top-down order over the set of keys, while flagging $k_2,k_4,k_5$ at the level $1$ split, $k_3$ at level $2$ split, and $k_1$ at level $3$ split. 
		}
		\vspace{-0.15in}
	\end{figure}
	
	\textbf{\em \tsearch: amortized search operations on octrees:}
	The common approach for building the maps \oTo ~and \oTn~ is to generate keys corresponding to the location of neighboring octants and to perform a parallel binary search on the octree and build the maps \cite{SundarSampathBiros08, BursteddeWilcoxGhattas11}. 
	%
	%
	Assuming the number of keys we need to search is $\mathcal{O}(n)$, where $n$ is the number of octants, the cost of performing binary searches for all the keys is $\mathcal{O}(n\log(n))$. The $\log(n)$ term corresponding to the binary search is inefficient due to poor memory access and can end up being very expensive for large $n$. 
	We present an alternative \tsearch, with better memory access 
	for performing search operations on an ordered octree. To the best of our knowledge this algorithm is new. 
	The approach used in \tsearch~ is influenced by radix sort, where we traverse 
	the set of search keys and the octree in the space filling curve (SFC) induced ordering. As shown 
	in Figure \ref{fig:sfcSearch}, 
	we start at level $1$, split and calculate bucket counts $|b|$ generated by the split while reordering the keys in the same traversal order dictated by the SFC. $|b|=1$ suggests that octant $e \in \tau$. 
	At this point, $b$ is an ancestor of all keys $k\in b$, and 
	we have found the index the octant. 
	In contrast with the other approaches, \tsearch~ performs a serial traversal over the set of keys 
	and the elements of the ordered octree 
	leading to better memory and cache performance (see Algorithm \ref{alg:tsearch}). Although the complexity for this approach is still $\mathcal{O}(n\log(n))$, it can be thought as performing $\log n$ streaming sweeps over the $\mathcal{O}(n)$ octants, leading to better performance.

	\begin{algorithm}
		\caption{\tsearch: Searching in octrees }\label{alg:tsearch}
		\footnotesize
		\begin{algorithmic}[1]
			\Require ordered octree $\tau$ on domain $\Gamma$, list keys $\mathcal{K} \in \Gamma$,
			\Ensure list keys $\mathcal{K} \in \Gamma$ with flagged search results.
			\State $oct\_counts [] \leftarrow 0$
			\State $key\_counts [] \leftarrow 0$
			\Comment $|oct\_counts|=|key\_counts| = 2^{d}$, 8 for $3D$
			\For{$e \in \tau$}
			\State increment $oct\_counts[child\_num(e)]$
			\EndFor
			\For{$k \in \mathcal{K}$}
			\State increment $key\_counts[child\_num(k)]$
			\State $k.result \leftarrow \emptyset$
			\EndFor    
			\State $oct\_counts [],$key\_counts []$ \leftarrow R_h(oct\_counts,key\_counts)$ 
			\Comment Permute counts using SFC ordering
			\State offsets\_oct [] $\leftarrow scan(oct\_counts)$
			\State offsets\_key [] $\leftarrow scan(key\_counts)$
			\For{$k \in \mathcal{K}$}
			\State $i\leftarrow child\_num(k)$
			\State append $k$ to $\mathcal{K}_i$ at offsets\_key$[i]$
			\State increment offset\_key$[i]$
			\EndFor 
			\For{$i:=1:2^{d}$}
			\If {$oct\_counts >1$}
			\State \tsearch ($\mathcal{K}_i,\tau_i$)
			\Else
			\For {$k \in \mathcal{K}_i$}
			\State $k.result \leftarrow$ offset\_oct$[i]$
			\EndFor
			\EndIf
			\EndFor 
			\State \Return $\mathcal{K}$
		\end{algorithmic}
	\end{algorithm}

	\textbf{Ghost/Halo octants:}
	Since the octree is partitioned into disjoint subtrees owned by different processes, we need access to a layer of octants belonging to other processes. These are commonly known as the {\em halo} or a {\em ghost} layer. The computation of ghost octants can be reduced to a distributed search problem where each octant in each partition generates a set of keys that can be searched for to determine the ghost layer. Note that after the ghost-exchange all search operations are local to each partition/process. 
	\textbf{Octant to octant map (\oTo):}
	%
	Once the ghost layer has been received, we compute the \oTo ~map by performing local searches using \tsearch~ for the neighbors (along $x,y$ and $z$ axes directions) of all octants and storing their indices. Therefore for an given octant $e \in \tau$, $\oTo(e)=\{e_1,...,e_8\}$ will return $8$ neighbor octants of $e$. The algorithm for the $\oTo$ map construction is listed in Algorithm~\ref{alg:e2e}.

	\begin{algorithm}
		\caption{\teToe: compute \oTo  }\label{alg:e2e}
		\footnotesize
		\begin{algorithmic}[1]
			\Require an ordered 2:1 balanced distributed octree $\mathcal{T}$ on $\Gamma$, $\mathit{comm}$, $p$,
			$p_r$ of current task in $\mathit{comm}$.
			\Ensure compute \oTo 
			\State $\hat{\tau_{p_r}} \leftarrow \tghost(\mathcal{T},comm,p,p_r)$
			\State \oTo $~\leftarrow \emptyset$ 
			\State $keys [] \leftarrow$ compute $\mathcal{K}(\hat{\tau_{pr}})$ 
			\State $\tsearch(\hat{\tau_{p_r}},keys)$ 
			\For {$key \in keys$}
			\If {$key$ is found}
			\State \oTo [key.owner][key.neighbor]=key.result
			\EndIf
			\EndFor
			\State \Return \oTo
		\end{algorithmic}
	\end{algorithm}

	\begin{algorithm}
		\caption{\teTon: Octant to nodal map generation - \oTn }\label{alg:e2n}
		\footnotesize
		\begin{algorithmic}[1]
			\Require an ordered 2:1 balanced distributed octree $\mathcal{T}$ on $\Gamma$, $\mathit{comm}$, $p$,
			$p_r$ of current task in $\mathit{comm}$.
			\Ensure compute \oTn 
			\State $\hat{\tau_{p_r}} \leftarrow \tghost(\mathcal{T},comm,p,p_r)$
			\State \oTn $~ [] \leftarrow $ $initialize(\mathcal{V}_L)$ \Comment{ \oTn ~initialized with \dgn} 
			\State $\mathcal{V}_S \leftarrow $ $\mathcal{V}_L \setminus \mathcal{V}_D$
			\For {$e\in \hat{\tau_k}$}
			\For {$v\in N_d(e)$} \Comment{$N_d(e)$ denotes $(d+1)^3$ nodes of $e$} 
			\State $owner\_idx\leftarrow $ compute $\mathcal{O}(v)$
			\State \oTn $~ \leftarrow owner\_idx$
			\EndFor
			\EndFor 	
			\State \Return \oTn
		\end{algorithmic}
	\end{algorithm}
	
	\textbf{Octant to nodal map (\oTn):}
	Since we use the \cgn~ ($V_S$)~ to store all the simulation variables, we need a mapping between the underlying octree and $V_S$. The \oTn~ map simply specifies the subset $v$ of \cgn~ nodes (where $|v|=(d+1)^3$) of a given octant $e\in \tau$. When computing \oTn~ we initially start with octant to the $V_D$ map which is trivially constructed by definition of $V_D$. In order to remove duplicate nodes (see Figure \ref{fig:dg_to_cg}), we need to define a globally consistent rule of nodal ownership. The ownership of nodes which lie on a hanging face or edge (see Figure \ref{fig:hangingElementsNodes}) will belong to the coarser octant (since they can be interpolated from coarser level) while ties and non-hanging nodal ownership are determined by the SFC ordering of octants. Duplicate nodes are removed from the \dgn~($V_D$) to obtain the \cgn~($V_S$) while modifying octant to $V_D$ map to generate octant to $V_S$ map (\oTn). The overview of computing \oTn~ map is presented in Algorithm \ref{alg:e2n}. By the assumption that the octree is 2:1 balanced, the owner nodes (see Figure \ref{fig:hangingElementsNodes}) of hanging nodes cannot be hanging, which simplifies the construction of the \oTn~ mapping.

	\begin{figure}[tbh]
		\centering
		\includegraphics[width=0.9\textwidth]{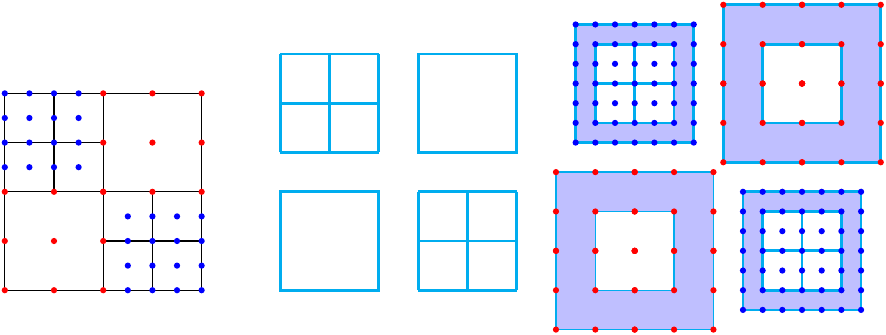}
		\caption{\label{fig:unzip} \small A simplistic example of octree to block decomposition and \unzip~ operation. The leftmost figure shows the considering adaptive octree with \cgn~ and its block decomposition is shown in the middle. Note that the given octree is decomposed into four regular blocks of different sizes. The rightmost figure shows the decomposed blocks padded with values coming from neighboring octants with interpolation if needed. In order to perform \unzip~ operation both $\oTo$ and $\oTn$ mappings are used. }
		\vspace{-0.15in}
	\end{figure}
	
	\begin{figure*}[tbh]
		\noindent\fbox{
			\begin{minipage}[t]{.51\textwidth}
				\small
				\begin{eqnarray*}
					\partial_t \alpha &=&  \mathcal{L}_\beta\alpha - 2 \alpha K, \\
					\partial_t \beta^i &=& \lambda_2 \beta^j\,\partial_j\beta^i + \frac{3}{4} f(\alpha) B^i\\
					\partial_t B^i  &=& \partial_t \tilde\Gamma^i  - \eta B^{i}   + \lambda_3 \beta^j\,\partial_j B^i -\lambda_4 \beta^j\,\partial_j \tilde\Gamma^i \\
					\partial_t \tilde \gamma_{ij} &=&  \mathcal{L}_\beta\tilde{\gamma}_{ij} -2 \alpha \tilde A_{ij}, \\
					\partial_t \chi &=& \mathcal{L}_\beta\chi + \frac{2}{3}\chi \left(\alpha K -  
					\partial_a \beta^a\right)\\
					\partial_t \tilde A_{ij} &=& \mathcal{L}_\beta\tilde{A}_{ij} + \chi \left(-D_i D_j \alpha +
					\alpha R_{ij}\right)^{TF} +\nonumber \\
					&\,&\alpha \left(K \tilde A_{ij} -
					2 \tilde A_{ik} \tilde A^{k}_{\,j}\right), \label{eq:at_evol}\\
					\partial_t K &=& \beta^k\partial_kK- D^i D_i \alpha + \\
					&\,&\alpha \left(\tilde A_{ij}\tilde
					A^{ij} +\frac{1}{3}K^2\right),\\
					\partial_t \tilde \Gamma^i &=& \tilde \gamma^{jk} \partial_j
					\partial_k \beta^i + \frac{1}{3} \tilde \gamma^{ij} \partial_j
					\partial_k \beta^k + \beta^j \partial_j \tilde \Gamma^i - \nonumber \\
					&\,&\tilde
					\Gamma^j \partial_j \beta^i + 
					\frac{2}{3}\tilde \Gamma^i \partial_j
					\beta^j - 2 \tilde A^{i j}\partial_j \alpha + \nonumber \\
					&\,& 2 \alpha \left(\tilde
					{\Gamma^i}_{jk} \tilde A^{jk} -\frac{3}{2\chi} \tilde A^{ij}\partial_j \chi -
					\frac{2}{3} \tilde \gamma^{ij} \partial_j K\right) \\
				\end{eqnarray*}
			\end{minipage}
		}
\begin{minipage}[t]{0.45\textwidth}
\footnotesize
\begin{Verbatim}[frame=single,commandchars=\\\{\}]
 \PYG{k+kn}{from} \PYG{n+nn}{DENDRO\PYGZus{}sym} \PYG{k+kn}{import} \PYG{o}{*}
 \PYG{n}{a\PYGZus{}rhs} \PYG{o}{=} \PYG{n}{Dendro}\PYG{o}{.}\PYG{n}{Lie}\PYG{p}{(}\PYG{n}{b}\PYG{p}{,} \PYG{n}{a}\PYG{p}{)} \PYG{o}{\PYGZhy{}} \PYG{l+m+mi}{2}\PYG{o}{*}\PYG{n}{a}\PYG{o}{*}\PYG{n}{K}
 \PYG{n}{b\PYGZus{}rhs} \PYG{o}{=} \PYG{p}{[}\PYG{l+m+mi}{3}\PYG{o}{/}\PYG{l+m+mi}{4} \PYG{o}{*} \PYG{n}{f}\PYG{p}{(}\PYG{n}{a}\PYG{p}{)} \PYG{o}{*} \PYG{n}{B}\PYG{p}{[}\PYG{n}{i}\PYG{p}{]} \PYG{o}{+}
 	\PYG{n}{l2}\PYG{o}{*}\PYG{n}{vec\PYGZus{}j\PYGZus{}del\PYGZus{}j}\PYG{p}{(}\PYG{n}{b}\PYG{p}{,} \PYG{n}{b}\PYG{p}{[}\PYG{n}{i}\PYG{p}{])}
 	\PYG{k}{for} \PYG{n}{i} \PYG{o+ow}{in} \PYG{n}{e\PYGZus{}i}\PYG{p}{]}
 	\PYG{n}{l2}\PYG{o}{*}\PYG{n}{vec\PYGZus{}j\PYGZus{}del\PYGZus{}j}\PYG{p}{(}\PYG{n}{b}\PYG{p}{,} \PYG{n}{b}\PYG{p}{[}\PYG{n}{i}\PYG{p}{])}
	\PYG{k}{for} \PYG{n}{i} \PYG{o+ow}{in} \PYG{n}{e\PYGZus{}i}\PYG{p}{]}

 \PYG{n}{B\PYGZus{}rhs} \PYG{o}{=} \PYG{p}{[}\PYG{n}{Gt\PYGZus{}rhs}\PYG{p}{[}\PYG{n}{i}\PYG{p}{]} \PYG{o}{\PYGZhy{}} \PYG{n}{eta} \PYG{o}{*} \PYG{n}{B}\PYG{p}{[}\PYG{n}{i}\PYG{p}{]} \PYG{o}{+}
	\PYG{n}{l3} \PYG{o}{*} \PYG{n}{vec\PYGZus{}j\PYGZus{}del\PYGZus{}j}\PYG{p}{(}\PYG{n}{b}\PYG{p}{,} \PYG{n}{B}\PYG{p}{[}\PYG{n}{i}\PYG{p}{])} \PYG{o}{\PYGZhy{}}
	\PYG{n}{l4} \PYG{o}{*} \PYG{n}{vec\PYGZus{}j\PYGZus{}del\PYGZus{}j}\PYG{p}{(}\PYG{n}{b}\PYG{p}{,} \PYG{n}{Gt}\PYG{p}{[}\PYG{n}{i}\PYG{p}{])}
	\PYG{k}{for} \PYG{n}{i} \PYG{o+ow}{in} \PYG{n}{e\PYGZus{}i}\PYG{p}{]}

 \PYG{n}{gt\PYGZus{}rhs} \PYG{o}{=}  \PYG{n}{Dendro}\PYG{o}{.}\PYG{n}{Lie}\PYG{p}{(}\PYG{n}{b}\PYG{p}{,} \PYG{n}{gt}\PYG{p}{)} \PYG{o}{\PYGZhy{}} \PYG{l+m+mi}{2}\PYG{o}{*}\PYG{n}{a}\PYG{o}{*}\PYG{n}{At}

 \PYG{n}{chi\PYGZus{}rhs} \PYG{o}{=} \PYG{n}{Dendro}\PYG{o}{.}\PYG{n}{Lie}\PYG{p}{(}\PYG{n}{b}\PYG{p}{,} \PYG{n}{chi}\PYG{p}{)} \PYG{o}{+}
	\PYG{l+m+mi}{2}\PYG{o}{/}\PYG{l+m+mi}{3}\PYG{o}{*}\PYG{n}{chi}\PYG{o}{*}\PYG{p}{(}\PYG{n}{a}\PYG{o}{*}\PYG{n}{K} \PYG{o}{\PYGZhy{}} \PYG{n}{del\PYGZus{}j}\PYG{p}{(}\PYG{n}{b}\PYG{p}{))}

 \PYG{n}{At\PYGZus{}rhs} \PYG{o}{=} \PYG{n}{Dendro}\PYG{o}{.}\PYG{n}{Lie}\PYG{p}{(}\PYG{n}{b}\PYG{p}{,} \PYG{n}{At}\PYG{p}{)} \PYG{o}{+} \PYG{n}{chi} \PYG{o}{*}
		\PYG{n}{Dendro}\PYG{o}{.}\PYG{n}{TF}\PYG{p}{(}\PYG{o}{\PYGZhy{}}\PYG{n}{DiDj}\PYG{p}{(}\PYG{n}{a}\PYG{p}{)} \PYG{o}{+}
		\PYG{n}{a}\PYG{o}{*}\PYG{n}{Dendro}\PYG{o}{.}\PYG{n}{Ricci}\PYG{p}{)} \PYG{o}{+}
		\PYG{n}{a}\PYG{o}{*}\PYG{p}{(}\PYG{n}{K}\PYG{o}{*}\PYG{n}{At} \PYG{o}{\PYGZhy{}}\PYG{l+m+mi}{2}\PYG{o}{*}\PYG{n}{At\PYGZus{}ikAtKj}\PYG{p}{)}

 \PYG{n}{K\PYGZus{}rhs} \PYG{o}{=} \PYG{n}{vec\PYGZus{}k\PYGZus{}del\PYGZus{}k}\PYG{p}{(}\PYG{n}{K}\PYG{p}{)} \PYG{o}{\PYGZhy{}} \PYG{n}{DIDi}\PYG{p}{(}\PYG{n}{a}\PYG{p}{)} \PYG{o}{+}
	\PYG{n}{a}\PYG{o}{*}\PYG{p}{(}\PYG{l+m+mi}{1}\PYG{o}{/}\PYG{l+m+mi}{3}\PYG{o}{*}\PYG{n}{K}\PYG{o}{*}\PYG{n}{K} \PYG{o}{+} \PYG{n}{A\PYGZus{}ij\PYGZus{}A\PYGZus{}IJ}\PYG{p}{(}\PYG{n}{At}\PYG{p}{))}
\end{Verbatim}
\end{minipage}

		\caption{\label{fig:symb} \small The left panel shows the \BSSN ~formulation of the 
			Einstein equations. These are tensor equations, with indices $i,j,\ldots$
			taking the values $1, 2, 3$. On the right we show the \texttt{{\dendro\_sym}}
			code for these equations. \texttt{\dendro\_sym} uses \texttt{SymPy} and other tools
			to generate optimized C++ code to evaluate the equations. Note that $\mathcal{L}_\beta,\ D,\ \partial$ denote Lie derivative, covariant derivative and partial derivative respectively, and we have excluded $\partial_t\Gamma^i$ from \texttt{\dendro\_sym} to save space. (See \cite{Baumgarte:1998te,Alcubierre:1138167} for more information about the equations and the differential operators.)}
		\label{fig:bssneqs}
		\vspace{-0.15in}
	\end{figure*}

	\subsubsection{\unzip~ and \zip~ Operations:}
	\label{sec:unzip_and_zip}
	All simulation variables are stored in their most compact or {\em zipped} representation, i.e., without any duplication.  Due to the use of 2:1 balanced adaptive octrees, performing FD computation on the octree is non-trivial. In order to overcome the above, we use {\em unzip} representation (a representation in between \cgn~ and \dgn).  Any given adaptive octree $\tau_k$ can be decomposed into a set of regular grid blocks of different sizes--basically a set of octants that are all at the same level of refinement. Due to the memory allocation and performance, we enforce block sizes to be powers of two. In order to perform stencil operations on these blocks, we need information from neighboring blocks, similar to the ghost layer which is required by the distributed case. In the context of blocks, we refer to this layer as the \textit{padding}. During meshing, we compute and save the octree-to-block decomposition, i.e., which octants are grouped together as a block. The computation of octree-to-block decomposition primarily involves a top-down traversal over the local octants and stopping when all elements in the block are at the same level. In order to convert the {\em zipped}  to the {\em unzipped} representation, we copy the data from the {\em zipped} representation to the blocks with padding region. This involves copying the data within the block, and copying--potentially with interpolation--from neighboring octants. Nodes on the block boundary which are hanging need to be interpolated during the copy. The 2:1 balance condition guarantees that at most a single interpolation is performed for any given octant. 
	\par The stencil and other update operations are only performed on the block internal while the padding region is read-only. At the end of the update, the simulation variables are {\em zipped} back, i.e., injected back to the {\em zipped} representation. This step does not involve any interpolations or communication and is very fast. Note that several key operations such as RK update and inter-process communications operate using the \textit{zip} representation, and are extremely efficient which is depicted in strong and weak scaling results (see Figures \ref{fig:ws_g1000} and \ref{fig:ss_r10}). There are several additional advantages for {\em unzipped} representation. 1). {\em unzipped} representation decouples the octree adaptivity from the FD computations. 2). The block representation enables code portability and enables to perform architecture specific optimizations.
	
	\par Although, several similar approaches \cite{Chombo,CARPET} exist for {\em zip} and {\em unzip} operations, these approaches rely on structured or block structured adaptivity. In contrast to existing approaches we have designed efficient scalable data structures to perform {\em zip} and {\em unzip} operations on fully adaptive $2:1$ balanced grids.


	\subsubsection{\remesh~ and \igt~operations}
	\label{sec:remesh_and_gridTransfer}
	As the BHs orbit around each other, we need to remesh so that maximum refinement occurs around the singularities. We do not enforce maximum refinement at singularities artificially, this is automatically performed by WAMR due to the fact that, \BSSN~ variables might not even be $C^{0}$ continuous at BH locations. 
	The \unzip~ representation at the end of the time-step is used to determine the wavelet coefficients for each block  based on a user-specified threshold. This allows us to {\em tag} each octant with \textit{refine}, \textit{coarsen} or \textit{no change}. This is used to remesh, followed by a repartition to ensure load-balance. 
	Once the \remesh ~operation is performed we transfer the solution from old mesh to the newly generate mesh using interpolations as needed. We refer to this is \igt ~and 
	this is done via interpolations or injections at the block level. 
	


	\subsection{Symbolic interface and code generation}
	\label{sec:symbolic}

	The Einstein equations are a set of non-linear, coupled, partial 
	differential equations. Upon discretization, one can end up with 24 or 
	more equations with thousands of terms.
	Sustainability, code optimizations and keeping it relevant for architectural changes are additional difficulties. To address these issues, we 
	have developed a symbolic interface for \dendrogr. Note that there are several significant attempts 
	such as Kranc\cite{Husa:2004ip} and NRPy \cite{Ruchlin:2017com} on symbolic code generation for computational relativity due to the complexity of the \BSSN~ equations.
	We leverage symbolic Python (\texttt{SymPy}) as the backend for this along with the Python package 
	\texttt{cog} to embed Python code within our application-level \texttt{C++} 
	code. The \texttt{\dendro\_sym} package allows us to write the discretized 
	versions of the equations similar to how they are written mathematically 
	and enable improved usability for \dendrogr~ users. 
	An example 
	for the \BSSN~equations are shown in Figure \ref{fig:symb}, with the 
	equations on the left and the corresponding Python code on the right. 
	

	There are several
	advantages to using a symbolic interface like \texttt{\dendro\_sym}
	for the application-specific equations. First, it improves the portability of the code by separating the 
	high-level description of the equations from the low-level optimizations,
	which can be handled by architecture-specific code generators. 
	We support \texttt{avx2} code generators and are working on developing
	a \texttt{CUDA} generator as well. Since these are applied at a
	block level, it is straightforward to schedule these blocks across
	cores or GPUs. Note that the auto-generated code consists of several
	derivative terms that are spatially dependent as well as other point-wise
	update operations. We perform common subexpression elimination (CSE) \cite{cse,sympy} to 
	minimize the number of operations.  
	Additionally, we auto-vectorize the pointwise operations and have
	specialized implementations based on the stencil-structure for the
	derivative terms.  
	
	
	\begin{algorithm}
		\caption{\small Overview of our approach}\label{alg:overview}
		\footnotesize
		\begin{algorithmic}[1]
			\State $M \leftarrow$ initialize mesh \Comment{\S\ref{sec:meshing}}
			\State $u \leftarrow$ initialize variables $(M)$
			\While{$t < T$}
			\For{$r = 1:3$} \Comment{RK stages}
			\State $B, \hat{u} \leftarrow \text{Unzip}(M, u)$ \Comment{\S\ref{sec:unzip_and_zip}}
			\For{$ b \in B$} 
			\State Compute derivatives \Comment Machine generated code \S\ref{sec:symbolic}
			\State Compute $\hat{u}_{rhs}(b)$ \Comment Machine generated code \S\ref{sec:symbolic}
			\EndFor 
			\State $u_{rhs} \leftarrow \text{Zip}(M, B, \hat{u}_{rhs})$ \Comment{\S\ref{sec:unzip_and_zip}}
			\State RK update
			\EndFor  
			\State $t\leftarrow t+dt$
			\If{need remesh $M$}  \Comment{\S\ref{sec:remesh_and_gridTransfer}}
			\State $M' \leftarrow$ remesh($M$) 
			\State $u' \leftarrow$ Intergid\_Transfer$(M, M', u)$ \Comment{\S\ref{sec:remesh_and_gridTransfer}}
			\EndIf
			\EndWhile  
		\end{algorithmic}
	\end{algorithm}
	
	\subsection{Putting everything together}
	\label{sec:alltogether}
	We use an RK time stepper to perform the time evolution. The algorithmic choices we have made in \dendrogr~ support arbitrary $d^{th}$ order RK time integration. The initial octree is constructed based on the WAMR method until the generated grid convergers to capture specified initial data, which is followed by the $2:1$ octree balancing and mesh generation phase which result in all the distributed data structures that are needed to perform ghost/halo exchange, \unzip~ and \zip~ operations. A given RK stage is computed by performing \unzip~ operations with overlapped exchange of the ghost layer for the evolution variables, computation of the derivatives and right-hand-side (rhs) using the code generated by the symbolic framework for all local blocks and finally performing \zip~ operation to get the computed {\em zipped} rhs variables. The RK update is then performed on the {\em zipped} variables. After a specified number of timesteps, we compute the wavelet coefficient for the current solution represented on the grid, and perform \textit{remesh} and \textit{inter-grid transfer} operations if the underlying octree grid needs to be changed. Note that wavelet computation for the grid is a local to each process while \textit{remesh} and \textit{inter-grid transfer} need interprocess communication.  An complete outline of our approach for simulating binary BH mergers demonstrating how the various components come together is listed in Algorithm \ref{alg:overview} and illustrated in Figure \ref{fig:overview}.
	
	In the Appendix \ref{sec:nlsm} we present an example on how to use \dendrogr~ framework to solve simpler (compared to \BSSN) \NLSM~ equations. \NLSM~ example also serves as an additional test to ensure all the \dendrogr~ modules are working and integrated correctly. We also performed additional tests with \NLSM~ with zero source term result in the standard linear wave equation which enables to perform convergence testing with the analytical solution.

	\section{Results} 
	\label{sec:results}
	In this section we perform a thorough evaluation of our code, including detailed comparisons with the \ET. We first describe the machines used for these experiments followed by results demonstrating the improvements to \dendro\ and comparisons with \ET. Finally, we push our code to the limit of extreme adaptability to demonstrate its capability, using cases that are currently--to the best of our knowledge--beyond the capability of \ET.
	
\begin{sidewaysfigure}
	\centering
	\includegraphics[width=\textwidth]{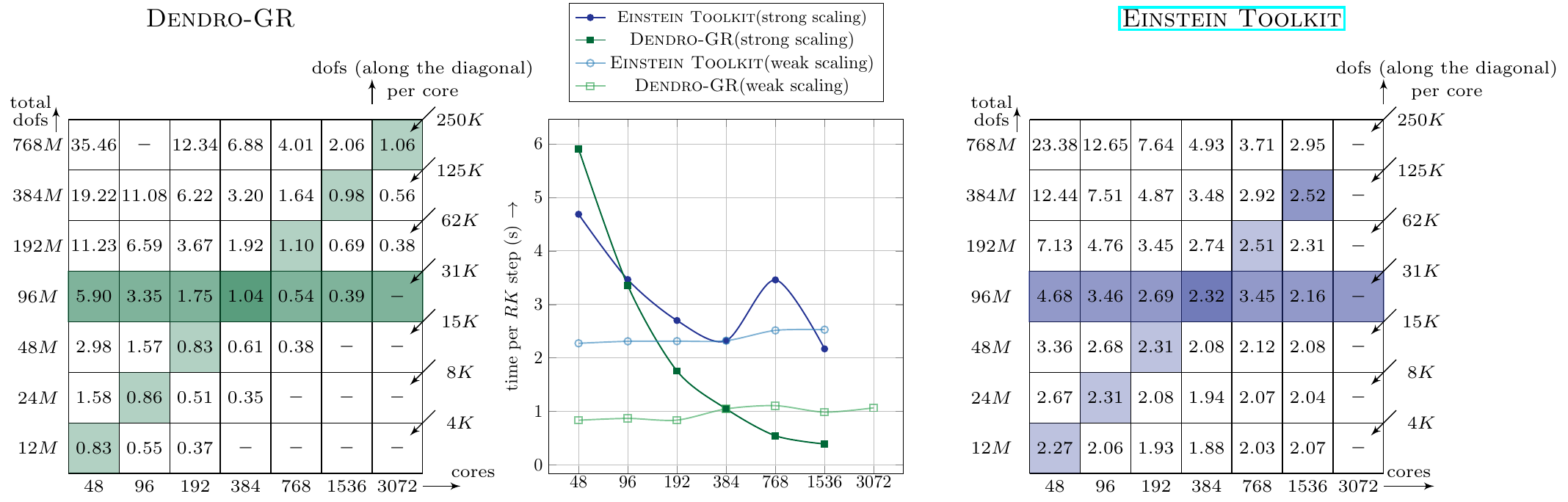}
	\caption{\label{fig:et_sans_adaptivity} \small Comparison between \ET~ and \dendrogr\ without factoring in adaptivity (i.e. both \ET~ and \dendrogr~ support uniform grids.). For a fixed tolerance, we expand the domain for a $1:1$ mass-ratio simulation such that both \ET~ and \dendrogr\ have roughly the same number of dofs. We present both weak and strong scaling results using both codes. On the left table are results from \dendrogr\ and from \ET\  on the right. In the middle, we plot a representative strong and weak scaling curve for each code. The \dendrogr\  scaling is plotted in green (lighter shade for weak) and blue for \ET. The corresponding data entries are also marked in the tables. Note that the rows represent strong scaling and the diagonal entries represent weak scaling results and runtime is reported in seconds$(s)$. }
	\vspace{-0.2in}
\end{sidewaysfigure}

\vspace{0.05in}
\noindent {\bf Experimental Setup:}
The large scalability experiments reported in this paper were
performed on \Titan~and \Stampede. \Titan~is a Cray XK7 supercomputer
at Oak Ridge National Laboratory (ORNL) with a total of 18,688
nodes, each consisting of a single 16-core AMD Opteron 6200 series
processor, with a total of 299,008 cores. Each node has 32GB of memory.
It has a Gemini interconnect and 600TB of memory across all nodes.
%
%
\Stampede~  is the flagship supercomputer at the Texas Advanced
Computing Center (TACC), University of Texas at Austin. 
It has $1,736$ Intel Xeon Platinum 8160 (SKX) compute nodes with $2\times 24$ cores and 192GB of RAM
per node. Stampede2 has a 100Gb/sec Intel Omni-Path
(OPA) interconnect in a fat tree topology. We used the SKX
nodes for the experiments reported in this work.

\vspace{0.05in}
\noindent {\bf Implementation Details:} 
The \dendrogr  ~framework is written in \texttt{C++} using \texttt{MPI}. 
The symbolic interface and code generation module uses symbolic Python (\texttt{SymPy}).
In the comparisons with the \ET, we have used
Cactus \texttt{v4.2.3} and the Tesla release of the Einstein Toolkit. We integrated the
\BSSN~equations with third-order RK for all comparisons in this paper.
The constraint analysis, apparent horizon finder, and output were turned off for these runs.




\begin{table}[tbh]
	\centering
	\scalebox{.8}{
		\begin{tabular}{|c|c|c|c|} 
			\hline
			unbalanced octants & balanced octants & 2:1 balance (\cite{SundarSampathBiros08}) (s) & 2:1 balance (\dendrogr~) (s) \\
			\hline
			3K &	5K &	0.0087	&	0.0043 \\
			33K &	59K &	0.0908 & 0.0541 \\
			338K &	553K &	0.7951 &	0.5461 \\
			3M &	5M &	7.7938 &		6.9313 \\
			6M &	11M & 16.1828 &	 14.7374 \\
			\hline
		\end{tabular}
	}
	\caption{Comparison study for 2:1 balancing approach used in \cite{SundarSampathBiros08} and the new balancing approach in a single core in \Stampede~(\texttt{SKX} node), with varying input octree sizes ranging from 
		$3K$ to $6M$ octants.}
	\label{tb:balcomparison}
	\vspace{-0.2in}
\end{table}

\begin{figure}[tbh]
	\centering
	\includegraphics[width=0.9\textwidth]{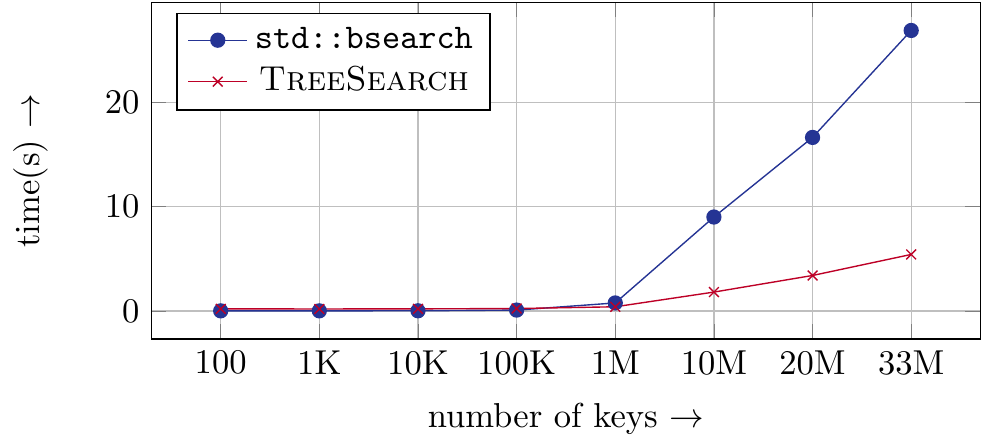}
	\caption{Comparison of \texttt{std::bsearch} with partial ordering operator $<$ and comparison free \tsearch~ approach for performing, varying number of keys on $33M$ sorted complete octree using single core in \Stampede~ \texttt{SKX} node.}
	\label{fig::balcomparison}
\end{figure}

\begin{figure}[tbh]
	\centering
	\includegraphics[width=0.9\textwidth]{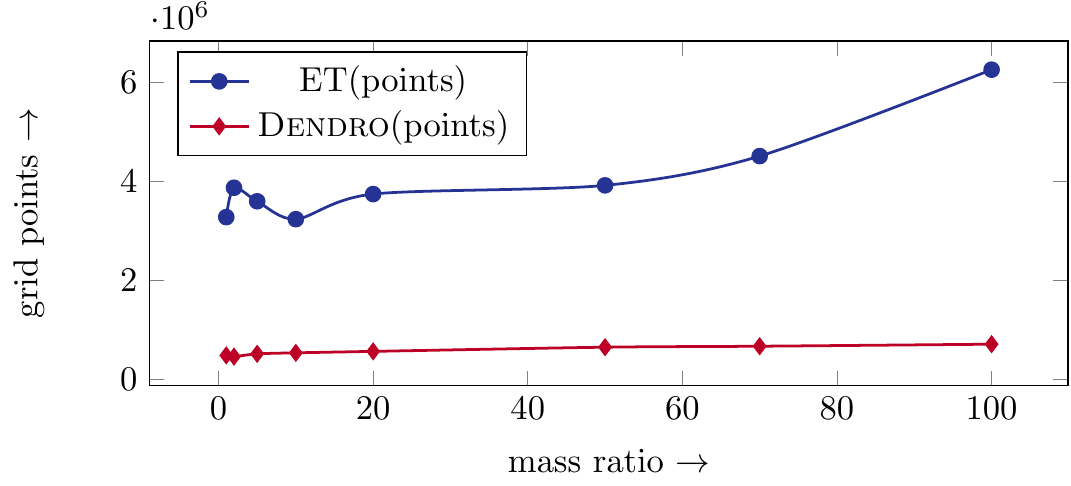}
	\caption{\label{fig:et_avec_adaptivity} \small Comparison between \et~ and \dendrogr\ for number of spatial points with increasing mass ratios. Note that these are not from complete simulations and the size of the problem as well as the time per RK-step is likely to increase, but it illustrates the rate of increase for both approaches. Parameters for the above experiment generated such that total mass of black holes equals to $1$ and the separation distance is $32$ for all cases and \maxDepth~ is set in a way that the spatial discretization $dx<\frac{\min(m_1,m_2)}{16}$ where $m1,m2$ denotes the individual masses of black holes. 
	}
	\vspace{-0.2in}
\end{figure}



\subsection{Meshing Performance}

In this section, we briefly present results for the improved scalability and performance of the proposed balancing and meshing algorithms. In Table~\ref{tb:balcomparison}, we list the improvement in enforcing 2:1 balancing by using \tsort\ instead of the ripple propagation used in \cite{SundarSampathBiros08}. We present only single core results, as the algorithmic changes are for the sequential portions of the algorithms. In Table~\ref{tb:balcomparison} we demonstrate significant savings for a range of problems sizes.

Similarly, significant savings are also obtained by the use of \tsearch\ compared to the use of binary searches for the various search operations needed for meshing. In this experiment, we searched for $k$ keys in an array of size $n=33M$ octants. This size was chosen based on the average grain size we used in our experiments. Note that for meshing, $k=\mathcal{O}(n)$, and therefore we plot results up to $k=n$. Having a large array that is being searched in (large $n$), results in the first few steps of the binary search resulting in cache misses affecting overall performance. \tsearch\ utilizes the deep memory hierarchy in a more effective fashion, but because of the additional work involved in sorting, requires a minimum number of keys to be searched for before it is cheaper. This can be seen in our results plotted in Figure~\ref{fig::balcomparison}, where \tsearch\ scales better than \texttt{std::bsearch}, and is faster for $k>1M$. Given the number of keys being searched for during meshing, \tsearch\ improves the overall meshing performance and scalability significantly.

\subsection{Correctness of Code}
We performed a number of tests to assess the correctness of our
code, including simulations of static (Appendix~\ref{sec:AE_sbh})
and boosted black holes (Appendix~\ref{sec:AE_sbhboost}), as well as
comparisons to other codes. To verify the automatic generation of the
computer code for the \BSSN~equations, we evaluated these equations
using arbitrary analytic functions of order unity over a grid of 
points, and compared the results to a known solution. The L2-norms
of the error are equivalent to machine zero, 
given the limitations of finite-precision arithmetic. 
Additional results on the correctness 
are presented in Appendix~\ref{sec:AE}.

\begin{sidewaysfigure}
	\centering
	\includegraphics[width=0.9\textwidth]{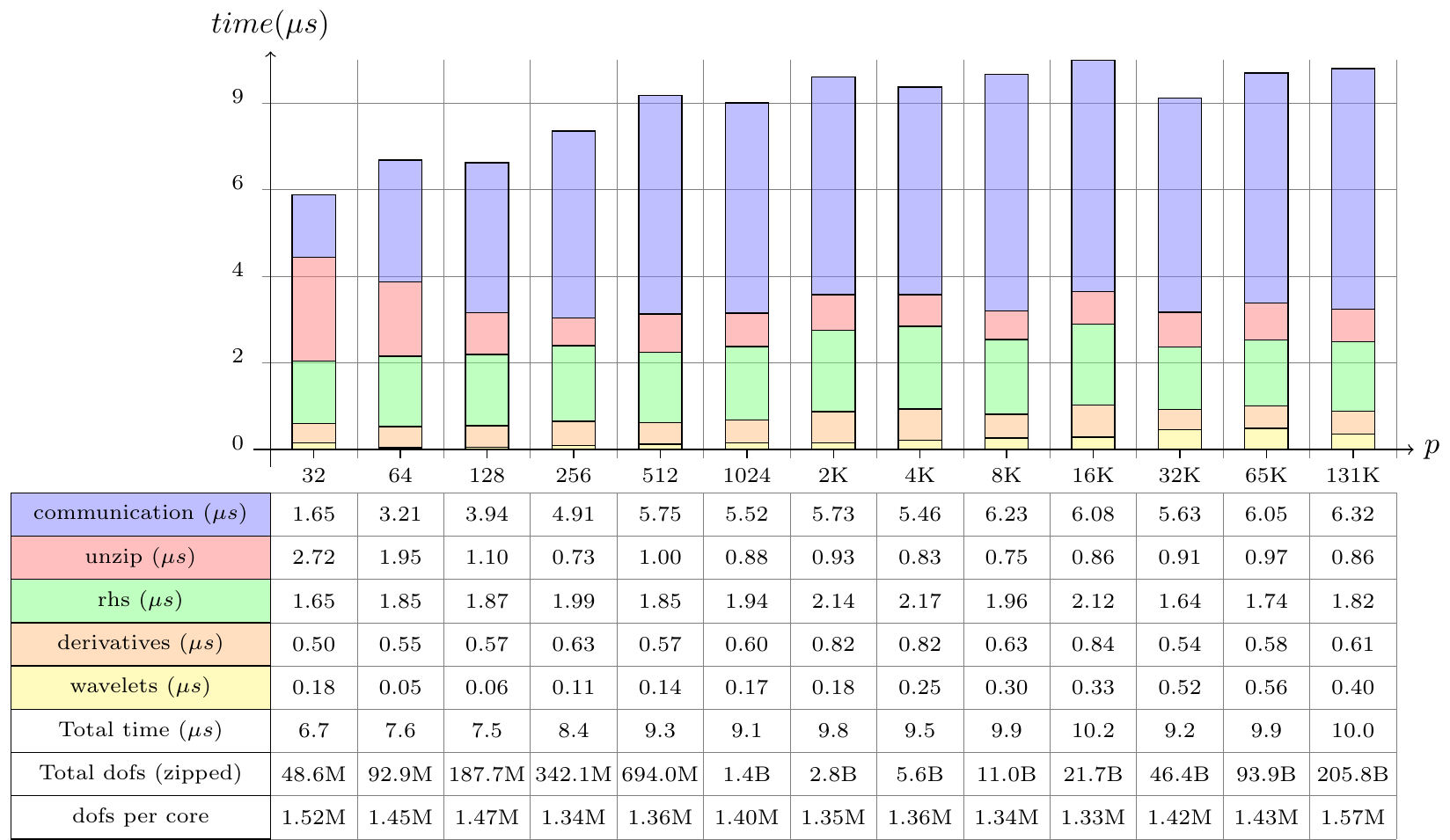}
	\caption{\small Weak scaling results in ORNL's \Titan~for  $RK/(dof/p)$ (averaged over 10 steps) where $RK,dof,p$ denotes the time for single $RK$ step, degrees of freedom, and number of cores respectively,  with derivative computation (\texttt{deriv}), right hand side ({\texttt rhs}) computation, \texttt{unzip} cost, wavelet computation (\texttt{wavelets}) and communication cost (\texttt{comm}) with the average of 1.41M unknowns per core where the number of cores ranging from $32$ to $131,072$ cores on $8,192$ nodes where the largest problem having $206$ Billion unknowns. Above results are generated with mass ratio $\mu=10$ with \maxDepth~ 18 and wavelet tolerance of $10^{-6}$. Note that the unknowns per core have a slight variation since with WAMR we do not have explicit control over the grid size and WAMR decides the refinement region on the mesh based on the how wavelets behave during the time evolution. This is why we have reported normalized $RK$ with $dof/p$ metrics to report accurate weak scaling results.   \label{fig:ws_g1000} }
	\vspace{-0.2in}
\end{sidewaysfigure}

\begin{figure}[tbh]
	\centering
	\includegraphics[width=0.7\textwidth]{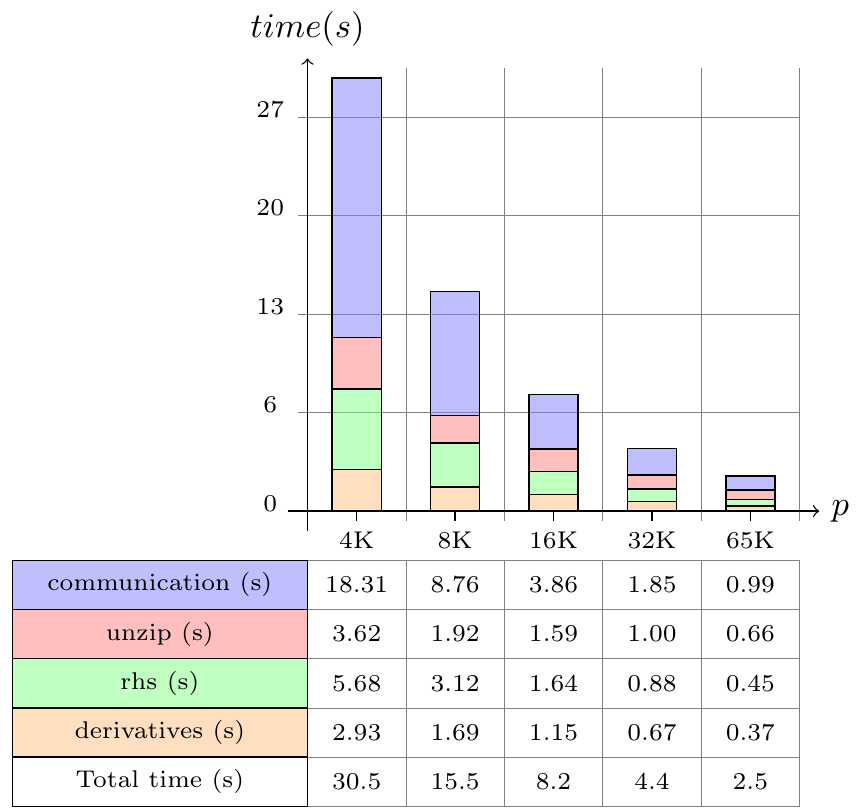}
	\caption{\small Strong scaling results in ORNL's \Titan~for a single RK step (averaged over 10 steps) with derivative computation (\texttt{deriv}), right hand side ({\texttt rhs}) computation, \texttt{unzip} cost 
		and communication cost (\texttt{comm}) for a fixed problem size of $10.5B$ unknowns where the number of cores ranging from $4,096$ to $65,536$ cores on $4096$ nodes. Note that for strong scaling results re-meshing is disabled in order to keep the problem size fixed.  \label{fig:ss_r10} }
	\vspace{-0.2in}
\end{figure}

\subsection{Comparison with \ET}
\label{sec:et_sans_adaptivity}

We compare our code with the \BSSN~formulation implemented in \et.
The AMR driver for Cactus, \textsc{Carpet}~\cite{PhysRevD.71.081301}, 
only supports block 
adaptivity. Therefore, we would
expect \dendrogr\  to require fewer degrees of freedom (dof) for a
given simulation and consequently be faster. Although \et~
uses vectorization for improved performance~\cite{Cactus_Goodale03a}, 
\dendrogr\  outperforms \et. Note that for all comparison studies with \ET~, we have used the non-vectorized version of the \dendrogr\ generated code. To highlight
the improvements over \et, we perform two independent experiments.
First, we compare the performance of both codes without adaptivity,  
and then in a separate experiment we show that \dendrogr\  provides a more
efficient adaptivity and scaling than \textsc{Carpet}.

\vspace{0.05in}
\noindent{\bf Uniform Grid Tests:}
In Figure~\ref{fig:et_sans_adaptivity}, we present a comparison
between \dendrogr\  and \et. This uses a regular grid for \et, and serves to highlight the efficiency of the \dendrogr\ code including the overhead of adaptivity (i.e., \zip\ and \unzip).
These runs were
performed on \Stampede, and we show strong and weak
scaling for both codes. Although both codes demonstrate good
scaling, the performance (as measured by the time for one
complete timestep) is better for \dendrogr ~at higher core counts.
\ET ~has better performance for large number of unknowns per core at low core counts. This effectively captures the overhead of \zip\ and \unzip\ compared to an efficient and mature code. 
For cases with higher core counts as well as smaller number of unknowns per core, \dendrogr 
~performs better. This is largely due to better cache utilization due to blocking, that largely compensates for the overhead of \zip/\unzip.
Additionally, 
\dendrogr\  demonstrates better strong and weak scaling. The plot in 
Figure~\ref{fig:et_sans_adaptivity} highlights the strong and weak
scaling of both codes for a representative grain and
problem size. \dendrogr\  scales well far beyond the $3072$ cores, 
as shown in Figure~\ref{fig:ws_g1000} and discussed in \S\ref{sec:scale}.

\vspace{0.05in}
\noindent {\bf Octree Adaptivity vs. Block Adaptivity:}
As motivated earlier, we wish to perform simulations of binary black hole 
mergers with large mass-ratios, $q \simeq 100$. 
Large mass-ratios require extensive refinement, increasing the
number of spatial degrees of freedom.
For an example let's assume an equal mass binary requires a certain resolution for the BSSN equations, about 100 points per BH in each dimension. 
For a BH 100 times smaller, 
we need a resolution equivalent to 1/10,000 times the total mass 
and $100\times$ more time steps. The consequent high computational cost is the primary reason that current catalogs of numerical waveforms
contain templates with a maximum mass ratio of 
$q < 10$~\cite{PhysRevD.93.084031,0264-9381-34-22-224001,0264-9381-33-20-204001}.

\begin{figure}
	\centering
	\includegraphics[width=\textwidth]{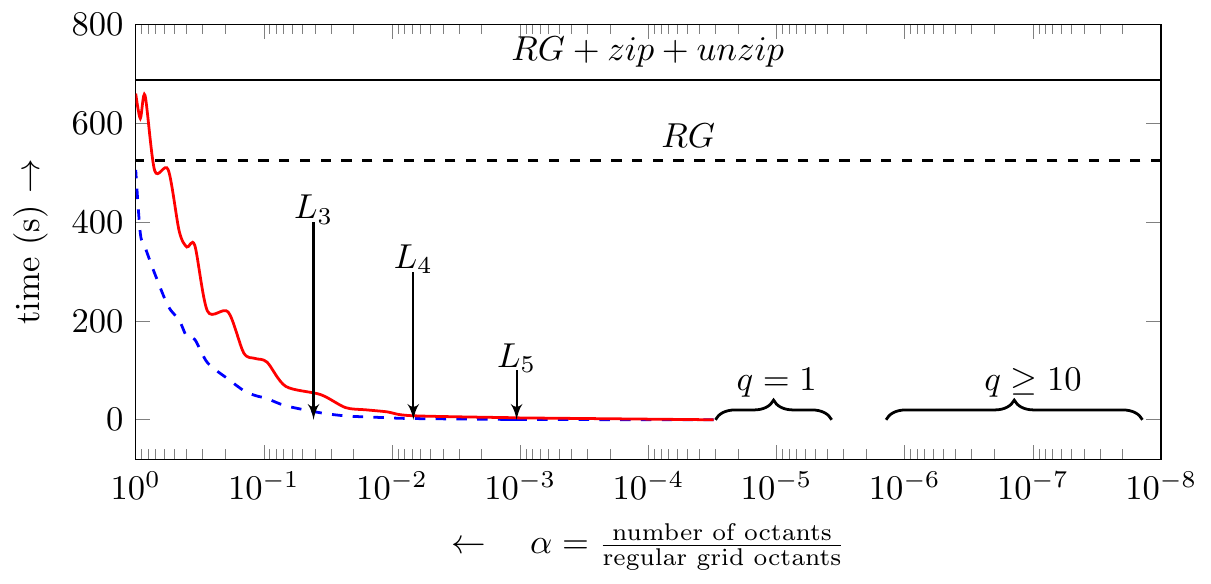}
	\caption{An illustrative single core example to evaluate the overhead of zip/unzip operations to evaluate BSSN equations on a sequence of octree grids over 10 timesteps. The parameter $\alpha$ denotes the ratio between the number of octants to the number of regular grid octants for \textsc{MaxDepth} $8$. Hence moving towards the right direction on $x$ axis the octree grids converge towards a regular grid. $RG$ and $RG+zip+unzip$ denotes the baseline performance for regular grid computations and regular grid computation with zip/unzip overhead respectively. The shaded $q=1$ region denotes the $\alpha$ value for an equal mass ratio simulation until the merger event using $\textsc{MaxDepth}$ $12$ over $2\times 10^5$ timesteps. For larger mass ratio runs the $\textsc{MaxDepth}$ will be determined by the smaller black hole, hence for $q=10$ with $\textsc{15}$, $\alpha$ reached maximum value of $1.4\times 10^{-8}$ over $6000$ time steps and for $q=100$ with \textsc{MaxDepth} 20 maximum value of $\alpha$ reached $5.51\times 10^{-13}$ over 1000 timesteps. Note that value of $\alpha$ can increase during the simulation, for the largest problem (see Figure \ref{fig:ws_g1000}) that was run in Titan with $131K$ cores for $q=10$ case the $\alpha$ value was $4.8\times 10^{-7}$. In the plot, we have marked the $L_k$ values for $k=3,4,5$ where $L_k$ is the computed $\alpha$ ratio for an adaptive octree where an equal number of octants spanning across $k$ levels. 
	}
	\label{fig:zipUnzip}
	
\end{figure}

The efficient adaptivity of \dendro\  is a big advantage over
the block adaptivity of the \et. To assess the effect of
adaptivity on code performance,
we compared both codes for increasing mass-ratios from $q=1$--$100$, 
measuring the {\em dofs}, as shown in
Figure~\ref{fig:et_avec_adaptivity}. 
As the mass ratio increases, the number of {\em dofs} for \et  ~increases 
much more rapidly than for \dendrogr.
While these results capture
the differences at the beginning of the simulation\footnote{It would
	be very expensive to run full simulations with \et~ due to its
	scalability for large $q$.}, they are representative
of the full simulation in terms of comparing the two codes. Because
of better adaptivity and better scalability, \dendrogr\  keeps the cost
of a single RK-step fairly flat as we scale up from $q=1$--$100$ mass-ratio.
In Table \ref{tb:mr_vs_cores} we present the number of cores needed to maintain a $24K$ unknowns per core with
increasing mass ratio using \dendrogr~ framework. Finally, to illustrate the advantages of octree-adaptivity over block-adaptivity, even with the overhead of \zip/\unzip, we plot the reduction in runtime for 10 timesteps for different levels of adaptivity in Figure \ref{fig:zipUnzip}. 
Based on these results, it is clear that \zip/\unzip~ with adaptivity can benefit especially for simulation of large mass ratio configurations and that the overhead is about $20\%$ even for a full regular grid simulation. These results further support our findings in Figure~\ref{fig:et_sans_adaptivity}.

\begin{table}[t]
	\centering
	\scalebox{1.0}{
		\begin{tabular}{|c | c | c | c | c | c | c | c | c |} 
			\hline
			mass ratio $q$ & 1 & 2 & 5 & 10 & 20 & 50 & 70 & 100 \\
			\hline
			cores & 27 & 31 & 64 & 79 & 89 & 90 & 99 & 96 \\
			\hline
		\end{tabular}
	}
	\caption{\small The number of cores required to maintain $24K$ unknowns per core with different mass ratios with \maxDepth~ $12$ wavelet tolerance of $10^{-4}$ and black hole separation distance of $32$.}
	\label{tb:mr_vs_cores}
	\vspace{-0.2in}
\end{table}

\begin{figure}[h]
	\begin{subfigure}{0.32\textwidth}
		\includegraphics[width=0.9\textwidth]{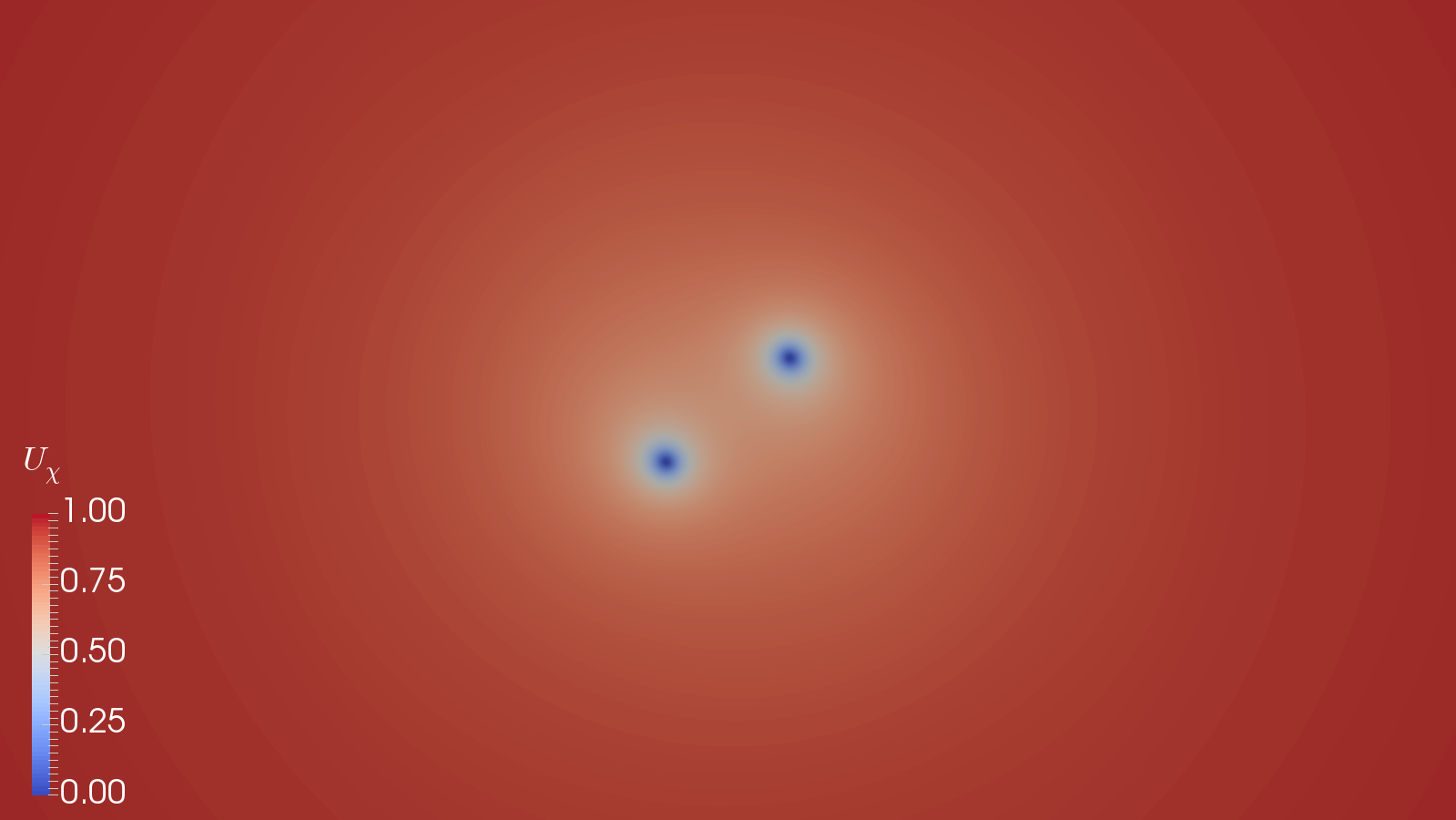}
		\caption{$q=1$}
	\end{subfigure}
	\begin{subfigure}{0.32\textwidth}
		\includegraphics[width=0.9\textwidth]{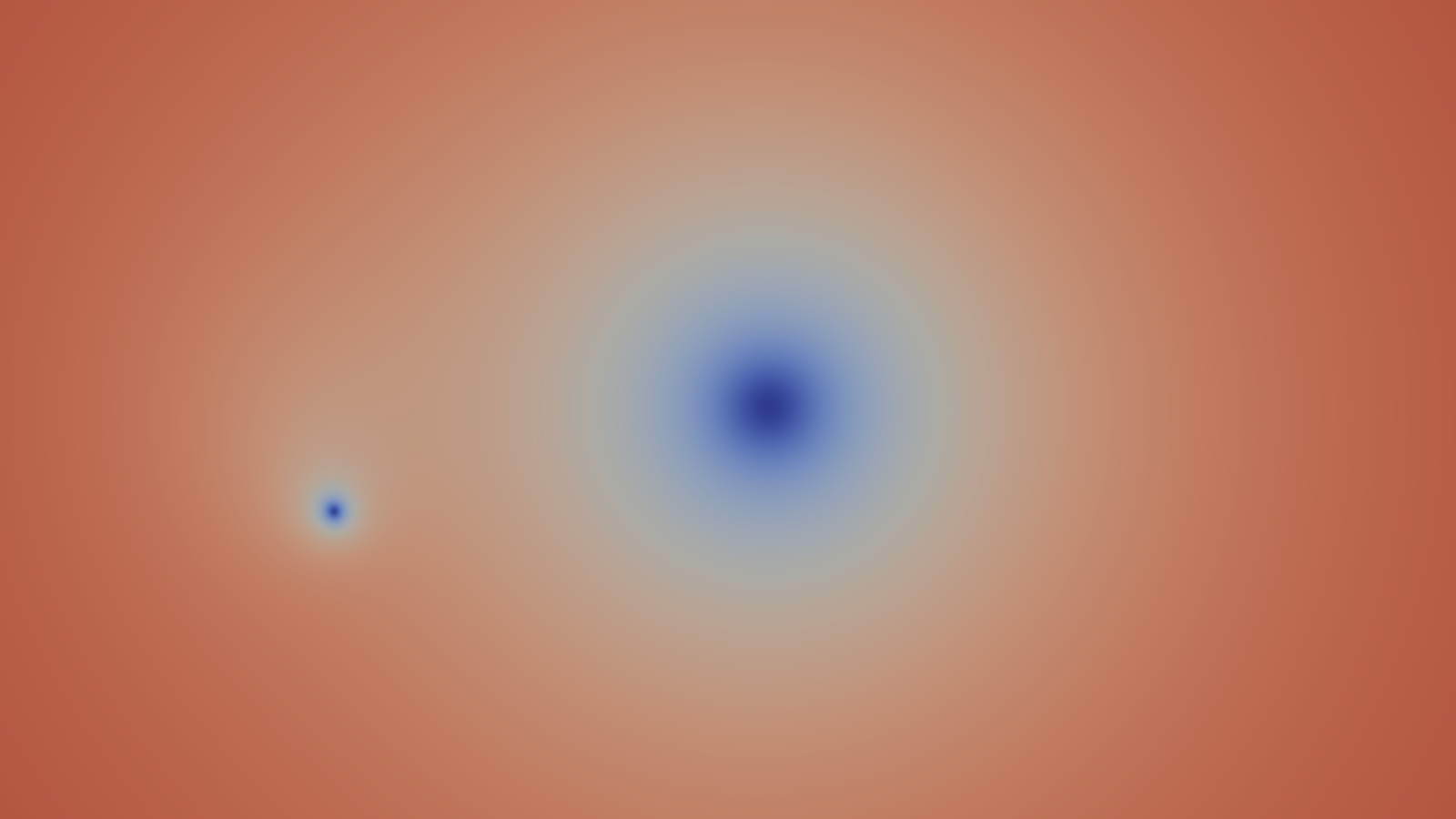}
		\caption{$q=10$}
	\end{subfigure}
	\begin{subfigure}{0.32\textwidth}
		\includegraphics[width=0.9\textwidth]{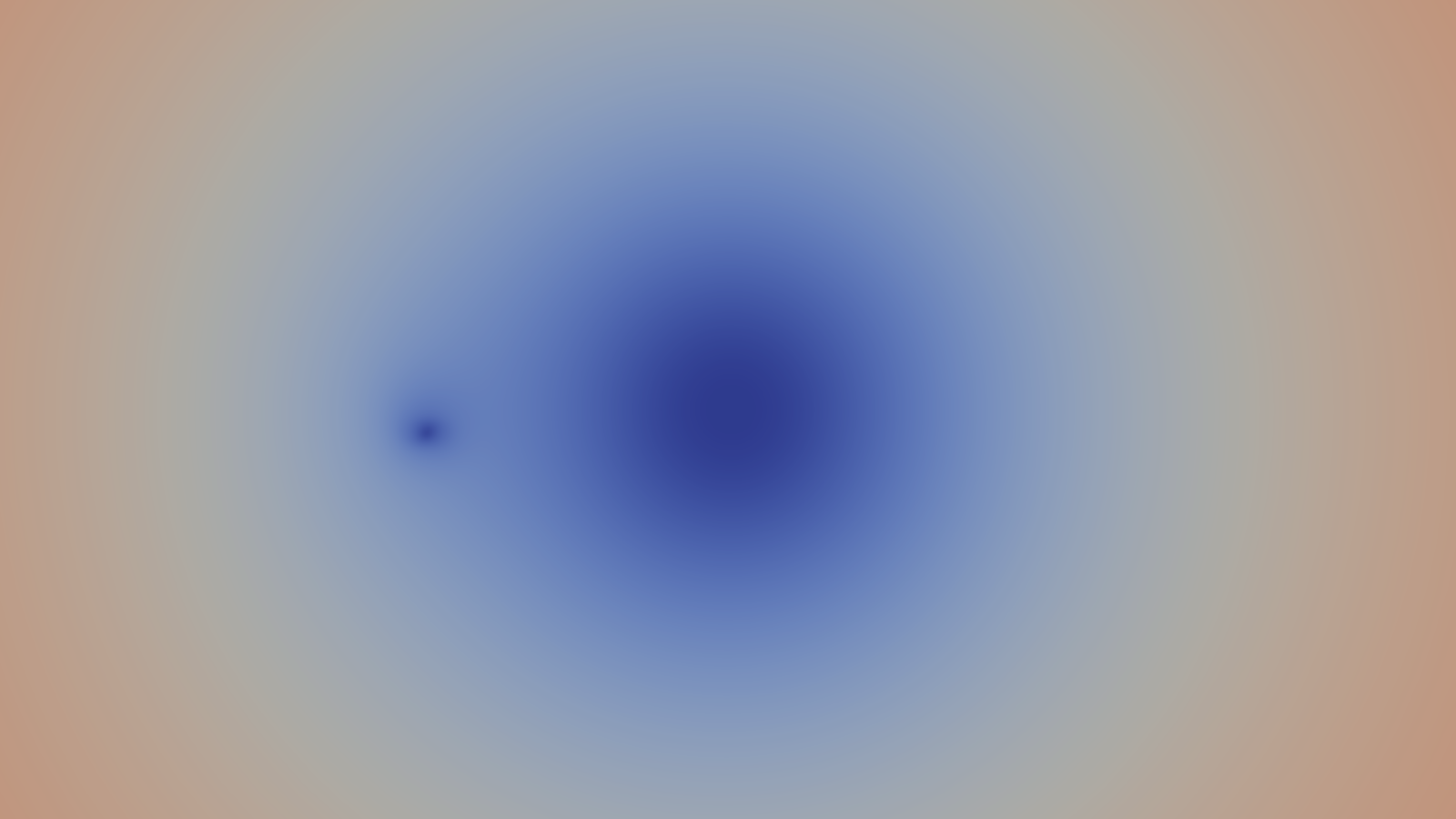}
		\caption{$q=100$}
	\end{subfigure} \hfill
	\begin{subfigure}{0.32\textwidth}
		\includegraphics[width=0.9\textwidth]{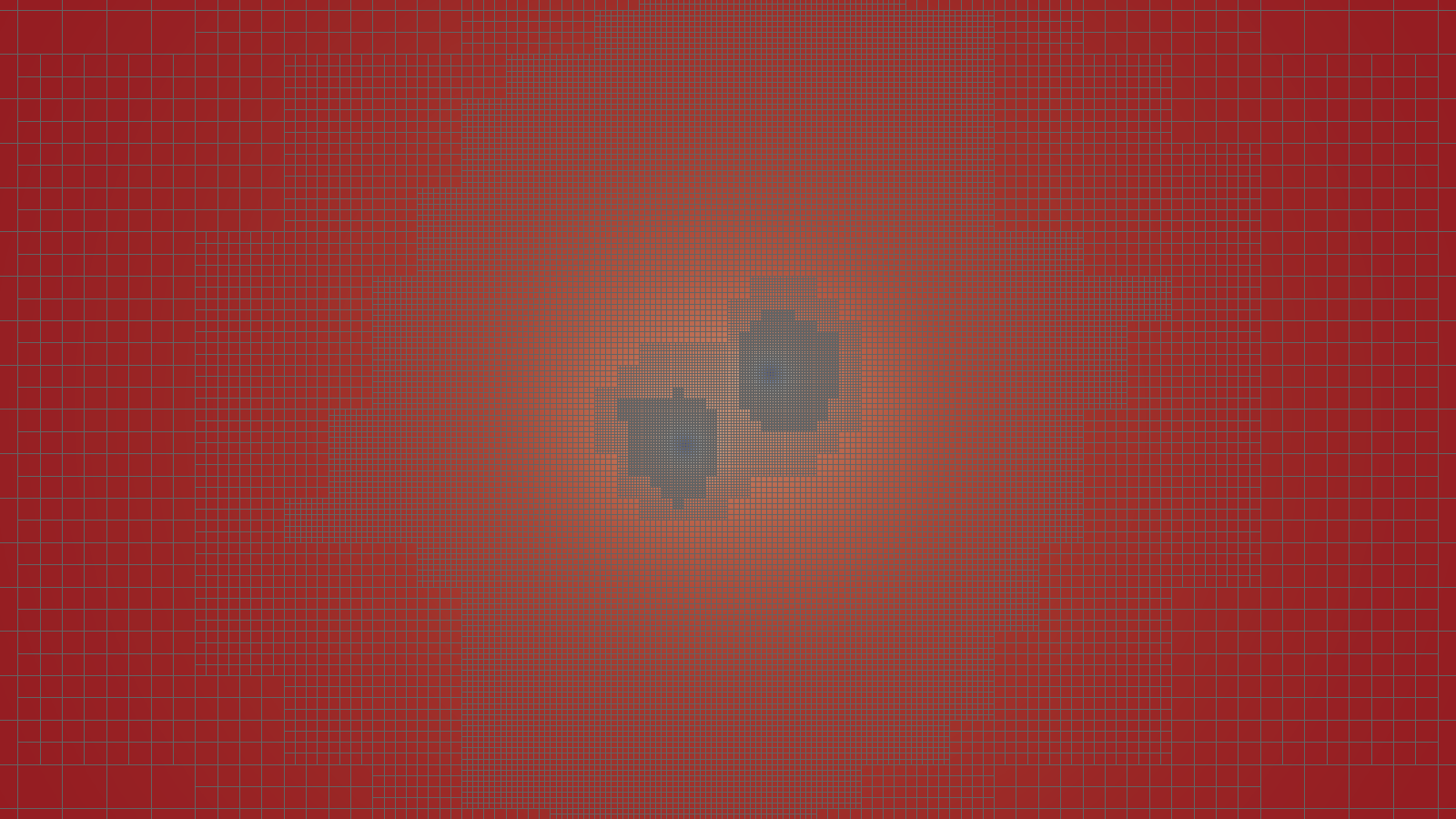}
		\caption{$q=1$}
	\end{subfigure}
	\begin{subfigure}{0.32\textwidth}
		\includegraphics[width=0.9\textwidth]{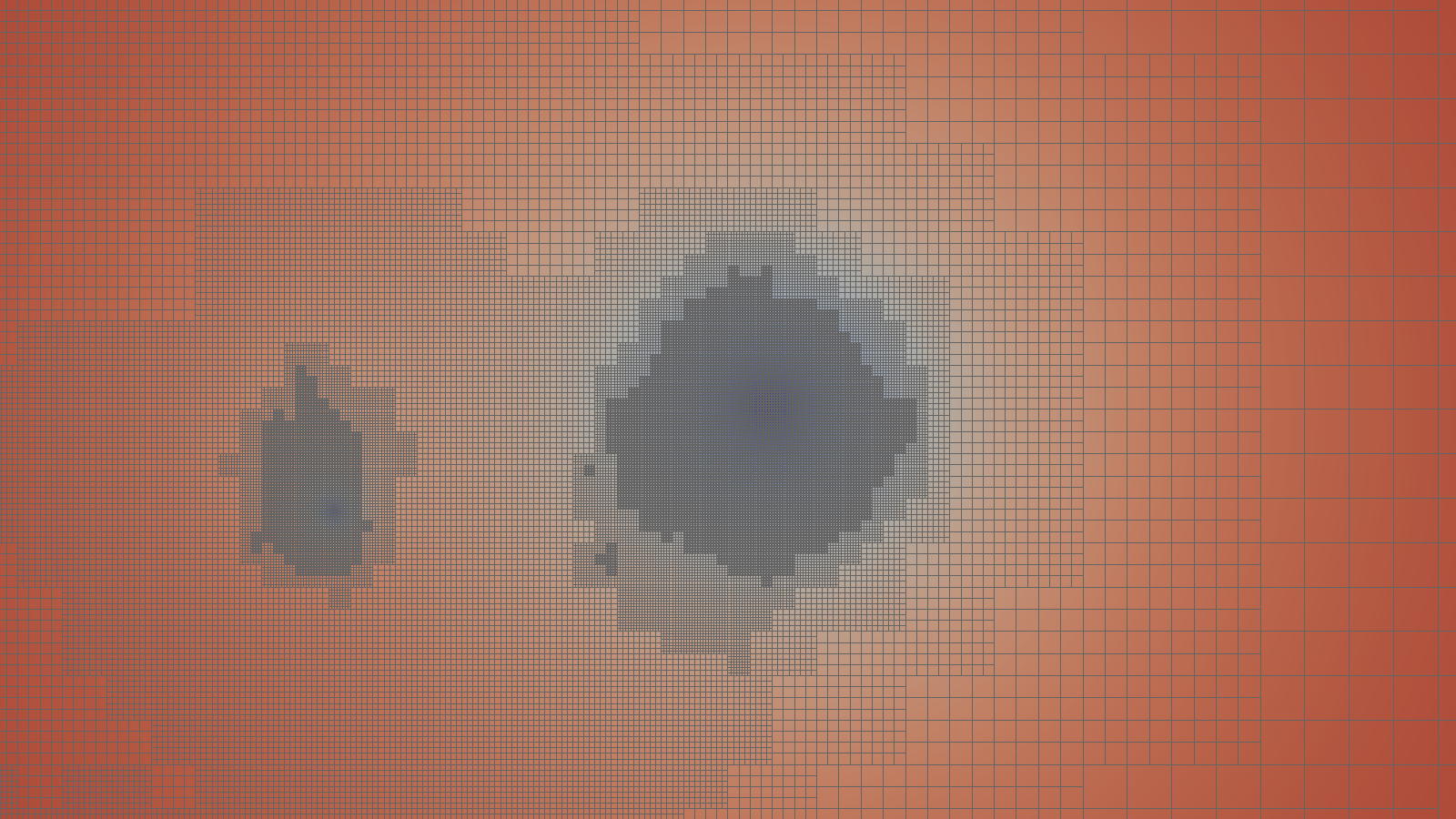}
		\caption{$q=10$}
	\end{subfigure}
	\begin{subfigure}{0.32\textwidth}
		\includegraphics[width=0.9\textwidth]{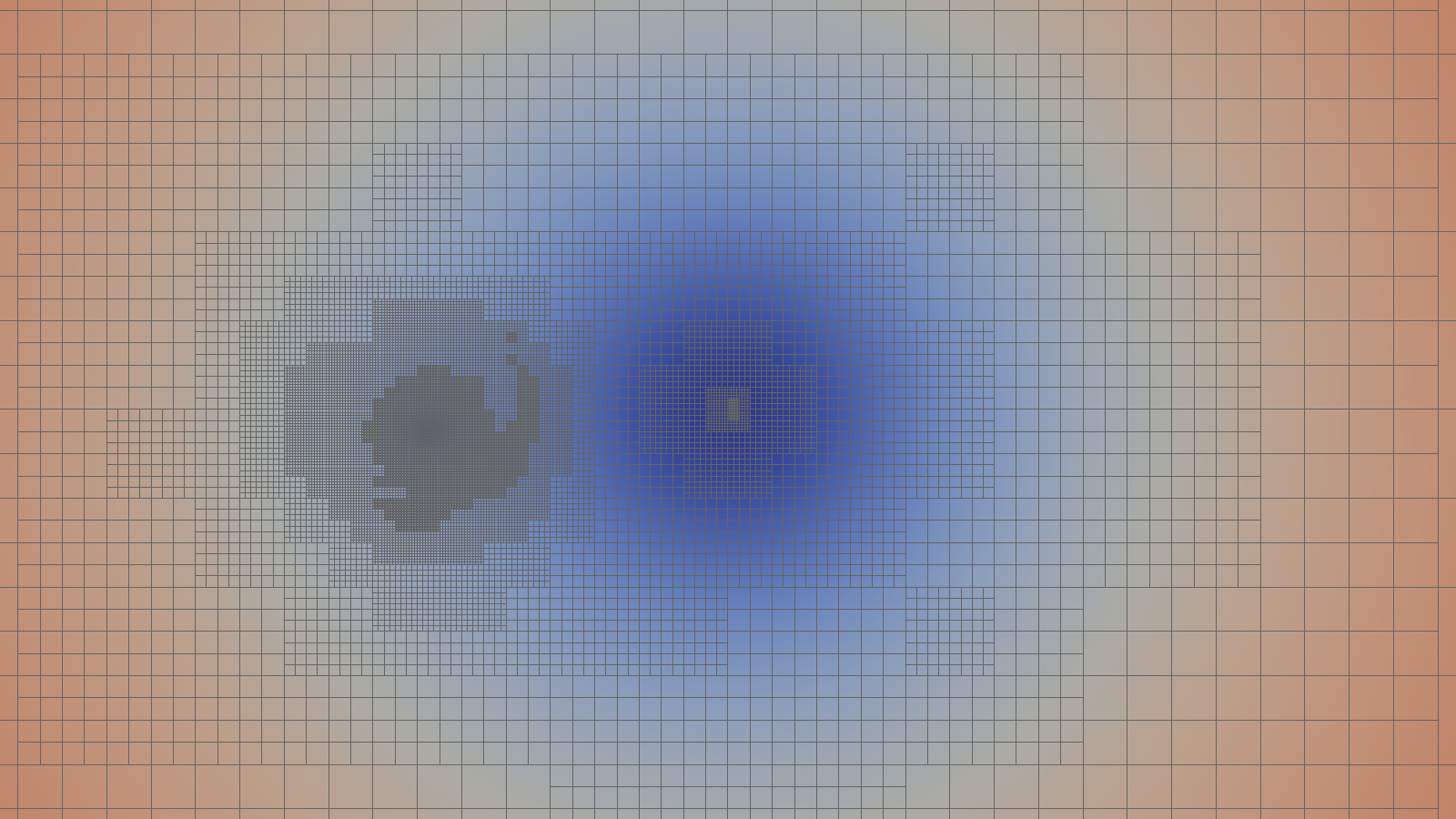}
		\caption{$q=100$}
	\end{subfigure}
	\caption{\small Time step snapshots of the binary black hole problem of black hole mass ratios $1, 10$ and $100$ where we in the top row we plot the \BSSN~variable $\chi$ in the lower row we plot the WAMR grids for each case at that specific instance. \label{fig:large_q}}
	\vspace{-0.2in}
\end{figure}



\subsection{Parallel Scalability}
\label{sec:scale}

Scalability is a key requirement for large-scale codes such as
\dendro. In the context of our target application this could be
in order to run simulations of black hole mergers with large mass ratios,
or to be able to run these simulations much faster. These cases
correspond to weak and strong scaling respectively.
As demonstrated in
\S\ref{sec:et_sans_adaptivity}, \dendrogr\ scales well in both cases.
In this section, we demonstrate the ability of \dendrogr\  to scale
to much larger core-counts for weak scaling. In Figure~\ref{fig:ws_g1000}, 
we demonstrate weak scalability to
$131,072$ cores on \Titan~using $1.5\times 10^6$
dofs/core for a largest problem of $206\times 10^9$ dofs. These
correspond to a simulation of $q=10$, with increasing
levels of refinement. There is a slight
fluctuation in the number of unknowns per core at each problem size. Therefore, we
report the average time for a single RK-step$/dof/p$. We can see
that the average time per RK-step$/dof/p$ remains less than $10\mu
s$. We note that the unzip costs are high at the lower core-counts
and stabilize at higher core counts. This is due to a larger grain
size at the smaller core-counts. The communication for a smaller number of processes is lower due to architectural reasons, as the majority of the communication happens primarily within the same node. This is common for most codes, and stabilizes for $p>512$ as the communication starts to get dominated by inter-node communication rather than intra-node communication.

In Figure \ref{fig:ss_r10}, we present the strong scaling results for \dendrogr~ to perform single RK step (i.e. averaged over 10 steps) for a fixed problem size of $10.5B$ unknowns in \Titan~ up to 4096 nodes. 
This experiment is carried out for a mass ratio of 10 binary merger problem with wavelet tolerance of $10^{-6}$. Note that for the strong scaling experiment we have disabled the \textit{re-mesh} and \textit{inter-grid} transfer operations in order to keep the problem size constant.

\subsection{Large Mass Ratios}

Finally, we present some representative images from simulations at mass rations of $q=1,10$, and $100$ in Figure~\ref{fig:large_q}. Here we plot the one of the evolved variables $\chi$ along with the adaptively refined mesh. We can observe the increased refinement needed to handle the increased mass ratio.

	\section{Conclusion}
	\label{sec:conclusion}
	
	In this work we presented a high-adaptive and highly scalable framework for relatavistic simulations. By combining a parallel octree-refined adaptive mesh with wavelet adaptive multiresolution and a physics module to solve the Einstein equations of general relativity in the \BSSN~formulation, we were able to perform simulations of IMRIs of binary black holes with mass ratios on the order of 100:1. In designing our framework and methods, we have focused on ensuring portability and extensibility by the use of automatic code generation from symbolic notation. This enables portability, since new architectures can be supported by adding new generators rather than rewriting the application. The code is extensible as new applications can be created by using the symbolic interface without having to focus on writing scalable code.
	Since both of these are achieved using Python, this reduces the cost of extending and porting the framework, especially by non-specialist programmers. We also made improvements to fundamental algorithms required for generating octree-based meshes, specifically an improved 2:1 balancing algorithm and a scalable search algorithm fundamental to constructing meshing data-structures. Finally, we performed extensive comparisons with the current state-of-the-art codes and demonstrated excellent weak and strong scalability.    
	
	In the short time that LIGO and Virgo
	have been searching for
	gravitational waves, we have already learned exciting things about
	neutron stars~\cite{Most:2018hfd,Shibata:2017xdx}, 
	the production of heavy elements 
	(such as gold)~\cite{0004-637X-855-2-99}, 
	and the population of black holes in the 
	universe~\cite{TheLIGOScientific:2016htt}. 
	The full scientific
	impact of multi-messenger astronomy is only realized when the
	observations are informed by sophisticated computer models of the underlying
	astrophysical phenomena. 
	\dendro\ provides the ability to run these models in a scalable way, 
	with local adaptivity criteria using WAMR.
	While AMR codes with block-adaptivity typically lose performance as the number
	of adaptive levels increases, \dendro\ achieves impressive scalability on a real 
	application even with many levels of refinement. The combination of scalability
	and adaptivity will allow us to study the gravitational radiation from
	IMRIs without simplifying approximations in direct numerical simulations.

	\section*{Acknowledgment}
	The authors would like to thank the reviewers for valuable comments and suggestions that greatly improved this manuscript. 
	\bibliographystyle{siamplain}
	\bibliography{bssn}

	\newpage
	\pagebreak
	\setcounter{page}{1}
	\appendix
	\section{Overview of \BSSN~ Equations}
	\label{sec:bssn}
	In this section, we briefly present the specific form of the \BSSN\ 
	equations used in this work. While a complete 
	description of these equations can not be given here, 
	many detailed descriptions can be found in the 
	literature~\cite{GourgoulhonBSSN, Alcubierre:1138167, Shibata:2015:NR:2904075}.
	
	The spacetime metric is written using the ADM 3+1 decomposition
	\begin{equation}
	ds^2 = -\alpha^2\, dt^2 + \gamma_{ij}\left(dx^i + \beta^i\, dt\right)\left(dx^j + \beta^j\, dt\right),
	\end{equation}
	where $\alpha$ is the lapse function, $\beta^i$ is the shift vector,
	and $\gamma_{ij}$ is the 3-metric on a space-like hypersurface.
	We write $\gamma_{ij}$ in terms of a conformally flat metric 
	$\tilde\gamma_{ij}$ and a conformal factor $\chi$ as
	\begin{equation}
	\gamma_{ij} = \frac{1}{\chi}\tilde\gamma_{ij}, \qquad
	\chi = \left(\det\gamma_{ij}\right)^{-1/3}, \qquad \det\tilde\gamma_{ij} = 1.
	\end{equation}
	The extrinsic curvature $K_{ij}$ is decomposed into its trace $K$ and the
	conformal, trace-less extrinsic curvature $\tilde A_{ij}$
	\begin{equation}
	K = K^i{}_i,\qquad 
	\tilde A_{ij} = \chi\left( K_{ij} - \frac{1}{3}\gamma_{ij}K\right).
	\end{equation}
	Covariant derivatives on the hypersurface, $D_i$, use
	connection coefficients $\Gamma^i{}_{jk}$ computed with respect to
	$\gamma_{ij}$. Conformal connection coefficients, $\tilde\Gamma^{i}{}_{jk}$,
	are computed with respect to $\tilde\gamma_{ij}$, and the
	conformal connection functions are defined to be 
	$\tilde\Gamma^i = \tilde\gamma^{jk}\tilde\Gamma^i{}_{ij}$.
	
	Similar to the Maxwell equations in electromagnetism, 
	the Einstein equations contain both hyperbolic and elliptic equations; the
	former are the evolution equations and the latter constraint equations. 
	The vacuum \BSSN\ evolution equations are
	\begin{align*}
		\partial_t \tilde \gamma_{ij} &=  \mathcal{L}_\beta\tilde{\gamma}_{ij} -2 \alpha \tilde A_{ij} \\
		\partial_t \chi &= \mathcal{L}_\beta\chi + \frac{2}{3}\chi \left(\alpha K -  
		\partial_a \beta^a\right)\\
		\partial_t \tilde A_{ij} &= \mathcal{L}_\beta\tilde{A}_{ij} + \chi \left(-D_i D_j \alpha +
		\alpha R_{ij}\right)^{TF} + \alpha \left(K \tilde A_{ij} - 2 \tilde A_{ik} \tilde A^{k}_{\,j}\right)\\
		\partial_t K &= \beta^k\partial_kK- D^i D_i \alpha + \alpha \left(\tilde A_{ij}\tilde
		A^{ij} +\frac{1}{3}K^2\right)\\
		\partial_t \tilde \Gamma^i &= \tilde \gamma^{jk} \partial_j
		\partial_k \beta^i + \frac{1}{3} \tilde \gamma^{ij} \partial_j
		\partial_k \beta^k + \beta^j \partial_j \tilde \Gamma^i -\tilde
		\Gamma^j \partial_j \beta^i + 
		\frac{2}{3}\tilde \Gamma^i \partial_j
		\beta^j - 2 \tilde A^{i j}\partial_j \alpha \nonumber \\
		& \quad  + 2 \alpha \left(\tilde
		{\Gamma^i}_{jk} \tilde A^{jk}  - \frac{3}{2\chi} \tilde A^{ij}\partial_j \chi -
		\frac{2}{3} \tilde \gamma^{ij} \partial_j K\right),
	\end{align*}
	where $R_{ij}$ is the Ricci tensor on the hypersurface, $(\ldots)^{\rm TF}$
	indicates the trace-free part of the quantity in parentheses, 
	$\partial_i$ is the partial derivative with respect to $x^i$, 
	and $\mathcal{L}_\beta$ is Lie derivative with respect to $\beta^i$.
	We use the ``1+log'' slicing condition and the $\Gamma$-driver shift condition
	to evolve the gauge (or coordinate) variables
	\begin{align*}
		\partial_t \alpha &= \mathcal{L}_\beta\alpha - 2 \alpha K\\
		\partial_t \beta^i &= \lambda_2 \beta^j\,\partial_j\beta^i + \frac{3}{4} f(\alpha) B^i\\
		\partial_t B^i  &= \partial_t \tilde\Gamma^i  - \eta B^{i}   + \lambda_3 \beta^j\,\partial_j B^i -\lambda_4 \beta^j\,\partial_j \tilde\Gamma^i,
	\end{align*}
	where $f(\alpha)$ is an arbitrary function, and $\eta$ and
	$\{\lambda_1, \lambda_2, \lambda_3, \lambda_4\}$ are parameters.
	We set $f(\alpha) = \lambda_A=1$ in this work.
	
	The Hamiltonian and momentum constraints are elliptic equations that must
	be satisifed at all times during the evolution. These constraint equations 
	in vacuum are
	\begin{align}
		R - K_{ij}K^{ij} + K^2 &= 0\\
		D_j (K^{ij} - \gamma^{ij} K) &= 0.
		\label{eq:constraints}
	\end{align}
	We follow the common practice in numerical relativity by solving the Einstein
	equations in a ``free evolution,'' which means that the we evolve the
	the hyperbolic evolution equations for the \BSSN\ variables and the 
	gauge. The independent constraint equations are evaluated during the
	evolution to monitor the quality of the solution, but they are otherwise
	not used in the update.

	\subsection{\NLSM: Non-linear Sigma Model}
	\label{sec:nlsm}
	In this section, we introduce a simple model to demonstrate 
	some capabilities of \dendrogr, the classical
	wave equation. We can also add a nonlinear source to this equation
	to explore the Non-Linear Sigma Model with the hedgehog Ansatz
	(NLSM)~\cite{NLSM}.
	We write the classical wave equation in a form similar to the 
	the \BSSN\ equations, i.e., with first derivatives in time and second
	derivatives in space. This allows us to test the \dendrogr\ infrastructure
	for the more complicated \BSSN\ equations.
	
	
	The NLSM for a scalar function $\chi(t,x^i)$ is the classical wave equation
	with a nonlinear source term. For simplicity we assume a flat spacetime
	and Cartesian coordinates $(t,x,y,z)$. The equation of motion is
	\begin{equation}
	\frac{\partial^2\chi}{\partial t^2} 
	- \left(\frac{\partial^2}{\partial x^2} + \frac{\partial^2}{\partial y^2}
	+ \frac{\partial^2}{\partial z^2}\right) \chi  =
	-\frac{\sin(2\chi)}{r^2},
	\end{equation}
	where $r = \sqrt{x^2 + y^2 + z^2}$. We write the equation as a first order
	in time system by introducing the variable $\phi$ as
	\begin{align}
		\frac{\partial\chi}{\partial t} &= \phi \label{eq:chi} \\
		\frac{\partial\phi}{\partial t} &= \left(\frac{\partial^2}{\partial x^2} 
		+ \frac{\partial^2}{\partial y^2} 
		+ \frac{\partial^2}{\partial z^2}\right) \chi 
		-\frac{\sin(2\chi)}{r^2} \label{eq:phi}.
	\end{align}
	We choose outgoing radiative boundary conditions for this 
	system~\cite{Alcubierre:1138167}. We assume that the variables $\chi$ and
	$\phi$ approach the form of spherical waves as $r\to\infty$, which decay
	as $1/r^k$. The radiative boundary conditions then have the form
	\begin{equation}
	\frac{\partial f}{\partial t} = \frac{1}{r}
	\left(     x\frac{\partial f}{\partial x} 
	+ y\frac{\partial f}{\partial y} 
	+ z\frac{\partial f}{\partial z}\right) - k(f - f_0),
	\end{equation}
	where $f$ represents the functions $\chi$ and $\phi$, and $f_0$ is
	an asymptotic value. We assume $k=1$ for $\chi$ and $k=2$ for $\phi$.
	
\begin{figure}[H]
		\begin{subfigure}{0.32\textwidth}
			\includegraphics[width=0.9\textwidth]{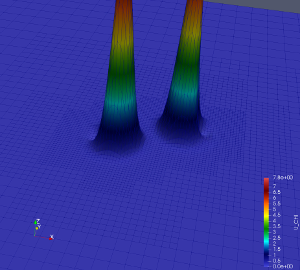}
			\caption{step=0}			
		\end{subfigure}
		\begin{subfigure}{0.32\textwidth}
			\includegraphics[width=0.9\textwidth]{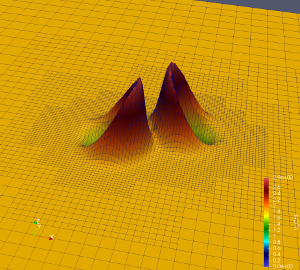}
			\caption{step=7}			
		\end{subfigure}
		\begin{subfigure}{0.32\textwidth}
			\includegraphics[width=0.9\textwidth]{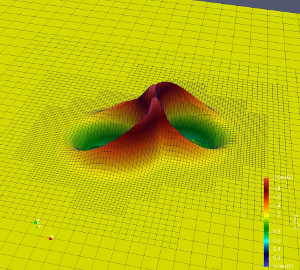}
			\caption{step=11}
		\end{subfigure}\hfil
		\begin{subfigure}{0.32\textwidth}
			\includegraphics[width=0.9\textwidth]{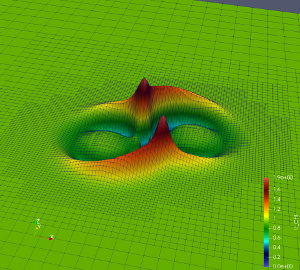}
			\caption{step=16}			
		\end{subfigure}
		\begin{subfigure}{0.32\textwidth}
			\includegraphics[width=0.9\textwidth]{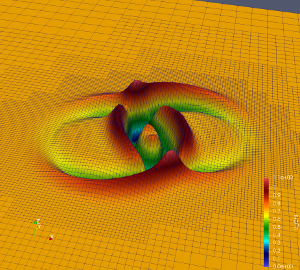}
			\caption{step=23}
		\end{subfigure}
		\begin{subfigure}{0.32\textwidth}
			\includegraphics[width=0.9\textwidth]{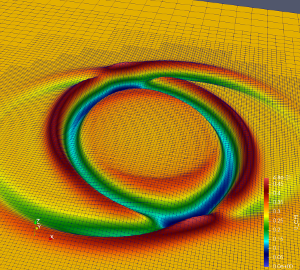}
			\caption{step=44}
		\end{subfigure}
		\caption{\small Frames from the evolution of the 3-dimensional \NLSM\ that 
			show the solution in the $z=0$ plane. The initial data are from family~(b)
			defined in~\cite{NLSM}, which consist of two Gaussian functions for $\chi$ 
			and velocities used to define $\phi$.
			Note how the mesh refines (based on the WAMR) as the pulses first 
			interact nonlinearly at the center of the grid and then begin to 
			propagate away from the origin.   \label{fig:nlsmB}}
		\vspace{-0.2in}
	\end{figure}

	Using the \dendrogr\ symbolic code 
	generation framework we generate right-hand side evaluation for the 
	equations \ref{eq:chi} and \ref{eq:phi} used in the time stepping scheme. 
	Figure~\ref{fig:nlsmB} shows frames from an evolution of the NLSM model
	using initial data from family (b) described in Table I of~\cite{NLSM}.
	This test demonstrates that the \dendrogr\ components for octree construction,
	mesh generation, spatial derivative operators, and time integration 
	work accurately. 
	
	
	
	\section{Code Evaluation and Verification}
	\label{sec:codeEvalAndVerify}
	\label{sec:AD}

	\subsection{Getting and Compiling \dendro~}
	
	The \dendro\ simulation code is freely available at GitHub (\href{https://github.com/paralab/Dendro-GR}{https://github.com/paralab/Dendro-GR}) under the MIT License. 
	The latest version of the code can be obtained by cloning the repository
	\begin{lstlisting}[language=bash]
	$ git clone git@github.com:paralab/Dendro-GR.git
	\end{lstlisting}
	
	The following dependencies are required to compile \dendro~
	\begin{itemize}
		\item C/C++ compilers with C++11 standards and OpenMP support
		\item MPI implementation (e.g. openmpi, mvapich2 )
		\item ZLib compression library (used to write \texttt{.vtu} files in binary format with compression enabled)
		\item BLAS and LAPACK are optional and not needed for current version of \dendro~
		\item CMake 2.8 or higher version
	\end{itemize}
	
	\textbf{Note}: We have tested the compilation and execution of \dendro\ with \texttt{intel}, \texttt{gcc} 4.8 or higher, \texttt{openmpi}, \texttt{mpich2} and \texttt{intelmpi} and \texttt{craympi} (in \Titan) using the linux operating systems. 
	
	To compile the code, execute these commands
	\begin{lstlisting}[language=bash]
	$ cd <path to DENDRO directory >
	$ mkdir build
	$ cd build
	$ ccmake ../     
	\end{lstlisting}
	The following options for \dendro\  can then be set in cmake:
	\begin{itemize}
		\item \texttt{DENDRO\_COMPUTE\_CONSTRAINTS} : Enables the computation of Hamiltonian and momentum constraints
		\item \texttt{DENDRO\_CONSEC\_COMM\_SELECT}  : If \texttt{ON} sub-communicators are selected from consecrative global ranks, otherwise sub-communicators are selected complete binary tree of global ranks (note that in this case global communicator size need to a power of 2).
		\item \texttt{DENDRO\_ENABLE\_VTU\_CONSTRAINT\_OUT} : Enables constraint variable output \linebreak while time-stepping 
		\item \texttt{DENDRO\_ENABLE\_VTU\_OUTPUT} : Enables evolution variable output while time-stepping
		\item \texttt{DENDRO\_VTK\_BINARY} : If \texttt{ON} vtu files are written in binary format, else ASCII format (binary format recommended).
		\item \texttt{DENDRO\_VTK\_ZLIB\_COMPRES} : If \texttt{ON} binary format is compressed (only effective if \texttt{DENDRO\_VTK\_BINARY} is \texttt{ON})
		\item \texttt{HILBERT\_ORDERING} : Hilbert SFC used if \texttt{ON}, otherwise Morton curve is used. (Hilbert curve is recommended to reduce the communication cost.)
		\item \texttt{NUM\_NPES\_THRESHOLD} : When running in large scale set this to $\sqrt{p}$ where $p$ number of mpi tasks for better performance.
		\item \texttt{RK\_SOLVER\_OVERLAP\_COMM\_AND\_COM} : If \texttt{ON} non blocking communication is used and enable overlapping of communication and computation \textit{unzip} (recommended option), otherwise blocking synchronized \textit{unzip} is used.  	
	\end{itemize}
	
	After configuring \dendro, generate the Makefile (use \texttt{c} to configure and \texttt{g} to generate). Then execute \texttt{make all} to build all the targets.  On completion, \bsolver~will be the main executable as related to this paper. 
	
	\subsection{Getting Started: Running \bsolver}
	
	\bsolver\  can be run as follows. 
	\begin{lstlisting}[language=bash]
	$ mpirun -np <number of mpi tasks>\
	./bssnSolver \
	<parameter file name>.par
	\end{lstlisting}
	Example parameter files can be found in \texttt{BSSN\_GR/pars/}. The following 
	is an example parameter file for equal mass ratio binary inspirals. 
	\begin{lstlisting}[basicstyle=\tiny]
	{
	"DENDRO_VERSION": 5.0,
	"BSSN_RESTORE_SOLVER":0,
	"BSSN_IO_OUTPUT_FREQ": 10,
	"BSSN_REMESH_TEST_FREQ": 5,
	"BSSN_CHECKPT_FREQ": 50,
	"BSSN_VTU_FILE_PREFIX": "bssn_gr",
	"BSSN_CHKPT_FILE_PREFIX": "bssn_cp",
	"BSSN_PROFILE_FILE_PREFIX": "bssn_r1",
	"BSSN_DENDRO_GRAIN_SZ": 100,
	"BSSN_ASYNC_COMM_K": 4,
	"BSSN_DENDRO_AMR_FAC": 1e0,
	"BSSN_WAVELET_TOL": 1e-4,
	"BSSN_LOAD_IMB_TOL": 1e-1,
	"BSSN_RK_TIME_BEGIN": 0,
	"BSSN_RK_TIME_END": 1000,
	"BSSN_RK_TIME_STEP_SIZE": 0.01,
	"BSSN_DIM": 3,
	"BSSN_MAXDEPTH": 12,
	"ETA_CONST": 2.0,
	"ETA_R0": 30.0,
	"ETA_DAMPING": 1.0,
	"ETA_DAMPING_EXP": 1.0,
	"BSSN_LAMBDA": {
	"BSSN_LAMBDA_1": 1,
	"BSSN_LAMBDA_2": 1,
	"BSSN_LAMBDA_3": 1,
	"BSSN_LAMBDA_4": 1
	},
	"BSSN_LAMBDA_F": {
	"BSSN_LAMBDA_F0": 1.0,
	"BSSN_LAMBDA_F1": 0.0
	},
	"CHI_FLOOR": 1e-4,
	"BSSN_TRK0": 0.0,
	"KO_DISS_SIGMA": 1e-1,
	"BSSN_BH1": {
	"MASS":0.48528137423856954,
	"X": 4.00000000e+00,
	"Y":0.0,
	"Z": 1.41421356e-05,
	"V_X": -0.00132697,
	"V_Y": 0.1123844,
	"V_Z": 0,
	"SPIN": 0,
	"SPIN_THETA":0,
	"SPIN_PHI": 0
	},
	"BSSN_BH2": {
	"MASS":0.48528137423856954,
	"X":-4.00000000e+00,
	"Y":0.0,
	"Z":1.41421356e-05,
	"V_X": 0.00132697,
	"V_Y": -0.1123844,
	"V_Z": 0,
	"SPIN": 0,
	"SPIN_THETA":0,
	"SPIN_PHI": 0
	}
	}
	\end{lstlisting}
	Here we list the key options for \bsolver\ with a short description.
	\begin{itemize}
		\item \texttt{BSSN\_RESTORE\_SOLVER} : Set $1$ to restore $RK$ solver from latest checkpoint. 
		\item \texttt{BSSN\_IO\_OUTPUT\_FREQ} : IO (i.e. \texttt{vtu} files) output frequency
		\item \texttt{BSSN\_CHECKPT\_FREQ} : Checkpoint file output frequency
		\item \texttt{BSSN\_REMESH\_TEST\_FREQ} : Remesh test frequency (i.e. frequency in time steps that is being tested for re-meshing) 
		\item \texttt{BSSN\_DENDRO\_GRAIN\_SZ} : Number of octants per core 
		\item \texttt{BSSN\_ASYNC\_COMM\_K} : Number of variables that are being processed during an asynchronous \textit{unzip} ($<24$)
		\item \texttt{BSSN\_DENDRO\_AMR\_FAC} : Safety factor for coarsening i.e. coarsen if and only if $W_c \leq AMR\_FAC \times WAVELET\_TOL$ where $W_c$ is the computed wavelet coefficient.  
		\item \texttt{BSSN\_WAVELET\_TOL} : Wavelet tolerance for WAMR. 
		\item \texttt{BSSN\_MAXDEPTH} : Maximum level of refinement allowed ($\leq 30$)
		\item \texttt{KO\_DISS\_SIGMA} : Kreiss-Oliger dissipation factor for \BSSN~formulation
		\item \texttt{MASS} : Mass of the black hole
		\item \texttt{X} : $x$ coordinate of the black hole
		\item \texttt{Y} : $y$ coordinate of the black hole
		\item \texttt{Z} : $z$ coordinate of the black hole
		\item \texttt{V\_X} : momentum of the black hole in $x$ direction
		\item \texttt{V\_Y} : momentum of the black hole in $y$ direction
		\item \texttt{V\_Z} : momentum of the black hole in $z$ direction
		\item \texttt{SPIN} : magnitude of the spin of the black hole
		\item \texttt{SPIN\_THETA} : magnitude of the spin of the black hole along $\theta$
		\item \texttt{SPIN\_PHI} : magnitude of the spin of the black hole along $\phi$
	\end{itemize}
	
	\subsubsection{Generating your own parameters}
	The intial data parameters for a binary black holes ~\cite{Tichy:2010qa} depend
	on the total mass ($M=m1+m2$), the mass ratio $q$ and the separation distance $d$. These parameters are calculated  using the Python script \texttt{BSSN\_GR/scripts/id.py}. The command to generate parameters for $q=10$, total mass $M=5$ and separation $d=16$ is 
	\begin{lstlisting}
	$ python3 id.py -M 5 -r 10 16
	\end{lstlisting}
	\begin{lstlisting}[basicstyle=\tiny]
	----------------------------------------------------------------
	PUNCTURE PARAMETERS (par file foramt)
	----------------------------------------------------------------
	"BSSN_BH1": {
	"MASS":4.489529,
	"X":1.454545,
	"Y":0.000000,
	"Z": 0.000014,
	"V_X": -0.020297,
	"V_Y": 0.423380,
	"V_Z": 0.000000,
	"SPIN": 0.000000,
	"SPIN_THETA":0.000000,
	"SPIN_PHI": 0.000000
	},
	"BSSN_BH2": {
	"MASS":0.398620,
	"X":-14.545455,
	"Y":0.000000,
	"Z":0.000014,
	"V_X": 0.020297,
	"V_Y": -0.423380,
	"V_Z": 0.000000,
	"SPIN": 0.000000,
	"SPIN_THETA":0.000000,
	"SPIN_PHI": 0.000000
	}
	The tangential momentum is just an estimate, and the value for a
	for a circular orbit is likely between (0.5472794147860968, 0.29947988193805547)
	\end{lstlisting}
	
	\subsection{Symbolic interface and code generation}
	The \BSSN~formulation is a decomposition of the Einstein equations into 
	24 coupled hyperbolic PDEs. Writing the computation code for the 
	\BSSN~formulation can be a tedious task. Hence we have written a symbolic 
	Python interface to generate optimized C code to compute the \BSSN~equations. 
	All the symbolic utilities necessary to write the \BSSN~formulation in symbolic 
	Python can be found in \texttt{GR/rhs\_scripts/bssn/dendro.py} and the symbolic \BSSN~code can 
	be found in \texttt{GR/rhs\_scripts/bssn/bssn.py}. 
	This could be modified for more advanced uses of the code such as including
	new equations to describe additional physics or for introducing a different 
	formulation of the Einstein equations.
	
	\subsection{Profiling the code}
	\dendro\ contains built-in profiler code which enables one to profile the 
	code extensively. On configuration, a user can enable/disable 
	the internal profiling flags using \texttt{ENABLE\_DENDRO\_PROFILE\_COUNTERS} 
	and the profile output can be changed between a human readable version and a 
	tab separated format using the flag \texttt{BSSN\_PROFILE\_HUMAN\_READABLE}.
	Note that in order to profile communication, internal profile flags need to 
	be enabled. The following is an example of profiling output for the first 
	10 time steps. 
	\begin{lstlisting}[basicstyle=\tiny]
	active npes : 16
	global npes : 16
	current step : 10
	partition tol : 0.1
	wavelet tol : 0.0001
	maxdepth : 12
	Elements : 4656
	DOF(zip) : 279521
	DOF(unzip) : 2078609
	============ MESH =================
	step                 min(#)    mean(#)    max(#)                        
	ghost Elements       634       824.062    1065                          
	local Elements       263       291        319                           
	ghost Nodes          43781     55671.7    71693                         
	local Nodes          14292     17470.1    20705                         
	send Nodes           18760     24872.9    36861                         
	recv Nodes           18113     24872.9    33777                         
	========== RUNTIME =================
	step                 min(s)    mean(s)    max(s)                        
	++2:1 balance        0         0          0                             
	++mesh               1.9753    1.98299    1.98946                       
	++rkstep             20.159    20.1856    20.1996                       
	++ghostExchge.       1.81442   3.15703    4.49568                       
	++unzip_sync         8.27839   9.67293    11.0991                       
	++unzip_async        0         0          0                             
	++isReMesh           0.04642   0.117357   0.207305                      
	++gridTransfer       1.53709   1.54899    1.56531                       
	++deriv              1.98942   2.34851    2.76695                       
	++compute_rhs        4.00119   4.61547    5.11566                       
	--compute_rhs_a      0.0137962 0.0245449  0.0351532                     
	--compute_rhs_b      0.0296426 0.0503471  0.069537                      
	--compute_rhs_gt     0.111898  0.12846    0.15463                       
	--compute_rhs_chi    0.0170642 0.0315392  0.044856                      
	--compute_rhs_At     2.40738   2.72922    3.05622                       
	--compute_rhs_K      0.358215  0.39879    0.457139                      
	--compute_rhs_Gt     0.774211  0.933581   1.05702                       
	--compute_rhs_B      0.0575426 0.071209   0.0855094                     
	++boundary con       0         0.0421986  0.134712                      
	++zip                0.23529   0.260862   0.291513                      
	++vtu                0.0872362 0.101362   0.128246                      
	++checkpoint         3.27e-06  3.85e-06   5.7469e-06     
	\end{lstlisting}
	
	\subsection{Visualizing the data}
	\dendro\  can be configured to output parallel unstructured grid files in 
	binary file format (\texttt{.pvtu}).   
	These files can be visualized using any visualization tool which supports 
	VTK file formats. All the images presented in this paper used Paraview 
	due to its robustness and scalability. Paraview allows Python based 
	scripting to perform \texttt{pvbatch} visualization, an example pvpython 
	script can be found in \texttt{scripts/bssnVis.py}

	\subsection{\BSSN: Verification Tests}
	\label{sec:AE}

	\def\TT{{\rm time}}
	\def\SS{{\rm step}}
	
	In this section, we present experimental evaluations that we performed to ensure the accuracy of the simulation code. 
	
	\subsection{Accuracy of stencil operators}
	
	In order to test the accuracy and convergence of WAMR and the derivative stencils, we used a known function to generate adaptive octree grid based on the wavelet expansion. 
	Then we compute numerical derivatives using finite difference stencils which are compared against the analytical derivatives of $f(x,y,z)$. $l_2$ and $l_{\infty}$ norms of the comparison is given in the Table \ref{tb:fd}. 
	
	\begin{table}
		\centering
		\scalebox{0.9}{
			\begin{tabular}{||c | c |c | c||} 
				\hline
				derivative & grid points & $\norm{.}_2$ & $\norm{.}_\infty$ \\
				\hline\hline
				$\partial_x$ & 4913 & 0.0201773 & 0.00144632 \\
				$\partial_x$ & 99221 & 0.000849063 & 2.74672e-05 \\
				\hline
			\end{tabular}
		}
		\caption{Normed difference in numerical derivative and analytical derivative evaluated at grid points for the function $f(x,y,z)=sin(2\pi x)sin(2\pi y)sin(2\pi z))$
			where in both cases wavelet tolerance 
			of $10^{-8}$ but increasing maximum level of refinement (i.e. \texttt{maxDepth}) from $4$ to $6$. Note that when \texttt{maxDepth} increases number of grid points increase hence normed difference between numerical and analytical derivatives goes down significantly. }
		\label{tb:fd}
	\end{table}

	\subsection{Accuracy of symbolic interface and code generation}
	
	Given the complexity of the \BSSN\ equations, writing code to
	evaluate these equations can be an error-prone and tedious task.
	For example, to evaluate the equations we need to calculate more than 300
	finite derivatives. Hence \dendro\ provides a symbolic framework
	written in \texttt{SymPy} for automatically generating C++ code for
	the equations. The user writes equations in a high-level representation 
	that more closely resembles their symbolic form. 
	We can then use the computational graph
	of the equations to generate optimized C++ code. The accuracy
	of the symbolic framework and code generation is certified by comparing 
	results from the generated C++ code to those from 
	the \HAD\ code, an established and tested code for numerical relativity.
	
	This table shows a comparison of \BSSN\ equations (i.e. all 24 equations),
	evaluated over a grid of $128^3$ points by arbitrary, non-zero functions
	by both the \dendro\ and \HAD\ codes. All spatial derivatives in the equations
	are evaluated using finite differences, for both the
	\dendro\ and \HAD\ codes.
	The table reports the $L_2$ norm of the difference in the equations
	as evaluated in both codes, 
	as well as the $L_2$ norms of the functions used to evaluate the equations.
	Equations with residual norms of order $10^{-15}$ are clearly at machine zero, 
	but this low level is reached for only the simplest equations. Residuals with
	norms of order $10^{-12}$ arise in complicated equations, where finite precision  
	errors can accumulate in hundreds of floating point operations. The optimized
	equations require about 4500 floating point operations to evaluate at a single point.
	
	\begin{lstlisting}[basicstyle=\tiny]
	L2 Norms differences in the HAD and DENDRO equations on 128^3 points.
	------------------------------------------------------------------------
	|| diff 0  || = 1.18413e-15, || rhs HAD || = 1.28114, || rhs DENDRO || = 1.28114
	|| diff 1  || = 7.53412e-16, || rhs HAD || = 2.18877, || rhs DENDRO || = 2.18877
	|| diff 2  || = 4.98271e-16, || rhs HAD || = 1.66315, || rhs DENDRO || = 1.66315
	|| diff 3  || = 1.03346e-15, || rhs HAD || = 0.720477,|| rhs DENDRO || = 0.720477
	|| diff 4  || = 5.82489e-16, || rhs HAD || = 1.40142, || rhs DENDRO || = 1.40142
	|| diff 5  || = 4.93128e-16, || rhs HAD || = 0.797567,|| rhs DENDRO || = 0.797567
	|| diff 6  || = 1.94194e-11, || rhs HAD || = 19.0107, || rhs DENDRO || = 19.0107
	|| diff 7  || = 2.14958e-11, || rhs HAD || = 19.5221, || rhs DENDRO || = 19.5221
	|| diff 8  || = 4.0673e-12,  || rhs HAD || = 8.96364, || rhs DENDRO || = 8.96364
	|| diff 9  || = 1.58532e-11, || rhs HAD || = 10.1459, || rhs DENDRO || = 10.1459
	|| diff 10 || = 4.31184e-12, || rhs HAD || = 8.70053, || rhs DENDRO || = 8.70053
	|| diff 11 || = 4.95696e-12, || rhs HAD || = 10.8644, || rhs DENDRO || = 10.8644
	|| diff 12 || = 4.27211e-15, || rhs HAD || = 19.3546, || rhs DENDRO || = 19.3546
	|| diff 13 || = 2.29542e-10, || rhs HAD || = 93.4829, || rhs DENDRO || = 93.4829
	|| diff 14 || = 1.7484e-11,  || rhs HAD || = 40.0804, || rhs DENDRO || = 40.0804
	|| diff 15 || = 4.79216e-11, || rhs HAD || = 30.9279, || rhs DENDRO || = 30.9279
	|| diff 16 || = 2.03434e-11, || rhs HAD || = 26.0603, || rhs DENDRO || = 26.0603
	|| diff 17 || = 1.07479e-15, || rhs HAD || = 15.5765, || rhs DENDRO || = 15.5765
	|| diff 18 || = 3.00151e-15, || rhs HAD || = 32.9891, || rhs DENDRO || = 32.9891
	|| diff 19 || = 7.79103e-16, || rhs HAD || = 8.10107, || rhs DENDRO || = 8.10107
	|| diff 20 || = 7.77113e-16, || rhs HAD || = 9.01369, || rhs DENDRO || = 9.01369
	|| diff 21 || = 1.74839e-11, || rhs HAD || = 46.3141, || rhs DENDRO || = 46.3141
	|| diff 22 || = 4.79217e-11, || rhs HAD || = 32.7981, || rhs DENDRO || = 32.7981
	|| diff 23 || = 2.03434e-11, || rhs HAD || = 32.7256, || rhs DENDRO || = 32.7256
	\end{lstlisting}
	
	\subsection{Single black hole}
	\label{sec:AE_sbh}
	
	Prior to simulating binary inspirals, we perform simple experiments
	with a single black hole to ensure the accuracy of the simulation code. 
	While a single black hole is a stable, static solution of the Einstein 
	equations, although there is some transient time dependence with:w
	our particular 
	coordinate conditions.
	The black hole
	parameters (parameter file can be found in
	\texttt{SC18\_AE/par/single\_bh1.par} in the repository) for this
	test is given below. Note that to generate initial data for a single black
	hole, we place one of the black holes in the binary far from the computational
	domain and set its mass to zero.
	
	\begin{lstlisting}[basicstyle=\small]
	"BSSN_BH1": { "MASS":1.0, "X":0.0, "Y":0.0, 
	"Z": 0.00123e-6, "V_X": 0.0, "V_Y": 0.0, 
	"V_Z": 0.0, "SPIN": 0, 
	"SPIN_THETA":0, "SPIN_PHI": 0 },
	
	"BSSN_BH2": { "MASS":1e-15,"X":1e15, "Y":0.0, 
	"Z":0.00123e-6, "V_X": 0.0, "V_Y": 0.0, 
	"V_Z": 0.0, "SPIN": 0, 
	"SPIN_THETA":0, "SPIN_PHI": 0 }
	\end{lstlisting}

	\subsection{Boosted Single Black Hole}
	\label{sec:AE_sbhboost}
	
	The next experiment is an extension of the single BH test;
	it ``boosts'' the BH with constant velocity in $x$-direction.
	The constant velocity of the BH should be apparent in the evolution. 
	The parameter file for this test can be found in the repository at 
	\texttt{SC18\_AE/par/single\_bh1\_boost.par}.
	The black hole parameters are given below (note
	that $BSSN\_BH1$ has a momentum of 0.114 in $x$-direction).
	
	\begin{lstlisting}[basicstyle=\small]
	"BSSN_BH1": { "MASS":1.0, "X":0.0, "Y":0.0, 
	"Z": 0.00123e-6, "V_X": 0.114, "V_Y": 0.0, 
	"V_Z": 0.0, "SPIN": 0, 
	"SPIN_THETA":0, "SPIN_PHI": 0 },
	
	"BSSN_BH2": { "MASS":1e-15,"X":1e15, "Y":0.0, 
	"Z":0.00123e-6, "V_X": 0.0, "V_Y": 0.0, 
	"V_Z": 0.0, "SPIN": 0, 
	"SPIN_THETA":0, "SPIN_PHI": 0 }
	\end{lstlisting}

	\begin{figure}[H]
		\begin{subfigure}{0.32\linewidth}
			\includegraphics[width=0.8\textwidth]{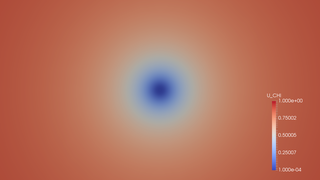}
			\caption{\small $\SS=0,\ \TT=0~M$}
		\end{subfigure}
		\begin{subfigure}{0.32\linewidth}
			\includegraphics[width=0.8\textwidth]{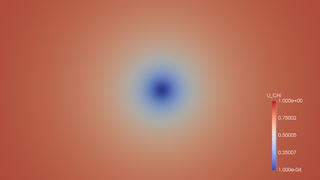}
			\caption{\small$\SS=500,\ \TT=21.97~M$}
		\end{subfigure}
		\begin{subfigure}{0.32\linewidth}
			\includegraphics[width=0.8\textwidth]{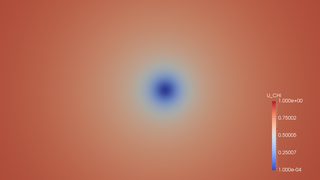}
			\caption{\small$\SS=1000,\ \TT=43.94~M$}
		\end{subfigure} \hfill
		
		\begin{subfigure}{0.32\linewidth}
			\includegraphics[width=0.8\textwidth]{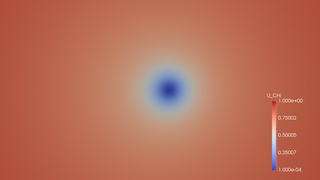}
			\caption{\small$\SS=1500,\ \TT=65.91~M$}
		\end{subfigure}
		\begin{subfigure}{0.32\linewidth}
			\includegraphics[width=0.8\textwidth]{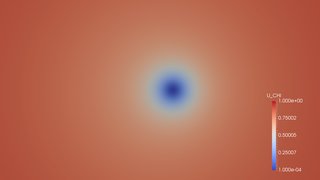}
			\caption{\small$\SS=2000,\ \TT=87.88~M$}
		\end{subfigure}
		\begin{subfigure}{0.32\linewidth}
			\includegraphics[width=0.8\textwidth]{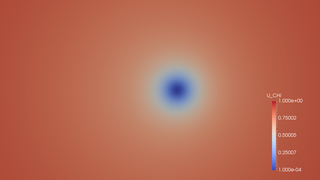}
			\caption{\small$\SS=2500,\ \TT=109.85~M$}
		\end{subfigure} \hfill
		
		\begin{subfigure}{0.32\linewidth}
			\includegraphics[width=0.8\textwidth]{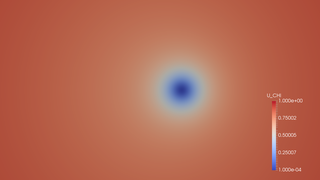}
			\caption{\small$\SS=3000,\ \TT=131.82~M$}
		\end{subfigure}
		\begin{subfigure}{0.32\linewidth}
			\includegraphics[width=0.8\textwidth]{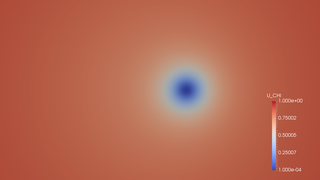}
			\caption{\small$\SS=3500,\ \TT=153.79~M$}
		\end{subfigure}
		\begin{subfigure}{0.32\linewidth}
			\includegraphics[width=0.8\textwidth]{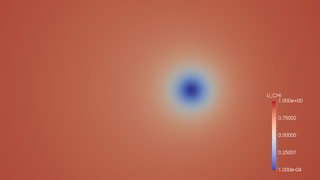}
			\caption{\small$\SS=4000,\ \TT=175.76~M$}
		\end{subfigure} \hfill

		\begin{subfigure}{0.32\linewidth}
			\includegraphics[width=0.8\textwidth]{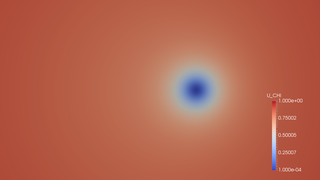}
			\caption{\small$\SS=4500,\ \TT=197.73~M$}
		\end{subfigure}
		\begin{subfigure}{0.32\linewidth}
			\includegraphics[width=0.8\textwidth]{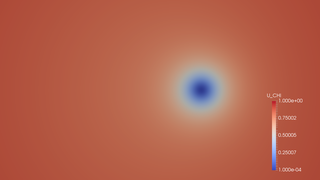}
			\caption{\small$\SS=5000,\ \TT=219.70~M$}
		\end{subfigure}
		\begin{subfigure}{0.32\linewidth}
			\includegraphics[width=0.8\textwidth]{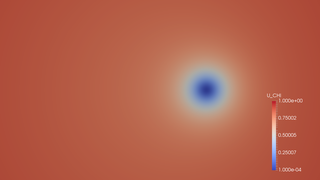}
			\caption{$\SS=5500,\ \TT=241.67~M$}
		\end{subfigure} \hfill
		
		\begin{subfigure}{0.32\linewidth}
			\includegraphics[width=0.8\textwidth]{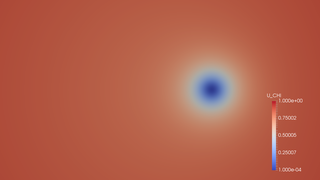}
			\caption{\small$\SS=6000,\ \TT=263.64~M$}
		\end{subfigure}
		
		\begin{subfigure}{0.32\linewidth}
			\includegraphics[width=0.8\textwidth]{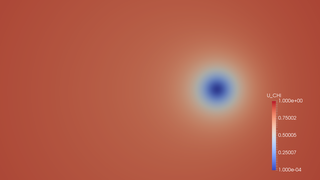}
			\caption{\small$\SS=6500,\ \TT=285.61~M$}
		\end{subfigure} \hfill
		
		\caption{A single black hole boosted in the $x$-direction, with \maxDepth=12 and wavelet tolerance of $10^{-3}$. Time is given in terms of the black hole
			mass, $M$. \label{fig:sbh_boost}}
	\end{figure}

	\subsection{Constraint equations}
	
	Similar to the Maxwell equations of electrodynamics, the Einstein equations
	contain both hyperbolic evolution equations and elliptic constraint equations,
	which must be satisfied at all times. Following the common practice in
	numerical relativity, we evolve the hyperbolic equations and monitor the 
	quality of the solution by checking that the constraint equations are 
	satisfied. The choice of coordinates for the BBH evolution (the puncture gauge) 
	does induce constraint violations in the vicinity of each black hole.
	The violations of the constraint equations in our runs are consistent with 
	the discretization error expected for the numerical derivatives in the
	constraint equations and the constraint violations near the black holes 
	(punctures). An example of monitored constraint violations are listed in the Table \ref{tb:constraints}.
	
	\begin{table}
		\centering
		\scalebox{0.9}{
			\begin{tabular}{||c | c | c | c| c||} 
				\hline
				Time (M) & $\norm{\mathcal{H}_{r>a}}_2$ & $\norm{M1_{r>a}}_2$ & $\norm{M2_{r>a}}_2$ & $\norm{M3_{r>a}}_2$ \\
				\hline\hline
				0 & 0.000777861 & 1.01855e-05 & 1.23443e-05 & 7.17572e-06 \\
				0.976562 & 0.000808294 & 3.05681e-05 & 2.91217e-05 & 2.79631e-05 \\
				1.95312 & 0.000793783 & 5.03912e-05 & 3.93872e-05 & 4.06693e-05 \\
				2.92969 & 0.00079551 & 7.54643e-05 & 5.068e-05 & 5.42466e-05 \\
				3.90625	& 0.000956987 & 0.000102901 & 7.38208e-05 & 7.47156e-05 \\
				4.88281 & 0.00247348 & 0.000200055	& 0.000140583 & 0.000139008 \\
				\hline
			\end{tabular}
		}
		\caption{Violation of constraint equations with time for an equal mass ratio binary merger simulation done using OT. Note that $\mathcal{H}$, $M1,M2,M3$ denotes the Hamiltonian and 3 momentum component constraints that is being monitored through the evolution.}
		\label{tb:constraints}
	\end{table}

	\subsection{Binary black holes with mass ratio $q=1$}
	
\begin{figure}
		\begin{subfigure}{0.32\linewidth}
			\includegraphics[width=0.8\textwidth]{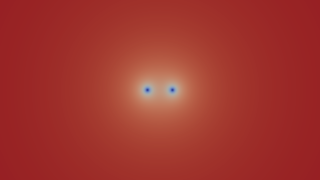}
			\caption{$\SS=0,\ \TT=0~M$}
		\end{subfigure}
		\begin{subfigure}{0.32\linewidth}
			\includegraphics[width=0.8\textwidth]{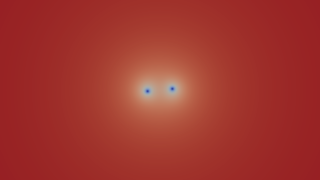}
			\caption{$\SS=500,\ \TT=21.97~M$}
		\end{subfigure}
		\begin{subfigure}{0.32\linewidth}
			\includegraphics[width=0.8\textwidth]{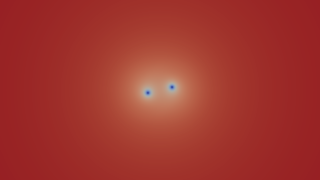}
			\caption{$\SS=1000,\ \TT=43.94~M$}
		\end{subfigure} \hfill
		
		\begin{subfigure}{0.32\linewidth}
			\includegraphics[width=0.8\textwidth]{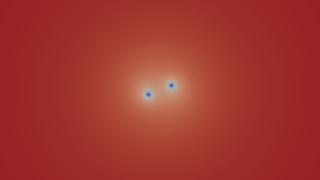}
			\caption{$\SS=1500,\ \TT=65.91~M$}
		\end{subfigure}
		\begin{subfigure}{0.32\linewidth}
			\includegraphics[width=0.8\textwidth]{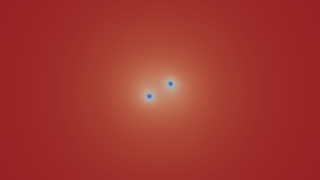}
			\caption{$\SS=2000,\ \TT=87.88~M$}
		\end{subfigure}
		\begin{subfigure}{0.32\linewidth}
			\includegraphics[width=0.8\textwidth]{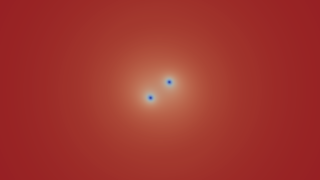}
			\caption{$\SS=2500,\ \TT=109.85~M$}
		\end{subfigure} \hfill
		
		\begin{subfigure}{0.32\linewidth}
			\includegraphics[width=0.8\textwidth]{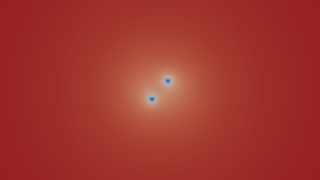}
			\caption{$\SS=3000,\ \TT=131.82~M$}
		\end{subfigure}
		\begin{subfigure}{0.32\linewidth}
			\includegraphics[width=0.8\textwidth]{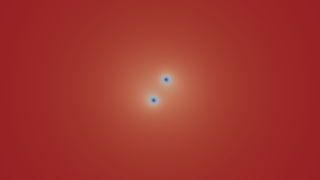}
			\caption{$\SS=3500,\ \TT=153.79~M$}
		\end{subfigure}
		\begin{subfigure}{0.32\linewidth}
			\includegraphics[width=0.8\textwidth]{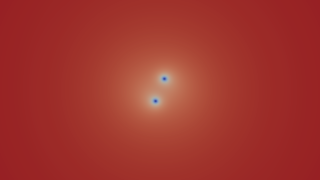}
			\caption{$\SS=4000,\ \TT=175.76~M$}
		\end{subfigure} \hfill

		\begin{subfigure}{0.32\linewidth}
			\includegraphics[width=0.8\textwidth]{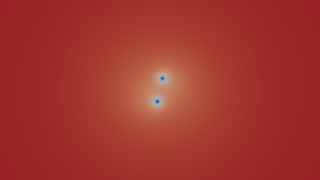}
			\caption{$\SS=4500,\ \TT=197.73~M$}
		\end{subfigure}
		\begin{subfigure}{0.32\linewidth}
			\includegraphics[width=0.8\textwidth]{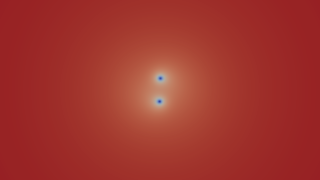}
			\caption{$\SS=5000,\ \TT=219.70~M$}
		\end{subfigure}
		\begin{subfigure}{0.32\linewidth}
			\includegraphics[width=0.8\textwidth]{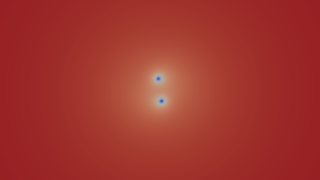}
			\caption{$\SS=5500,\ \TT=241.67~M$}
		\end{subfigure} \hfill
		
		\caption{This figure shows frames from the evolution of a black hole
			binary with an equal mass ratio, $q=1$. Time is measured in terms of the total
			black hole mass $M$.\label{fig:r1}	}
	\end{figure}

	We performed a series of short-term binary BH evolutions with different
	mass ratios. This run was used to validate the code by comparing the
	trajectories of the BHs calculated using \dendro\ to the trajectories calculated
	by \HAD. Frames from the evolution are shown in the Figure~\ref{fig:r1},
	and the BH parameters used for this run are listed below.
	
	\begin{lstlisting}[basicstyle=\small]
	"BSSN_BH1": {
	"MASS":0.485,
	"X": 4.00e+00, "Y":0.0, "Z": 1.41-05,
	"V_X": -0.00133, "V_Y": 0.112, "V_Z": 0,
	"SPIN": 0, "SPIN_THETA":0, "SPIN_PHI": 0 },
	"BSSN_BH2": {
	"MASS":0.485,
	"X":-4.00+00, "Y":0.0, "Z":1.41-05,
	"V_X": 0.00132, "V_Y": -0.112, "V_Z": 0,
	"SPIN": 0, "SPIN_THETA":0, "SPIN_PHI": 0 }
	\end{lstlisting}

	
	\begin{figure}[H]
		\centering
		\includegraphics[width=0.9\textwidth]{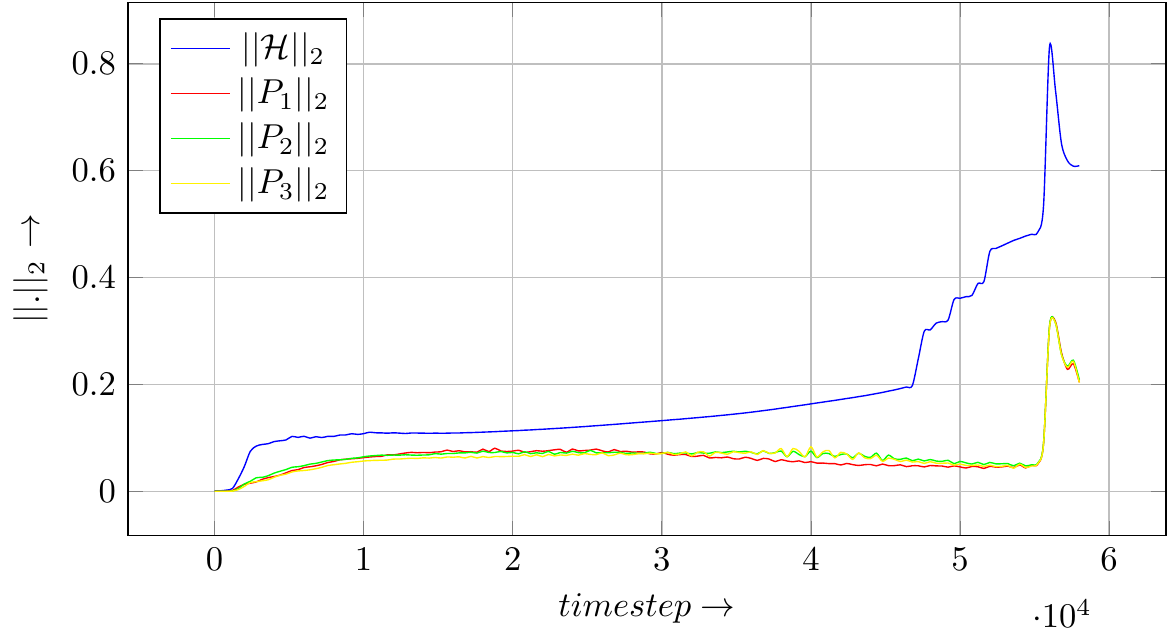}
		\caption{$l_2$ norm of the Hamiltonian and momentum constraint violation as hyperbolic equations are evolved in time for equal mass ratio binary black hole configuration test run performed using \dendrogr. Note that the final spike in constraint violation is happens during the merging even of binary black holes.}
		\label{fig::r1:constraints}
	\end{figure}

	\subsection{Binary black holes with mass ratio $q=10$}
	
	We performed a short simulation with a mass ratio $q=10$
	This is a short demonstration run to show that \dendro\
	easily handles large mass ratios and gives consistent
	results for the binary evolution.
	Frames from the evolution are shown in figure,
	and the BH parameters used for this run are listed below.
	
	\begin{lstlisting}[basicstyle=\small]
	"BSSN_BH1": {
	"MASS":0.903,
	"X":5.45-01, "Y":0.0, "Z": 1.41-05,
	"V_X": -3.90e-04, "V_Y": 0.0470, "V_Z": 0,
	"SPIN": 0, "SPIN_THETA":0, "SPIN_PHI": 0 },
	"BSSN_BH2": {
	"MASS":0.0845,
	"X":-5.45+00, "Y":0.0, "Z":1.41-05,
	"V_X": 3.90e-04, "V_Y": -0.0470, "V_Z": 0,
	"SPIN": 0, "SPIN_THETA":0, "SPIN_PHI": 0 }
	\end{lstlisting}

	\begin{figure}
		\begin{subfigure}{0.32\linewidth}
			\includegraphics[width=0.8\textwidth]{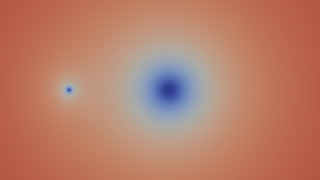}
			\caption{$\SS=0,\ \TT=0~M$}
		\end{subfigure}
		\begin{subfigure}{0.32\linewidth}
			\includegraphics[width=0.8\textwidth]{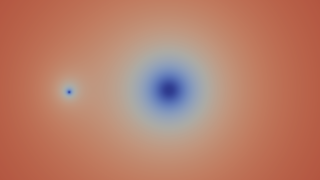}
			\caption{$\SS=500,\ \TT=21.97~M$}
		\end{subfigure}
		\begin{subfigure}{0.32\linewidth}
			\includegraphics[width=0.8\textwidth]{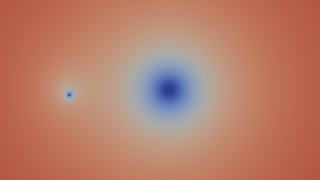}
			\caption{$\SS=1000,\ \TT=43.94~M$}
		\end{subfigure} \hfill
		
		\begin{subfigure}{0.32\linewidth}
			\includegraphics[width=0.8\textwidth]{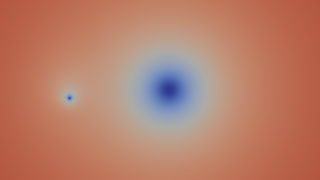}
			\caption{$\SS=1500,\ \TT=65.91~M$}
		\end{subfigure}
		\begin{subfigure}{0.32\linewidth}
			\includegraphics[width=0.8\textwidth]{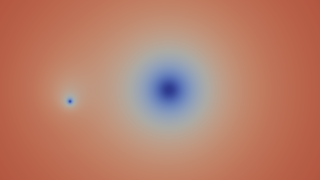}
			\caption{$\SS=2000,\ \TT=87.88~M$}
		\end{subfigure}
		\begin{subfigure}{0.32\linewidth}
			\includegraphics[width=0.8\textwidth]{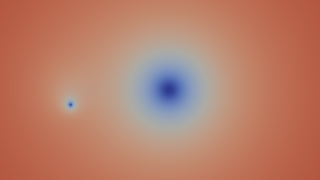}
			\caption{$\SS=2500,\ \TT=109.85~M$}
		\end{subfigure} \hfill
		
		\begin{subfigure}{0.32\linewidth}
			\includegraphics[width=0.8\textwidth]{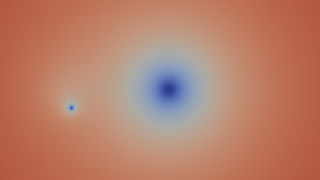}
			\caption{$\SS=3000,\ \TT=131.82~M$}
		\end{subfigure}
		\begin{subfigure}{0.32\linewidth}
			\includegraphics[width=0.8\textwidth]{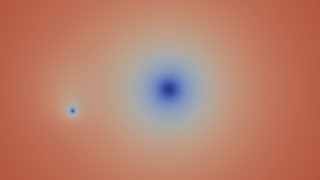}
			\caption{$\SS=3500,\ \TT=153.79~M$}
		\end{subfigure}
		\begin{subfigure}{0.32\linewidth}
			\includegraphics[width=0.8\textwidth]{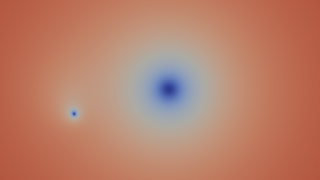}
			\caption{$\SS=4000,\ \TT=175.76~M$}
		\end{subfigure} \hfill

		\begin{subfigure}{0.32\linewidth}
			\includegraphics[width=0.8\textwidth]{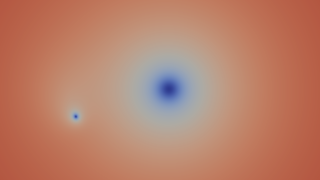}
			\caption{$\SS=4500,\ \TT=197.73~M$}
		\end{subfigure}
		\begin{subfigure}{0.32\linewidth}
			\includegraphics[width=0.8\textwidth]{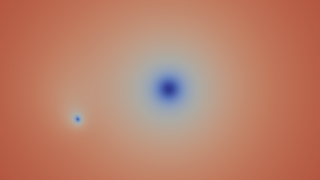}
			\caption{$\SS=5000,\ \TT=219.70~M$}
		\end{subfigure}
		\begin{subfigure}{0.32\linewidth}
			\includegraphics[width=0.8\textwidth]{figs/AE/r10/img_slice_000100.png}
			\caption{$\SS=5500,\ \TT=241.67~M$}
		\end{subfigure} \hfill

		\caption{This figure shows frames from the evolution of a black hole
			binary with mass ratio $q=10$. Time is measured in terms of the total
			black hole mass $M$.\label{fig:r10}}
	\end{figure}

	\subsection{Binary black holes with mass ratio $q=100$}
	
	We performed a short simulation with a mass ratio $q=100$
	This is a short demonstration run to show that \dendro\ 
	produces the proper grid structure for this system and
	reasonable results for a very challenging binary configuration.
	Frames from the evolution are shown in the figure
	and the BH parameters used for this run are listed below.

	\begin{lstlisting}[basicstyle=\small]
	"BSSN_BH1": {
	"MASS":0.989,
	"X":5.94-02, "Y":0.0, "Z": 1.41-05,
	"V_X": -5.60-06, "V_Y": 5.61-03, "V_Z": 0,
	"SPIN": 0, "SPIN_THETA":0, "SPIN_PHI": 0 },
	"BSSN_BH2": {
	"MASS":0.00914,
	"X":-5.94+00, "Y":0.0, "Z":1.41421356e-05,
	"V_X": 5.60-06, "V_Y": -5.61-03, "V_Z": 0,
	"SPIN": 0, "SPIN_THETA":0, "SPIN_PHI": 0 }
	\end{lstlisting}

	\begin{figure}
		\begin{subfigure}{0.32\linewidth}
			\includegraphics[width=0.8\textwidth]{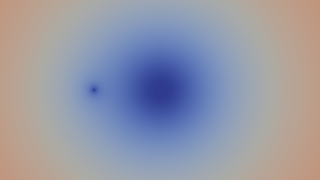}
			\caption{$\SS=0,\ \TT=0~M$}
		\end{subfigure}
		\begin{subfigure}{0.32\linewidth}
			\includegraphics[width=0.8\textwidth]{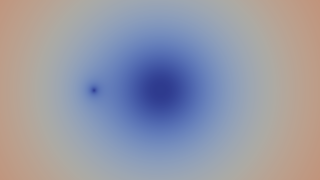}
			\caption{$\SS=500,\ \TT=21.97~M$}
		\end{subfigure}
		\begin{subfigure}{0.32\linewidth}
			\includegraphics[width=0.8\textwidth]{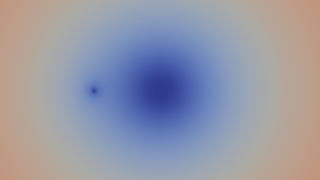}
			\caption{$\SS=1000,\ \TT=43.94~M$}
		\end{subfigure} \hfill
		
		\begin{subfigure}{0.32\linewidth}
			\includegraphics[width=0.8\textwidth]{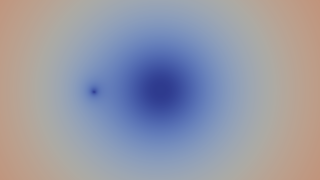}
			\caption{$\SS=1500,\ \TT=65.91~M$}
		\end{subfigure} 
		\begin{subfigure}{0.32\linewidth}
			\includegraphics[width=0.8\textwidth]{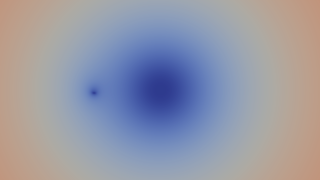}
			\caption{$\SS=2000,\ \TT=87.88~M$}
		\end{subfigure}
		\begin{subfigure}{0.32\linewidth}
			\includegraphics[width=0.8\textwidth]{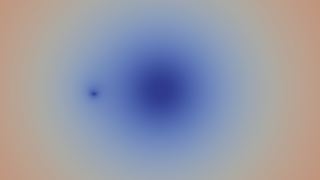}
			\caption{$\SS=2500,\ \TT=109.85~M$}
		\end{subfigure} \hfill
		
		\begin{subfigure}{0.32\linewidth}
			\includegraphics[width=0.8\textwidth]{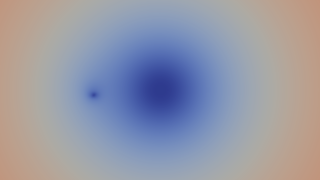}
			\caption{$\SS=3000,\ \TT=131.82~M$}
		\end{subfigure}
		\begin{subfigure}{0.32\linewidth}
			\includegraphics[width=0.8\textwidth]{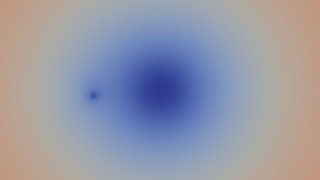}
			\caption{$\SS=3500,\ \TT=153.79~M$}
		\end{subfigure} 
		\begin{subfigure}{0.32\linewidth}
			\includegraphics[width=0.8\textwidth]{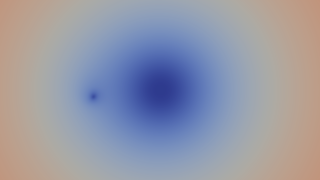}
			\caption{$\SS=4000,\ \TT=175.76~M$}
		\end{subfigure} \hfill
		
		\begin{subfigure}{0.32\linewidth}
			\includegraphics[width=0.8\textwidth]{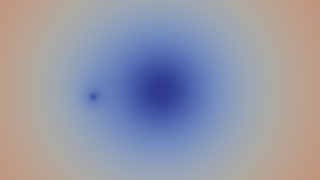}
			\caption{$\SS=4500,\ \TT=197.73~M$}
		\end{subfigure}
		\begin{subfigure}{0.32\linewidth}
			\includegraphics[width=0.8\textwidth]{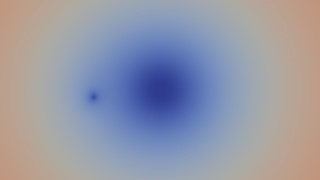}
			\caption{$\SS=5000,\ \TT=219.70~M$}
		\end{subfigure}
		\begin{subfigure}{0.32\linewidth}
			\includegraphics[width=0.8\textwidth]{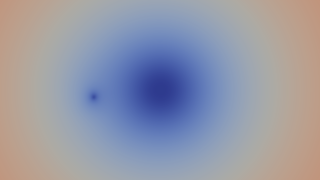}
			\caption{$\SS=5500,\ \TT=241.67~M$}
		\end{subfigure}\hfill
		
		\caption{This figure shows frames from the evolution of a black hole binary 
			with mass ratio $q=100$. Time is given
			in terms of the total mass $M$.\label{fig:r100}}
	\end{figure}

\end{document}